\documentclass[onecolumn,pre,floats,aps,amsmath,amssymb,nofootinbib]{revtex4-2}
\usepackage{comment}
\usepackage[utf8]{inputenc}
\usepackage{mathtools}
\usepackage{bbm}
\usepackage{cancel} 
\usepackage{dcolumn}
\usepackage{bm}
\usepackage{blindtext}
\usepackage{xcolor}
\usepackage{hyperref}
\usepackage{graphicx}
\usepackage{caption}
\usepackage{amsthm}
\usepackage{subfig}
\usepackage{bm}
\usepackage{verbatim}
\usepackage{microtype}
\usepackage{amsmath}
\usepackage{adjustbox}
\usepackage{siunitx}
\usepackage{physics}
\usepackage{amssymb}
\usepackage{slashed}
\newcommand{\be}{\begin{equation}}
\newcommand{\ee}{\end{equation}}
\newcommand{\beq}{\begin{equation}}
\newcommand{\eeq}{\end{equation}}
\newcommand{\bea}{\begin{eqnarray}}
\newcommand{\eea}{\end{eqnarray}}

\newcommand{\EM}{\mathrm{em}}

\newcommand{\bs}{\boldsymbol}

\newcommand{\dthree}{d^{\hspace{1pt}3}}
\newcommand{\dfour}{d^{\hspace{1.35pt}4}}

\newcommand{\Romatre}{Dipartimento di Matematica e Fisica, Universit\`a  Roma Tre and INFN, Sezione di Roma Tre,\\ Via della Vasca Navale 84, I-00146 Rome, Italy}
\newcommand{\RomatreINFN}{Istituto Nazionale di Fisica Nucleare, Sezione di Roma Tre,\\ Via della Vasca Navale 84, I-00146 Rome, Italy}
\newcommand{\Romadue}{Dipartimento di Fisica and INFN, Universit\`a di Roma ``Tor Vergata",\\ Via della Ricerca Scientifica 1, I-00133 Roma, Italy}
\newcommand{\LaSapienza}{Physics Department and INFN Sezione di Roma La Sapienza,\\ Piazzale Aldo Moro 5, 00185 Roma, Italy}
\newcommand{\soton}{Department of Physics and Astronomy, University of Southampton,\\ Southampton SO17 1BJ, UK}

\begin{document}
\title{ Lattice calculation of the $\boldmath{D_{s}}$ meson radiative form factors over the full kinematical range
}
\author{R.\,Frezzotti}\affiliation{\Romadue} 
\author{G.\,Gagliardi}\affiliation{\RomatreINFN}
\author{V.\,Lubicz}\affiliation{\Romatre} 
\author{G.\,Martinelli}\affiliation{\LaSapienza}
\author{F.\,Mazzetti}\affiliation{\Romatre}
\author{C.T.\,Sachrajda}\affiliation{\soton}
\author{F.\,Sanfilippo}\affiliation{\RomatreINFN}
\author{S.\,Simula}\affiliation{\RomatreINFN}
\author{N.\,Tantalo}\affiliation{\Romadue}
\date{\today}
\begin{abstract}
We compute the structure-dependent axial and vector form factors for the radiative leptonic decays $D_s\to \ell\nu_\ell\gamma$, where $\ell$ is a charged lepton, as functions of the energy of the photon in the rest frame of the $D_s$ meson. 
The computation is performed using gauge-field configurations with 2+1+1 sea-quark flavours generated by the European Twisted Mass Collaboration and the results have been extrapolated to the continuum limit.
For the vector form factor we observe a very significant partial cancellation between the contributions from the emission of the photon from the strange quark and that from the charm quark. 
The results for the form factors are used to test the reliability of various Anz\"atze based on single-pole dominance and its extensions, and
we present a simple parametrization of the form factors which fits our data very well and which can be used in future phenomenological analyses. 
Using the form factors we compute the differential decay rate and the branching ratio for the process $D_s\to e\nu_e\gamma$ as a function of the lower cut-off on the photon energy.
With a cut-off of 10\,MeV for example, we find a branching ratio of Br$(E_\gamma>10\,\mathrm{MeV})=4.4(3)\times 10^{-6}$ which, unlike some model calculations, is consistent with the upper bound from the BESIII experiment 
Br$(E_\gamma>10\,\mathrm{MeV})<1.3\times 10^{-4}$ at 90\% confidence level.
Even for photon energies as low as 10\,MeV, the decay $D_s\to e\nu_e\gamma$ is dominated by the structure-dependent contribution to the amplitude (unlike the decays with $\ell=\mu$ or $\tau$), confirming its value in searches for hypothetical new physics as well as in determining the Cabibbo-Kobayashi-Maskawa (CKM) parameters at $O(\alpha_\mathrm{em})$, where $\alpha_{\mathrm{em}}$ is the fine-structure constant.
\end{abstract}

\maketitle

\section{Introduction}
The comparison between experimental measurements and theoretical predictions for flavour changing processes accompanied by photon emission represents an important tool in the search of New Physics (NP) beyond the Standard Model (SM). In this paper we consider radiative weak leptonic decays of the form $P\to \ell \nu_{\ell}\gamma$, where $P$ is a pseudoscalar meson, $\ell\nu_{\ell}$ a lepton-neutrino pair, and $\gamma$ a real photon. 
For each meson $P$, in addition to the leptonic decay constant $f_P$, the computation of the corresponding decay rate requires the calculation of two Structure-Dependent (SD) hadronic form factors, $F_{V}$ and $F_{A}$, which depend on the energy of the photon in the meson rest frame.
If instead the structure dependence of the meson is neglected, i.e. in the ``point-like" approximation, the only non-perturbative input required to determine the decay rate is $f_P$.
An interesting feature of these decays is that, because of helicity suppression, the point-like contribution to the decay rate is suppressed with respect to the SD contribution by the square of the ratio $r_{\ell}= m_{\ell}/m_{P}$, where $m_\ell$ and $m_P$ are the masses of the charged lepton $\ell$ and meson $P$ respectively. 
For heavy mesons $P$ and light final-state charged leptons $\ell$, the SD contribution, which is sensitive to the internal structure of the decaying meson, can already be dominant at relatively low photon energies, and in particular those well below the typical energy cut-off imposed in experimental measurements. 
This makes such decay channels an ideal place to probe the internal structure of the meson and the presence of possible NP contributions.  
While for pion and kaon decays several experimental measurements of the axial and vector form factors exist (see e.g.~Ref.\,\cite{Bychkov:2008ws,E787:2000ehx,KLOE:2009urs, OKA:2019gav, ISTRA:2010smy, JPARCE36:2021yvz}), for heavy mesons only very little is known. For charmed meson decays the BESIII collaboration recently searched for signals of the Cabibbo-suppressed decay $D^+\to e^{+}\nu_{e}\gamma$~\cite{BESIII:2017whk} and of the Cabibbo-favoured one $D_{s}^+\to e^{+}\nu_{e}\gamma$\,\cite{BESIII:2019pjk}, finding no events with  emission of photons with energies $E_{\gamma} > 0.01\,{\rm GeV}$, and setting the following upper bounds on the branching ratios: ${\rm Br}[ D \to e\nu_{e}\gamma] < 3\times 10^{-5}$ and ${\rm Br}[ D_{s} \to e\nu_{e}\gamma] < 1.3\times 10^{-4}$ at 90\% confidence level. 
For the $B$ meson, the Belle collaboration has recently set the bounds\,\cite{Belle:2015mpp,Belle:2018jqd} ${\rm Br}[ B \to e\nu_{e}\gamma] < 4.3\times 10^{-6}$ and ${\rm Br}[ B \to \mu\nu_{\mu}\gamma] < 3.4\times 10^{-6}$, and observed photons with energies $E_{\gamma} > 1\,{\rm GeV}$. We believe that by providing accurate predictions from first principles for the axial and vector form factors for heavy mesons, we will  motivate further experimental studies.    \\

We have recently computed the rates for $P\to \ell \nu_{\ell}\gamma$ decays where $P$ is a light meson, $P=\pi$ or $K$,\,\cite{Desiderio:2020oej} and compared our results to experimental measurements finding some puzzling and interesting discrepancies yet to be resolved\,\cite{Frezzotti:2020bfa}. In Ref.\,\cite{Desiderio:2020oej} we have also computed the amplitude for the decays of the $D_s$ meson, but only over part of the physical phase space; specifically for photon energies up to $0.4\,{\rm GeV}$, as measured in the rest frame of the $D_{s}$ meson. 
In this paper, we return to the radiative decays of $D_s$ mesons and compute the relevant axial and vector form factors $F_{V}$ and $F_{A}$ over the full physical kinematic range and with high statistical accuracy, thus improving significantly upon the previous study of Ref.\,\cite{Desiderio:2020oej}.
The computation is performed using the $N_{f}=2+1+1$ Wilson-clover twisted-mass gauge ensembles generated by the Extended Twisted Mass Collaboration (ETMC) with quark masses tuned very close to their physical values, for almost all the ensembles~\cite{Alexandrou:2018egz,ExtendedTwistedMass:2021qui,ExtendedTwistedMass:2021gbo,Alexandrou:2022amy}. The ensembles correspond to four values of the lattice spacing $a$ in the range $[0.56, 0.9]~{\rm fm}$, with the spatial extent of the lattice, $L$, ranging from $4.4\,{\rm fm}$ to $7.6\,{\rm fm}$.\\

Our main results for the $D_s$ radiative form factors $F_{V}$ and $F_{A}$ are collected in Tab.\,\ref{tab:result_FA_FV} and plotted in Fig.\,\ref{fig:cont_spline}, and we provide their correlation matrices in Appendix \ref{app:B}.
Recently, numerical results for the lattice computation of $F_V$ and $F_A$ of the $D_s$ meson also have been published in Ref.\,\cite{Giusti:2023pot}. The primary focus of that paper however, is on developing and testing different strategies for the lattice computation of the form factors. Their numerical results are based on a single gauge ensemble at an unphysical pion mass, and for this reason a direct comparison with our results is not possible at present. In the future, it would be interesting to compare our results with those obtained from other lattice regularizations in the continuum limit.\\

We use our results for the form factors to compute the branching fraction of the $D_{s}\to e^{+}\nu_{e}\gamma$ decay, as a function of the lower cut-off, $\Delta E_{\gamma}$, on the photon energy, as measured in the meson rest frame. Our results are showed in Fig.\,\ref{fig:decay_rate}. For $\Delta E_{\gamma}=0.01$ GeV, our prediction for the branching fraction lies well below the experimental upper limit set by the BESIII collaboration \cite{BESIII:2019pjk}. Due to the strong helicity suppression, the branching fraction is dominated by the SD contribution, even for a lower cut on the photon energy as small as $0.01$ GeV.
In Fig.\,\ref{fig:diff_decay_rate}, we show the SD contribution to the differential $D_{s}\to e^{+}\nu_{e}\gamma$ decay rate, as a function of the photon energy in the meson rest frame; this goes to zero at the edge of phase space and reaches a maximum in the region kinematic $x_\gamma\equiv 2E_\gamma/M_{D_s}\simeq 0.6-0.7$.\\

Having calculated the form factors from first principles in a lattice computation, we test how well model calculations based on single-pole dominance and light-cone sum rules (LCSR) reproduce our results. Such a test is important because model calculations are commonly used to describe the form factors of heavy mesons, in particular the $B$ meson~\cite{Korchemsky:1999qb, Atwood:1994za, Yang:2012jp}, for which a direct lattice calculation is currently missing.
We fit our results for the vector and axial form factors of the $D_{s}$ meson to several pole-like Ans\"atze, finding that, in general, a pure vector-meson-dominance (VMD) Ansatz does not describe very well the momentum dependence of the data, particularly for the axial form factor $F_A$. However, by including the leading non-singular corrections in the Laurent expansion around the pole, we obtain a very good description of our lattice data. The resulting fit parameters are collected in Tab.\,\ref{tab:phen_fitparameters} and can be used for future phenomenological analyses.\\

From the pole-like fits to $F_V$ we extract the coupling $g_{D_s^* D_s\gamma}$, which is the form factor describing the $D_{s}^{*}\to D_{s}\gamma$ decay. Our estimate of $g_{D_s^* D_s\gamma}$ is in good agreement with the direct lattice determination of Ref.\,\cite{Donald:2014} but strongly disagrees with the value predicted by light-cone sum rules (LCSR) at next-to-leading order (NLO) \cite{Pullin:2021ebn}, from which it differs by a factor 5. LCSR at NLO order were also used to estimate the radiative form factors $F_V$ and $F_A$ of the $D_s$ meson, at one specific kinematical point\,\cite{Lyon:2012fk}.
Their estimates disagree very significantly with those from our direct lattice computation, differing by a factor $4$ for the vector form factor $F_V$, and by an order of magnitude and a relative minus sign, for the axial form factor $F_A$. We conclude that caution should be exercised when using such calculations based on LCSR to predict the heavy-meson radiative form factors. A similar message is conveyed in a recent paper on the $B_{s}\to \mu^{+}\mu^{-}\gamma$ decay\,\cite{Guadagnoli:2023zym}.\\

The plan for the remainder of this paper is as follows. In Sec.\,\ref{sec:formfactors} we explain how the two structure dependent form factors contributing to the amplitude for $D_s^+\to\ell^+\nu_\ell\gamma$ decays,  
$F_{V}$ and $F_{A}$, can be determined from suitable Euclidean lattice correlation functions. 
In Sec.\,\ref{sec:comp_details} we briefly describe the mixed-action lattice framework which we use and present detailed information on the technical aspects of the simulations. 
Sec.\,\ref{sec:Num_results} contains the determination of the form factors, together with our estimates of the systematic errors, including a description of the extrapolations to the continuum and infinite volume limits. In this section we also present the determination of the differential decay rate and branching fraction, as a function of $\Delta E_{\gamma}$, for the $D_s\to e\,\nu_e\,\gamma$ decay.
In Sec.\,\ref{sec:phenomenology} we provide a simple pole-like parameterization of our data for $F_{V}$ and $F_{A}$, which may be useful for those interested in using our data for phenomenological analyses. 
We also compare the results presented in Sec.\,\ref{sec:Num_results} with predictions from models based on pole dominance or light-cone sum-rules.  
Finally in Sec.\,\ref{sec:conclusions} we present our conclusions. There are three appendices to supplement the information in the main text. In Appendix\,\ref{app:A} we explain the reason for the observed deterioration of the signal-to-noise ratio at large photon energies. The results for the form factors $F_V$ and $F_A$, together with the corresponding correlation matrices are tabulated in Appendix\,\ref{app:B} so that they can be used in phenomenological studies. 
In Appendix\,\ref{app:C} we present a detailed analysis of single-pole parametrizations of our results for the form factors.
 
\section{Definition of the form factors}\label{sec:formfactors}

In order to make this paper self contained, in this section we briefly summarise our conventions and notation, and in particular recall the definition of the structure-dependent form factors which had previously been introduced in Refs.\,\cite{Carrasco:2015xwa,Bijnens:1992en,Desiderio:2020oej,Gagliardi:2022szw}.
The non-perturbative contribution to the radiative leptonic decay rate for the processes $D^+_s\to \ell^+\nu_{\ell}\gamma$ is encoded in the hadronic matrix-element
\begin{align}
H_{W}^{r\nu}(k, \bs{p})=\epsilon_\mu^r(k)\,H_{W}^{\mu\nu}(k, \bs{p})= \epsilon_\mu^r(k) \int \dfour y\, e^{ik\cdot y}\, \bra{0} \hat{\mathrm{T}}[\,j_W^\nu(0) j^\mu_\mathrm{em}(y)]\ket{D^+_s(\bs{p})}\;,  
\label{eq:H_munu}
\end{align}
where $\hat{\mathrm{T}}$ implies time-ordering of the two currents, $\epsilon_\mu^r$ is the polarisation vector of the outgoing photon with four-momentum $k$, $\bs{p}$ is the three-momentum of the $D_s$ meson, and $j_W^\nu(x)$ and $j^\mu_\mathrm{em}(x)$ are the weak and electromagnetic hadronic currents respectively:
\begin{align}\label{eq:currents}
j_W^\nu(x)=j_V^\nu(x)-j_A^\nu(x) = \bar \psi_s(x) \, (\gamma^\nu - \gamma^\nu \gamma_5) \, \psi_c(x)\,,
\qquad
j^\mu_\mathrm{em}(x)=\sum_f q_f \bar \psi_f(x) \gamma^\mu \psi_f(x)\,,
\end{align}
where $q_{f}$ is the electric charge of the flavour $f$. The hadronic tensor $H^{\mu\nu}_{W}$ can be decomposed in terms of a ``point-like" contribution $H_\mathrm{pt}^{\mu\nu}$ (i.e. the expression obtained by treating the $D_s$ meson as a point-like particle) and four structure-dependent (SD) scalar form factors, $F_V,\,F_A,\,H_1$ and $H_2$\,\cite{Carrasco:2015xwa,Bijnens:1992en,Desiderio:2020oej,Gagliardi:2022szw}\,
\footnote{Here, we use the dimensionless definitions of $H_{1,2}$ introduced in Ref.\,\cite{Gagliardi:2022szw} which differ by simple factors from those used in our earlier papers\,\cite{Carrasco:2015xwa,Desiderio:2020oej}. As explained below, the form factors $H_{1,2}$ do not contribute to the decays studied here, i.e. those with a real photon in the final state.}:
\begin{eqnarray}
H^{\mu \nu}_W(k,\bs{p}) &=& H^{\mu\nu}_{\rm{SD}}(k,\bs{p})+ H^{\mu\nu}_{\rm{pt}}(k,\bs{p}) \\
H^{\mu\nu}_{\rm{SD}}(k,\bs{p})&=&\frac{H_1(p\cdot k,k^2)}{M_{D_s}}\,\left[k^2 g^{\mu\nu}-k^\mu k^\nu\right]
+
\frac{H_2(p\cdot k,k^2)}{M_{D_s} }\, \frac{\left[(p\cdot k-k^2)k^\mu-k^2(p-k)^\mu\right]}{(p-k)^{2} -
M_{{D_s}}^{2}}(p-k)^\nu
\nonumber \\
&&\hspace{0.5in}
-i\frac{F_V(p\cdot k,k^2)}{M_{D_s}}\varepsilon^{\mu\nu\gamma\beta}k_\gamma p_\beta
+\frac{F_A(p\cdot k,k^2)}{M_{D_s}}\left[(p\cdot k-k^2)g^{\mu\nu}-(p-k)^\mu k^\nu\right]\,\\
H^{\mu\nu}_{\rm{pt}}(k,\bs{p}) &=&
f_{D_s}\left[g^{\mu\nu}+\frac{(2p-k)^\mu(p-k)^\nu}{2p\cdot k-k^2}\right]
\;,
\label{eq:ffdef}
\end{eqnarray}
where $M_{{D_s}}$ is the mass of the ${D_s}$ meson and $p= \left(E, \,\bs{p} \right)$ its four momentum, with $E=\sqrt{ M_{D_s}^{2}+\bs{p}^{2} }$. 
The point-like contribution $H_{\rm{pt}}^{\mu\nu}$ saturates the Ward-Identity (WI) satisfied by $H^{\mu\nu}_{W}$:
\begin{align}
k_\mu\, H^{\mu \nu}_W(k,\bs{p})= k_\mu\,H^{\mu \nu}_\mathrm{pt}(k,\bs{p})= i\bra{0}j_W^\nu(0)\ket{D_s^+(\bs{p})} = f_\mathrm{D_s}\, p^\nu .
\label{eq:contWI}
\end{align}
which implies that $k_\mu\,H^{\mu \nu}_\mathrm{SD}(k,\bs{p}) =0$. Moreover, when integrating over the full three-body phase space, it is only the square of the point-like term which is infrared divergent. At order $O(\alpha_{\textrm{em}})$, this infrared divergence is cancelled by the virtual photon correction to the purely leptonic decay.

Eq.\,(\ref{eq:ffdef}) is valid for generic (off-shell) values of the photon four-momentum $k$ and can also be used to study the four-body decay $D_s^+\to \ell^+ \nu_{\ell}\, \ell^{\prime +} \ell^{\prime -}$, where the $\ell^\prime$ are charged leptons, or more generally the decays $P\to \ell \nu_{\ell}\, \ell^{\prime +} \ell^{\prime -}$ of any pseudoscalar meson $P$, as we showed in an exploratory work with $P=K$\,\cite{Gagliardi:2022szw}. In this paper we study the emission of a real photon, so that $k^2=0$ and
$\epsilon^r\cdot k=0$, and therefore only the axial form factor $F_{A}(p\cdot k)$ and the vector form factor $F_{V}(p\cdot k)$, together with the point-like term,  contribute to the decay rate for the process $D_s^+\to \ell^+\nu_{\ell}\gamma$.

In the following, as in our previous study\,\cite{Desiderio:2020oej}, we find it convenient to evaluate the form factors $F_{V}$ and $F_{A}$ as functions of the dimensionless variable
\begin{align}\label{eq:xgammadef}
x_{\gamma} \equiv \frac{2p\cdot k}{ M_{D_s}^{2}};\,\qquad 0 \leq x_{\gamma} \leq 1 - \frac{m_{\ell}^{2}}{M_{D_s}^{2}} < 1\; ,      
\end{align}
where $m_{\ell}$ is the mass of the charged
lepton $\ell$. In the rest frame of the $D_s$ meson ($\bs{p}=\bs{0}$\hspace{2pt}) $x_\gamma = 2E_\gamma/M_{D_s}$ where $E_\gamma$ is the energy of the photon. The above discussion applies to other pseudoscalar mesons ($\pi$, $K$, $D$, $B_{(s)}$) with the natural replacement of $D_s$ in Eqs.\,(\ref{eq:H_munu})\,-\,(\ref{eq:xgammadef}) by the meson being studied and the corresponding change of the quark flavours in the weak current in Eq.\,(\ref{eq:currents}).

\subsection{Evaluating  $F_{V}$ and $F_{A}$ from Euclidean lattice correlation functions}

In Sec.\,III and Appendix B of Ref.\,\cite{Desiderio:2020oej} we showed in detail that for the emission of a real photon, the hadronic tensor $H_{W}^{\mu\nu}$ can be extracted for all values of $x_{\gamma}$ from the Euclidean three-point correlation function:
\begin{equation}\label{eq:Cmunudef}
C^{\mu\nu}_W(t;k,p) =-i\sum_{t_y=0}^{T} \sum_{\bs{y}}\sum_{\bs{x}}\left( \theta(T/2-t_{y}) + \theta(t_{y}-T/2)e^{-E_{\gamma}T}\right) ~e^{\hspace{1pt}t_{y}E_\gamma-i\bs{k}\cdot\bs{y}+i\bs{p}\cdot\bs{x}}~
\bra 0\hat{\mathrm{T}}\,[j^\nu_W(t,\bs{0})j_\mathrm{em}^\mu(t_{y}, \bs{y})\phi^\dagger_{D_s}(0,\bs{x})]\ket 0\,,
\end{equation}
where $T$ is the temporal extent of the lattice\,\footnote{$T$ is not to be confused with $\hat{\mathrm{T}}$ which represents "time-ordered".}, $\phi^\dagger_{D_s}$ is an interpolating operator with the quantum numbers to create the $D_s$ meson, $k=(E_{\gamma}, \bs{k})$, and $E_\gamma$ is the energy of the photon. In the forward half of the lattice $0\ll t \ll T/2$ for example, one has 
\begin{equation}
R^{\mu\nu}_W(t;E_{\gamma},\bs{k},\bs{p}) 
\equiv \frac{2E}{e^{-t(E-E_\gamma)}\, \bra{D_s(\bs{p})}\phi^\dagger_{D_s}(0)\ket{0}}\, C^{\mu \nu}_W(t, E_{\gamma}, \bs{k},\bs{p})
=
H^{\mu\nu}_W(k,\bs{p}) + \cdots ~,
\label{eq:Rinf}
\end{equation}
where the ellipsis indicates terms that vanish exponentially in the large $t$ limit and $E=\sqrt{\bs{p}^2+M_{D_s}^2}$\,. 
 Eq.~(\ref{eq:Cmunudef}) is valid for $t < T/2$, however, as explained in Appendix B of Ref.\,\cite{Desiderio:2020oej}, $H^{\mu\nu}_W(k,\bs{p})$ can be obtained also from the backward half of the lattice $T/2\ll t\ll T$ exploiting time-reversal symmetry. In order to determine the form factors $F_V$ and $F_A$ it is convenient to distinguish the contributions from the vector and axial-vector components of the weak current, $j_W^\nu=V^\nu-A^\nu$, and to write
\begin{equation}
R^{\mu\nu}_W(t;E_{\gamma},\bs{k},\bs{p})=R^{\mu\nu}_V(t;E_{\gamma},\bs{k},\bs{p})-R^{\mu\nu}_A(t;E_{\gamma},\bs{k},\bs{p})\,.
\end{equation}

  \begin{figure}[t]
    \begin{center}
    \includegraphics[width=0.42\textwidth]{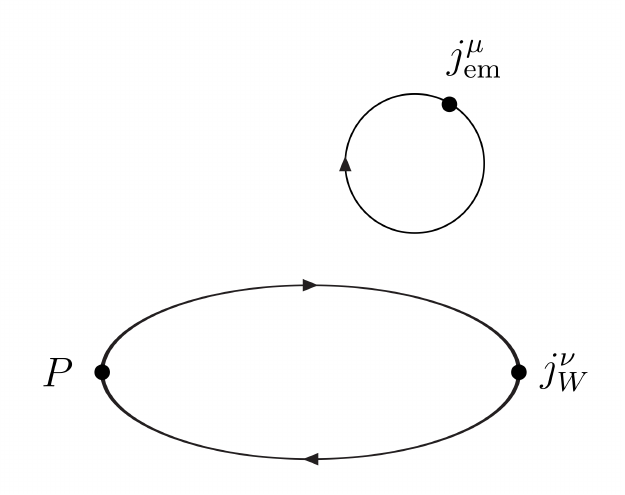}
    \hspace{0.1cm}
    \includegraphics[width=0.42\textwidth]{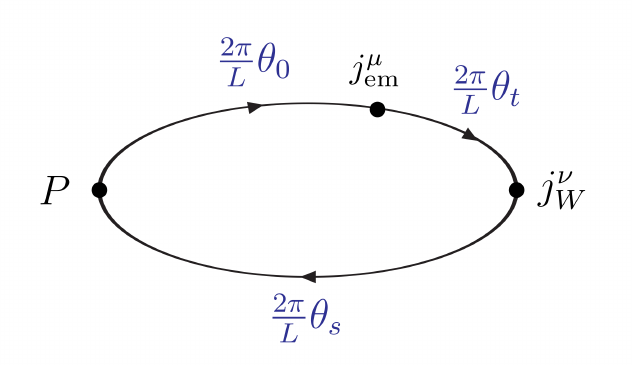}
    \end{center}
    \caption{\it The diagram on the left represents the quark-line disconnected contributions to the correlation function $C^{\mu\nu}_{W}$ in which the photon is emitted by a sea quark. The one on the right represents the  quark-line connected contributions and illustrates our choice of the spatial boundary conditions, which allow us to set arbitrary values for the meson and photon spatial momenta.
    The spatial momenta of the valence quarks, modulo $2\pi/L$, in terms of the twisting angles are as indicated. Each diagram implicitly includes all orders in QCD.
    \hspace*{\fill}
    \label{fig:Feyn_conn_disc}}
    \end{figure}
    
The Wick contractions of the correlation function in Eq.\,(\ref{eq:Cmunudef}) give rise to two distinct topologies of Feynman diagrams, namely to quark-line connected and quark-line disconnected diagrams; these are illustrated in Fig.\,\ref{fig:Feyn_conn_disc}. In the disconnected diagrams the photon is emitted from a sea quark. This contribution vanishes in the $\rm{SU}(3)$-symmetric limit and is neglected in the present study; this is the so-called electroquenched approximation. 
We focus instead on the calculation of the dominant, quark-connected contributions for which
as explained in Ref.\,\cite{Desiderio:2020oej}, it is possible to use twisted boundary conditions to assign arbitrary values to momenta of the photon and $D_s$-meson, 
$\bs{k}$ and $\bs{p}$ respectively, at the price of violations of unitarity which vanish exponentially with the lattice extent $L$
\,\cite{Sachrajda:2004mi,Flynn:2007ess}. This is achieved by treating the two quark propagators related to the electromagnetic current in the right-hand diagram of 
Fig.\,\ref{fig:Feyn_conn_disc} as corresponding to two distinct quark fields $\psi^{0}, \psi^{t}$ having the same mass and quantum number, but satisfying different spatial boundary conditions.
Defining $\psi^{s}$ to be the \textit{spectator} quark-field in the right-hand diagram of Fig.\,\ref{fig:Feyn_conn_disc}, we set the spatial boundary  conditions of the three quark fields $\psi^{0},\psi^{t},\psi^{s}$ as follows:
\begin{align}
\psi^{r}(x+\bs{n}L) = \exp{(2\pi i\bs{n}\cdot \bs{\theta}_{r})}\psi^{r}(x)\;,\qquad r=\{ 0, t, s \}~,
\end{align}
where $\bs{\theta}_{\{0,t,s\}}$ are arbitrary spatial-vectors of angles, in terms of which the photon and meson lattice momenta are given by
\begin{align}
\bs{p} = \frac{2}{a}\sin\left(\frac{a\pi}{L}\left( \bs{\theta}_0-\bs{\theta}_s\right)\right)\;,
\qquad
\bs{k} = \frac{2}{a}\sin\left(\frac{a\pi}{L}\left( \bs{\theta}_0-\bs{\theta}_t\right)\right)\;,
\label{eq:momenta}    
\end{align}
where $a$ is the lattice spacing.
The results presented in the following sections have been obtained in the rest frame of the $D_s$ meson ($\bs{p}=0$) and with the photon momentum chosen to be in the $z$-direction, $\bs{k}=(0,0,k_z)$, i.e. by setting
\begin{align}
\bs{\theta}_{0}= \bs{\theta}_{s} = \bs{0}\;,\qquad \bs{\theta_t} = (0,0,\theta_t).
\end{align}
With such a choice of kinematics, the two form factors $F_{V}$ and $F_{A}$ can be obtained from the large-time behaviour, $0\ll t \ll T/2$, of the following two estimators
\begin{align}
R_{V}(t,\bs{k}) &\equiv  \frac{1}{2k_{z}}\left( R_{V}^{12}(t,\bs{k},\bs{0}) - R_{V}^{21}(t,\bs{k},\bs{0})\right) ~  \xrightarrow[0\ll t \ll T/2]{} ~ F_{V}(x_{\gamma})\, ,  \\[10pt]
R_{A}(t, \bs{k}) &\equiv  \frac{1}{2E_{\gamma}}\left[ \left( R_{A}^{11}(t,\bs{k},\bs{0}) - R_{A}^{11}(t,\bs{0},\bs{0})\right) + \left( R_{A}^{22}(t,\bs{k},\bs{0}) - R_{A}^{22}(t,\bs{0},\bs{0})\right)\right] ~ \xrightarrow[0\ll t \ll T/2]{} ~ F_{A}(x_{\gamma})\,,\label{eq:RA}
\end{align}
where $x_{\gamma}= 2E_{\gamma}/M_{P}$ and, since for a real photon $E_\gamma$ is determined by $\bs{k}$, we have redefined
\begin{align}
\label{eq:redef_R}
R^{\mu\nu}_{V,A}(t,\bs{k},\bs{0}) \equiv R^{\mu\nu}_{V,A}(t, E_{\gamma}(\bs{k}),\bs{k},\bs{0})\,,\qquad E_{\gamma}(\bs{k}) = \frac{2}{a}\sinh^{-1}\left(\frac{ a|\bs{k}|}{2}\right)~,
\end{align}
where we used the lattice dispersion relation for the photon energy. 
Notice that in the estimator $R_{A}(t,\bs{k})$ of the axial 
form factor, the zero-momentum subtraction serves to remove the point-like contribution proportional to $f_{P}$. As discussed in Sec.\,IV of Ref.\,\cite{Desiderio:2020oej}, the subtractions in Eq.\,(\ref{eq:RA}) of $R_A^{11}$ and $R_A^{22}$  at $\bs{k}=\bs{0}$ allow us to isolate the SD form factor $F_{A}(x_{\gamma})
$ without generating infrared-divergent cut-off effects of order $\mathcal{O}(a^{2}/x_{\gamma})$. Such dangerous discretization effects, which could hinder the determination of $F_{A}(x_{\gamma})$ at small values of $x_{\gamma}$, are present instead if one subtracts the point-like contribution $H_{\rm{pt}}^{\mu\nu}$ in Eq.\,(\ref{eq:H_munu}), using the value of the decay constant $f_{D_s}$ determined from two-point correlation functions.

\section{Details of the computation} \label{sec:comp_details}
Our results have been obtained using the gauge field configurations generated by the Extended Twisted Mass Collaboration (ETMC) employing the Iwasaki gluon action~\cite{Iwasaki:1985we} and $N_{f}=2+1+1$ flavours of Wilson-Clover twisted-mass fermions at maximal twist~\cite{Frezzotti:2000nk}. This framework guarantees the automatic $\mathcal{O}(a)$ improvement of parity-even observables~\cite{Frezzotti:2003ni,Frezzotti:2004wz}. A detailed description of the ETMC ensembles can be found in Refs.~\cite{ExtendedTwistedMass:2021gbo,ExtendedTwistedMass:2021qui,Alexandrou:2022amy,Alexandrou:2018egz}, while essential informations on the ensembles we have used in the present work are collected in Table\,\ref{tab:simudetails}. The ensembles listed in Table\,\ref{tab:simudetails} correspond to four values of the lattice spacing $a$ in the range $[0.058,0.09]\,{\rm fm}$, and lattice extent $L$ in the range $[4.36, 5.46]\,{\rm fm}$. The mass of the light sea quarks on the three finest ensembles, has been tuned so as to give almost physical-mass pions, while on the coarsest ensemble the pion mass\footnote{For the present study of the $D_s$ meson, the presence of a heavier-than-physical pion, with mass $M_{\pi}\simeq 175\,{\rm MeV}$, on the coarsest ensemble does not require a 
chiral extrapolation since we expect that the 
form factors are largely insensitive to the value of the masses of the light sea quarks.} is $M_{\pi} \simeq 175\,{\rm MeV}$. For all the ensembles, the strange and charm sea-quark masses are set to within about 5\% of their physical values, defined through the requirement that (see Refs.~\cite{Alexandrou:2018egz,ExtendedTwistedMass:2021qui, Alexandrou:2022amy} for more details) 
\begin{align}
\frac{M_{D_{s}}}{f_{D_{s}}} = 7.9\pm0.1\, , \qquad 
\frac{m_{c}}{m_{s}} = 11.8\pm0.2\,.
\end{align}

\begin{table}
\begin{center}
    \begin{tabular}{||c||c|c|c|c|c||c||c||}
    \hline
    ~~~ ensemble ~~~ & ~~~ $\beta$ ~~~ & ~~~ $V/a^{4}$ ~~~ & ~~~ $a$\,(fm) ~~~ & ~~~ $a\mu_{\ell}$ ~~~ & ~ $M_{\pi}$\,(MeV) ~ & ~ $L$ (fm) & ~ $N_{g}$ ~ \\
  \hline
  cA211.12.48 & $1.726$ & $48^{3}\cdot 128$ & $0.09075~(54)$ & $0.00120$ & $174.5~(1.1)$ & $4.36$ & $109$  \\
  
  cB211.072.64 & $1.778$ & $64^{3}\cdot 128$ & $0.07957~(13)$ & $0.00072$ & $140.2~(0.2)$ & $5.09$ & $199$ \\
  
  cC211.060.80 & $1.836$ & $80^{3}\cdot 160$ & $0.06821~(13)$ & $0.00060$ & $136.7~(0.2)$ & $5.46$ & $72$ \\
  
  cD211.054.96 & $1.900$ & $96^{3}\cdot 192$ & $0.05692~(12)$ & $0.00054$ & $140.8~(0.2)$ & $5.46$ & $100$ \\
  \hline
    \end{tabular}
\end{center}
\caption{\it \small Parameters of the ETMC ensembles used in this work. We present the light-quark bare mass, $a \mu_\ell = a \mu_u = a \mu_d$, the lattice spacing $a$,  the pion mass $M_\pi$, the lattice size $L$, and the number of gauge configurations $N_{g}$ that have been used for each ensemble. The values of the lattice spacing are determined as explained in Appendix B of Ref.~\cite{Alexandrou:2022amy} using the 2016 PDG value $f_\pi^{phys} = f_\pi^{isoQCD} = 130.4(2)$\,MeV\,\cite{ParticleDataGroup:2016lqr} of the pion decay constant.}
\label{tab:simudetails}
\end{table} 

\begin{figure}
\includegraphics[scale=0.45]{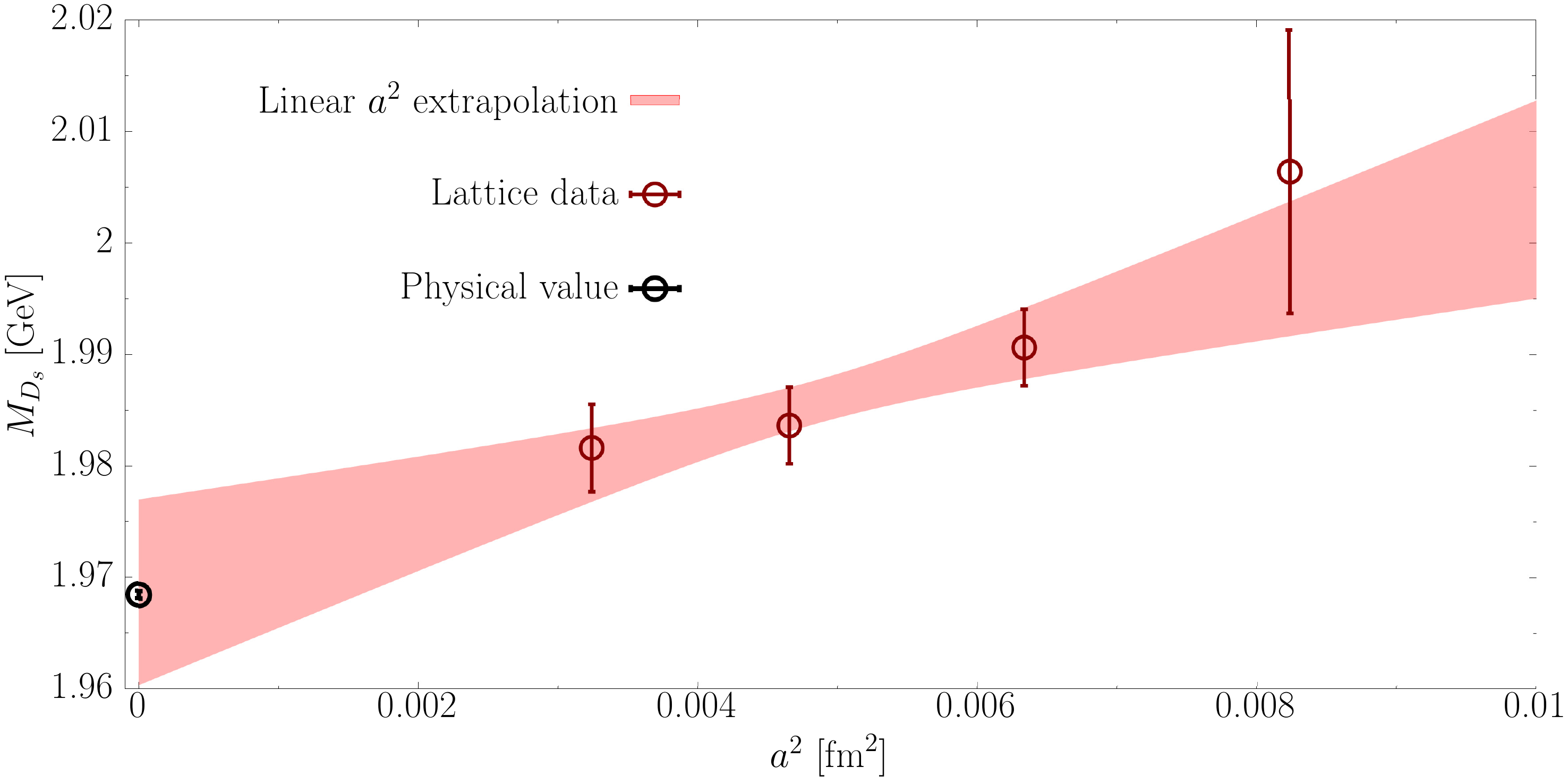}
\caption{The value of the $D_{s}$ meson mass on the ensembles of Table~\ref{tab:simudetails} is plotted as a function of squared lattice spacing. The red band corresponds to the result of the continuum extrapolation using a linear fit Ansatz in $a^{2}$. The black data point at $a^{2}=0$ is the experimental value $M_{D_{s}}^{\rm{exp}} = 1.96847\,(33)\,{\rm{GeV}}$.  }
\label{fig:Ds_meson_mass}
\end{figure}

We work in a mixed-action framework in which the valence strange and charm quarks are discretized as Osterwalder-Seiler fermions\,\cite{Osterwalder:1977pc, Frezzotti:2004wz}. The corresponding valence bare-quark mass parameters $\mu_{s}$ and $\mu_{c}$, have been tuned to reproduce the value of the pseudoscalar $\eta_{ss'}$ mass\,\footnote{The $\eta_{ss'}$ is a fictitious pseudoscalar meson made of two different mass-degenerate quarks $s$ and $s'$ having mass equal to that of the strange quark. Its mass is equivalent to that of the $\bar{s}\gamma^{5}s$ meson if one neglects quark-line disconnected contributions.} determined in Ref.~\cite{Borsanyi:2020mff} and the PDG value~\cite{ParticleDataGroup:2020ssz} of the pseudoscalar $\eta_{c}$ mass (see Appendix C of Ref.~\cite{Alexandrou:2022amy} for more 
details) 
\begin{align}
\label{eq:phys_val}
M_{\eta_{ss'}}^{\mathrm{phys}} = 689.89\,(49)\,{\rm MeV} \, , \qquad M_{\eta_{c}}^{\mathrm{phys}} = 2.984\,(4)\,{\rm GeV}\,, 
\end{align}
where the error in $M_{\eta_{c}}^{\mathrm{phys}}$ includes, in addition to the experimental uncertainty, an estimate of the contribution from the neglected disconnected diagrams\,\cite{Hatton:2020qhk,Zhang:2021xrs}. The values of $\mu_{s}$ and $\mu_{c}$ used for each of the ensembles of Table~\ref{tab:simudetails} are collected in Table~\ref{tab:musmud}. Since the strange and charm quark masses have been fixed using the mass of the $\eta_{ss'}$ and $\eta_{c}$ mesons, the mass of the $D_{s}$ meson deviates from the physical value by $\mathcal{O}(a^{2})$ cut-off effects, as we show in Figure\,\ref{fig:Ds_meson_mass}.\\

On each ensemble, we have analyzed $\mathcal{O}(100)$ gauge configurations and performed the inversions of the Dirac operator on 4 stochastic sources. The sources are randomly distributed over time, diagonal in spin and dense in the color. The interpolating operator $\phi^\dagger_{D_s}\!(0)$ has been smeared as in our previous works using Gaussian smearing (see e.g. Ref.~\cite{Carrasco:2014uya} for more details). We employ a local discretization of the weak current:  
\begin{flalign}
j_W^\nu(t,\bs{x}) = j_V^\nu(t,\bs{x})-j_A^\nu(t,\bs{x})= Z_{A}j_{V}^{\nu,{\rm bare}} - Z_{V}j_{A}^{\nu,{\rm bare}} =  Z_{A}\bar \psi_{s}(t,\bs{x})\gamma^\nu\psi_{c}(t,\bs{x})
-Z_{V}\bar \psi_{s}(t,\bs{x})\gamma^\nu\gamma_5\psi_{c}(t,\bs{x})\,. \label{eq:jWVA}
\end{flalign}
Note that at maximal-twist the Renormalization Constants (RCs) to be used for $j_{V}^{\nu,{\rm bare}}$ and $j_{A}^{\nu,{\rm bare}}$ are chirally-rotated w.r.t. the ones of standard Wilson-fermions, and the bare vector ($j_{V}^{\nu,{\rm bare}}$) and axial-vector ($j_{A}^{\nu,{\rm bare}}$) currents renormalize respectively with multiplicative renormalization constants $Z_{A}$ and $Z_{V}$. 
\begin{table}
    \centering
    \begin{tabular}{||c||c|c||}
    \hline
        Ensemble & $a\mu_{s}$ & $a\mu_{c}$ \\
        \hline
        cA211.12.48 & 0.0200 & 0.2725 \\
        cB211.072.64 & 0.0184 & 0.2370	 \\
        cC211.060.80 & 0.0162  & 0.2019	\\
        cD211.054.96 & 0.0136 & 0.1671	 \\
        \hline
    \end{tabular}
    \caption{\it \small Values of the bare valence-quark masses $\mu_{s}$ and $\mu_{c}$ in lattice units used for each ETMC ensemble considered in this work. They have been determined imposing the conditions in Eq.\,(\ref{eq:phys_val}).}
    \label{tab:musmud}
\end{table}
For the electromagnetic current $j_{\mathrm{em}}^{\mu}$, we use the exactly-conserved point-split current
\begin{align}
\label{eq:1PS_em}
j_{\mathrm{em}}^{\mu}(x) = -\sum_{f} q_{f} \left\{
\bar \psi_f(x)\frac{ir_{f}\gamma_5-\gamma^\mu}{2}\, U^\mu(x)\psi_f(x+\hat \mu)
-
\bar \psi_f(x+\hat \mu)\frac{ ir_{f}\gamma_5+\gamma^\mu}{2}U^\mu(x)^\dagger \psi_f(x) \right\}\,~,
\end{align}
where $U^{\mu}(x)$ are the QCD gauge links, and $r_{f}= \pm 1$ is the sign of the chirally-rotated Wilson term used for flavour $f$. In the electroquenched approximation only the $f=s$ and $f=c$ terms in $j^{\mu}_{\EM}$ contribute to the correlation function in Eq.~(\ref{eq:Cmunudef}). In our numerical simulations, we have chosen opposite signs for the chirally-rotated twisted term of the strange and charm valence quarks, $r_{c}=-r_{s}=1$. We evaluate the correlation functions $C^{\mu\nu}_{W}$ in Eq.\,(\ref{eq:Cmunudef}) and the estimators $R_{W}^{\mu\nu}$ in Eq.\,(\ref{eq:Rinf})  using the bare currents $j_{V,A}^{\nu,{\rm bare}}$ in Eq.\,(\ref{eq:jWVA}) and the point split electromagnetic current in Eq.\,(\ref{eq:1PS_em}). \\

The RC $Z_{V}$ which renormalizes the current $j_{A}^{\nu,{\rm bare}}$, is determined from the large-time behaviour of the following estimator
\begin{align}
\label{eq:est_ZV}
\bar{Z}_{V}(t) = -\frac{2f_{D_{s}}}{\sum_{i=1,2} R_{A}^{ii}(t,0,0)}  \xrightarrow[0\ll t \ll T]{}  Z_{V}\,.
\end{align}
In the twisted-mass framework which we are using, the decay constant $f_{D_s}$ can be determined from the large-time behaviour of the two-point correlation function $C_{PP}(t)$, without the need of additional renormalization, using
\begin{align}
\label{eq:PP_corr}
C_{PP}(t) \equiv \sum_{\bs{x}}\bra{0}  \left( \bar{\psi}_{s}(t,\bs{x})\gamma^{5}\psi_{c}(t,\bs{x}) \right)  \left(\bar{\psi}_{c}(0)\gamma^{5}\psi_{s}(0)\right)\ket{0} \xrightarrow[0\ll t \ll T]{} \big|\mathcal{Z}\big|^{2}~\frac{e^{-M_{D_{s}}T/2}}{2M_{D_{s}}}\cosh{( M_{D_{s}}(t - T/2))}\, ,
\end{align}
and
\begin{align}
f_{D_{s}} = \sqrt{\big|\mathcal{Z}\big|^{2}} \frac{a\mu_{s}+a\mu_{c} }{M_{D_{s}}\sinh aM_{D_{s}}}\,,
\end{align}
where the strange and charm quark fields entering $C_{PP}(t)$ in Eq.\,(\ref{eq:PP_corr}) carry opposite signs of the chirally-rotated twisted term. In practice, we find it convenient to define the following estimators to extract the physical form factors $F_{V}$ and $F_{A}$
\begin{align}
\bar{R}_{A}(t,\bs{k}) = \bar{Z}_{V}(t)R_{A}(t,\bs{k})\;,\qquad \bar{R}_{V}(t,\bs{k}) = \frac{Z_{A}}{Z_{V}} \bar{Z}_{V}(t)R_{V}(t,\bs{k})~, 
\end{align}
where the values of the ratio $Z_{A}/Z_{V}$ used for each of the ensembles in Table~\ref{tab:simudetails}, are taken from the analysis of Ref.~\cite{Alexandrou:2022amy}, and reported in Table\,\ref{tab:ratios_RCs}.\\ 
\begin{table}[htb!]
    \centering
    \begin{tabular}{||c||c||}
    \hline
    ensemble & $Z_A / Z_V $ \\
    \hline
    ~ cA211.12.48 ~ & ~ $1.0603~(26)$ ~  \\
    \hline
    ~ cB211.072.64 ~ & ~ $1.05176~(35)$ ~ \\
    \hline
    ~ cC211.060.80 ~ & ~ $1.04535~(22)$ ~ \\
    \hline
    ~ cD211.054.96 ~ & ~ $1.04011~(16)$ ~  \\
    \hline   
    \end{tabular}
    \caption{\it \small The values of $Z_A / Z_V$  used in the evaluation of $\bar{R}_{V}(t,\bs{k})$ are given for each of the gauge ensembles of Table\,\ref{tab:simudetails}.}
    \label{tab:ratios_RCs} 
\end{table}

\section{Numerical results} \label{sec:Num_results}

In this section we present the numerical results for the form
factors $F_A$ and $F_V$ at ten evenly spaced values of $x_\gamma$ (Subsec\,\ref{subsec:formfactorsnumerical}). 
We then use these results to calculate the differential decay rate  and branching fraction 
for the process $D_s^+ \to e^+ \nu_e\gamma$ (Subsec\,\ref{subsec:widths}).
 
\subsection{Results for the form factors}
\label{subsec:formfactorsnumerical}
\begin{figure}
\includegraphics[scale=0.50]{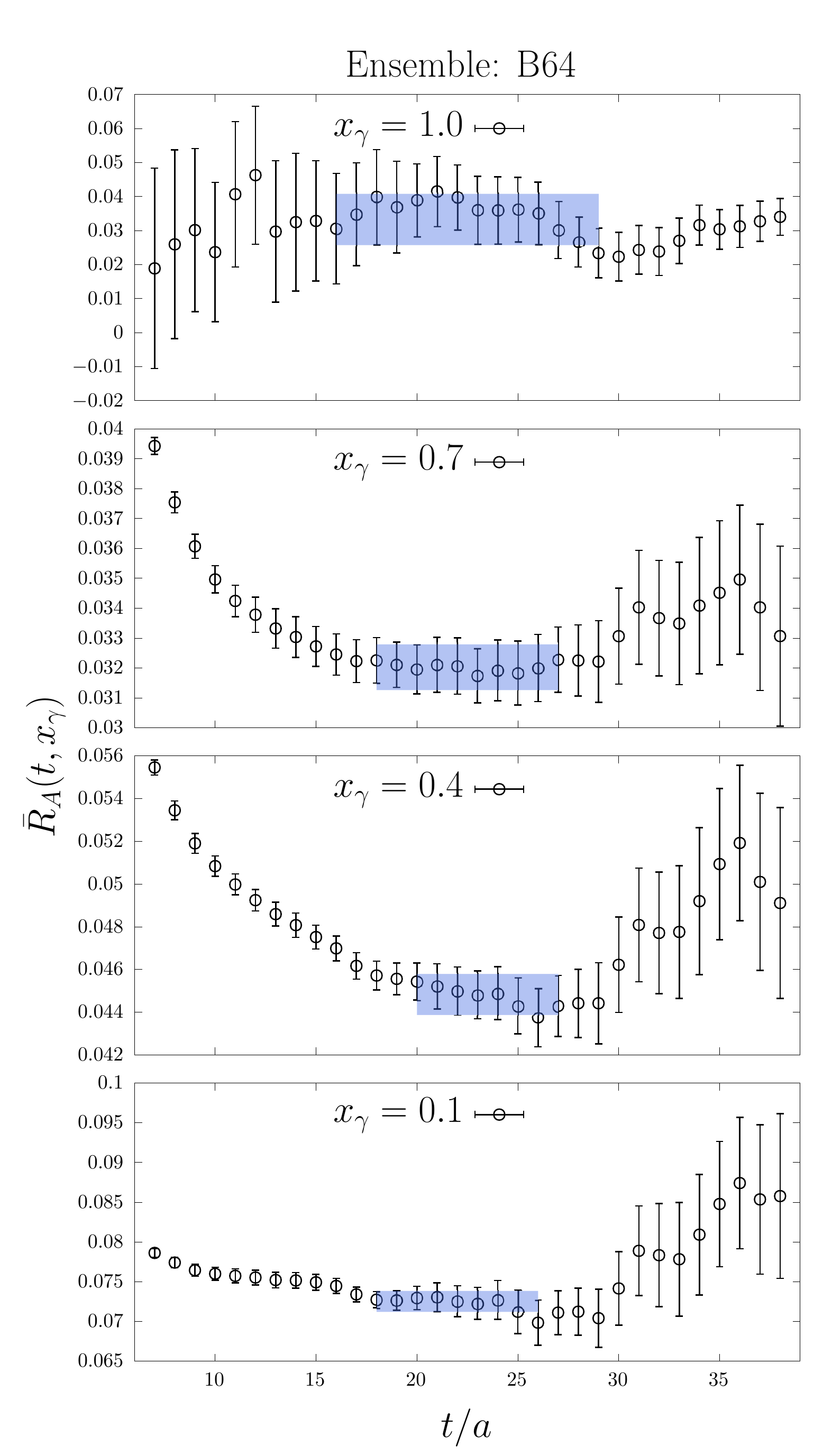}
\includegraphics[scale=0.50]{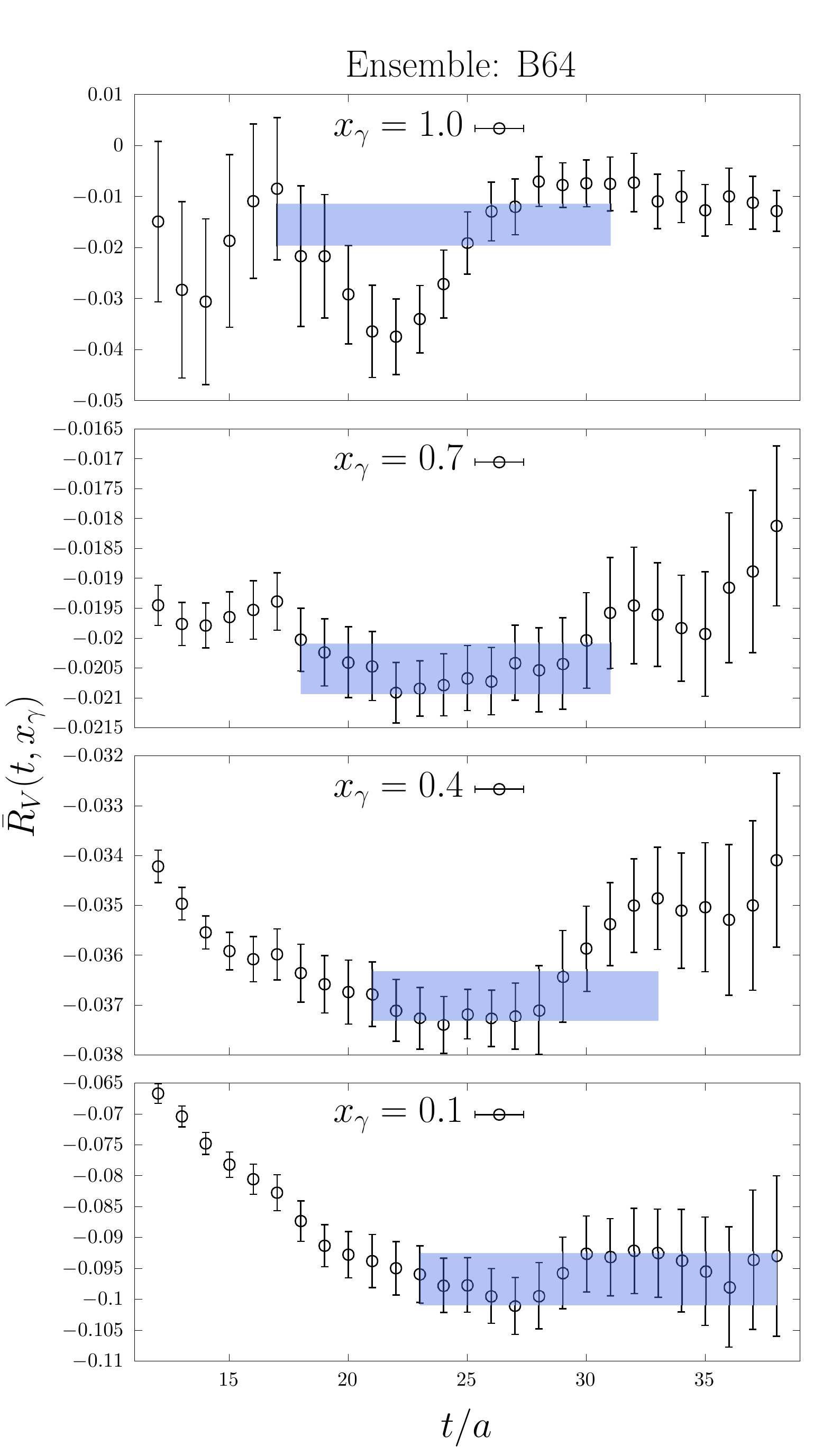}
\caption{ The estimators $\bar{R}_{A}(t,x_{\gamma})$ and $\bar{R}_{V}(t,x_{\gamma})$ as a function of $t/a$ for selected values of $x_{\gamma}$ on the ensemble B64. In each figure the blue band corresponds to the result of a constant fit over the indicated region.  }  
\label{fig:plat_B64}
\end{figure}

\begin{figure}
\includegraphics[scale=0.50]{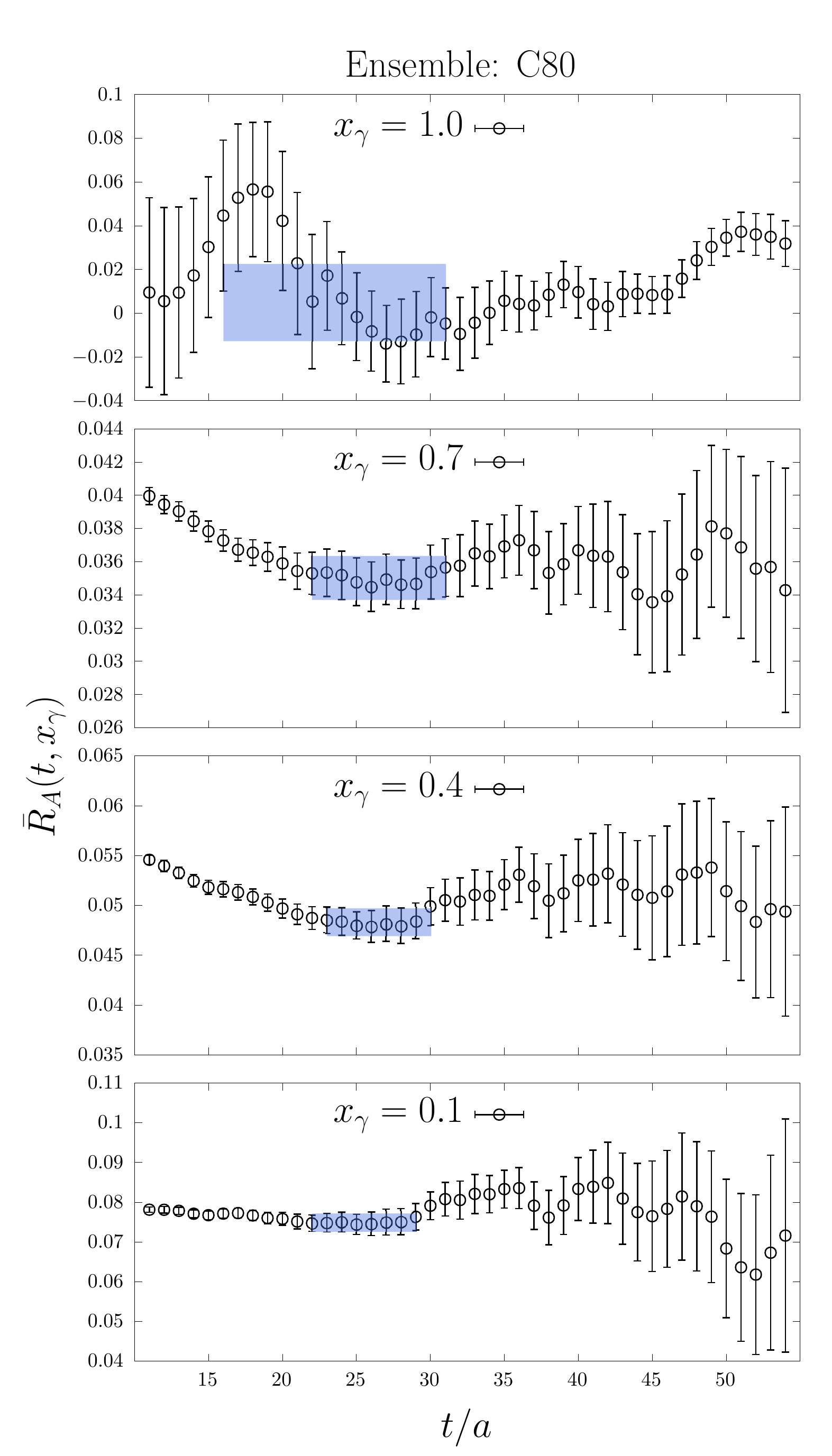}
\includegraphics[scale=0.50]{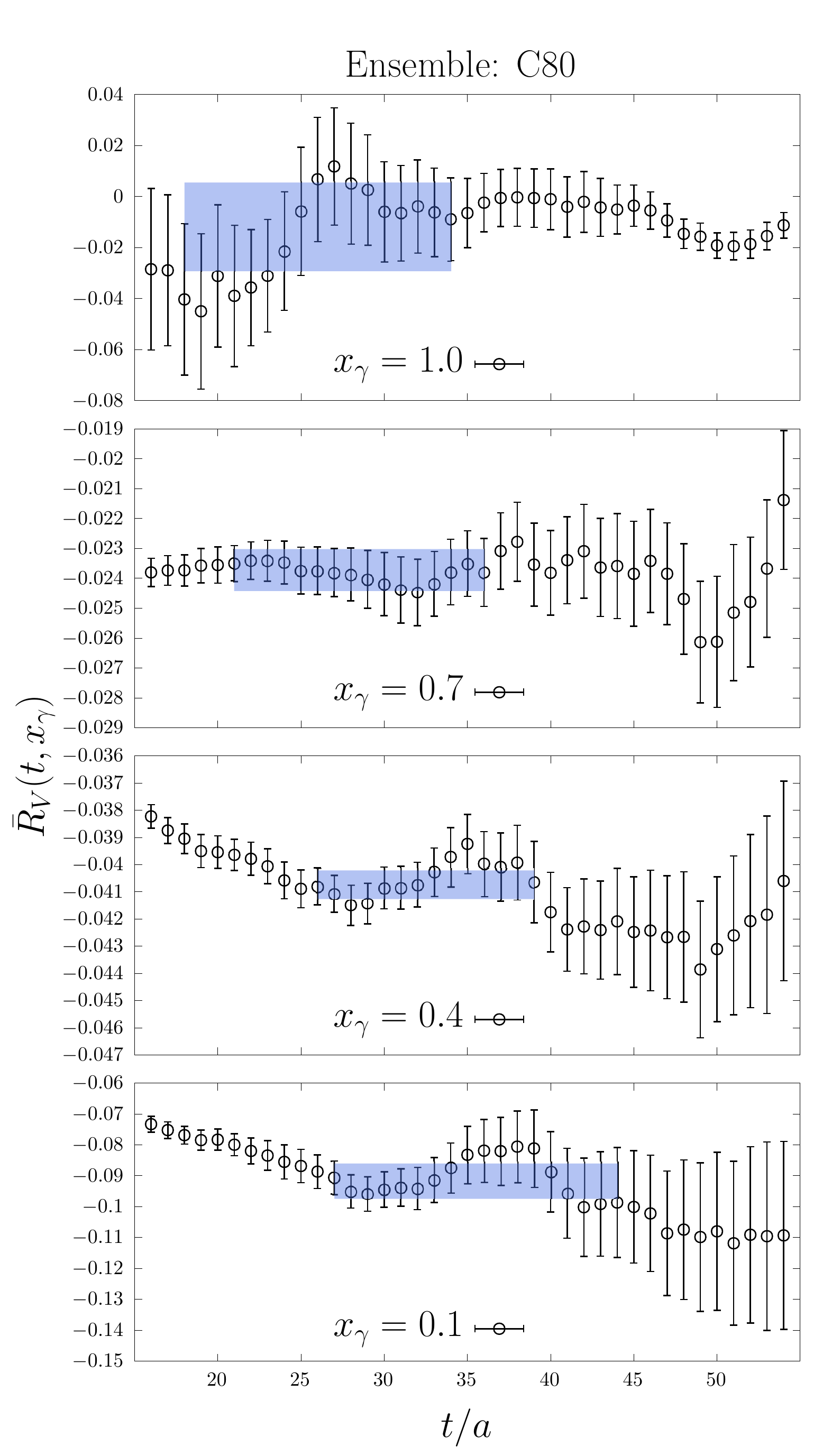}
\caption{The estimators $\bar{R}_{A}(t,x_{\gamma})$ and $\bar{R}_{V}(t,x_{\gamma})$ as a function of $t/a$ for selected values of $x_{\gamma}$ on the ensemble C80. In each figure the blue band corresponds to the result of a constant fit over the indicated region.  }
\label{fig:plat_D64}
\end{figure}

In order to determine the form factors, we evaluate the estimators $\bar{R}_{A,V}(t,\bs{k})$ at ten evenly-spaced values of the dimensionless variable $x_{\gamma}$:
\begin{align}
x_{\gamma} = \frac{2E_{\gamma}}{M_{D_{s}}} =  n\,\Delta x_{\gamma}\;,\qquad \Delta x_{\gamma}=0.1\;, \qquad n \in \{1,\ldots, 10\}\; ~. 
\end{align}
At finite lattice spacing $a$, the relations between the twist angle $\theta_{t}$, the photon momentum $\vec{k}=(0,0,k_{z})$ and $x_{\gamma}$ are obtained from Eqs.\,(\ref{eq:momenta}) and (\ref{eq:redef_R}):
\begin{align}
k_{z} = -\frac{2}{a}\sin{\left( \frac{a\pi}{L}\theta_{t}\right)}\;,\qquad x_{\gamma} = \frac{4}{aM_{D_{s}}}\sinh^{-1}{\left(\frac{a|k_{z}|}{2}\right)}~.
\label{eq:kzxgamma}
\end{align}
For each gauge ensemble, we obtain each of the values of $x_\gamma$ by tuning the twisting angle $\theta_{t}$ using the relations in Eq.\,(\ref{eq:kzxgamma}) and the value of $aM_{D_{s}}$. 
The resulting statistical uncertainty  on the values of $x_{\gamma}$ is negligibly small (typically below $\mathcal{O}(0.1\%)$). 
For an illustration of the quality of the plateaus, we present
in Figs.\,\ref{fig:plat_B64} and \ref{fig:plat_D64} the estimators $\bar{R}_{V,A}(t,x_{\gamma})\equiv \bar{R}_{V,A}(t, (0,0, k_{z}(x_{\gamma}))$ for selected values of $x_{\gamma}$, obtained on the ensembles cB211.072.64 (B64 for short) and cC211.06.80 (C80 for short) respectively. In each figure the blue band shows the values of $F_{V,A}$ obtained from a constant fit in the region where the estimators $\bar{R}_{A,V}(t,x_{\gamma})$ display a plateau\,\footnote{We have checked that the results are stable under small shifts, in both forward and backward direction, of the time intervals adopted in the constant fit.}.
As is clear from the figures, we observe a rapid deterioration of the signal for both $F_{V}$ and $F_{A}$ at large values of $x_{\gamma} \gtrsim 0.7$. In particular the statistical errors on $\bar{R}_{A,V}(t, x_{\gamma} > 0.8)$ turn out to be very large at small values of $t/a$, and then progressively decrease as $t/a$ increases. The origin of this  peculiar behaviour, which is discussed in detail in Appendix\,\ref{app:A}, is due to the existence of a threshold value of $E_{\gamma}$ above which the intrinsic statistical fluctuations of $R^{\mu\nu}_{W}$ start to grow asymptotically with $E_{\gamma}$. As argued in Appendix\,\ref{app:A}, this threshold value of $E_{\gamma}$ is given by the mass $M^{\rm{PS}}_{\bar{q}q}$ of the lightest pseudoscalar meson state $\bar{\psi}_{q}\gamma^{5}\psi_{q}$ with $q=U,D$. The outcome of the analysis is that for $E_{\gamma} > M^{\rm{ PS}}_{\bar{q}q}$,  the fluctuation $\sigma_{R^{\mu\nu}_{W}}$ of $R^{\mu\nu}_{W}$ scales asymptotically as
\begin{align}
\label{eq:scaling_error}
\sigma_{R^{\mu\nu}_{W}}(t,\bs{k},0)= \frac{B_{R^{\mu\nu}_{W}}}{|E_{\gamma} - M^{\rm{PS}}_{\bar{q}{q}}|}\exp\{ \left(E_{\gamma}- M^{\rm{ PS }}_{\bar{q}q}\right)(T/2-t)\} + \ldots~,
\end{align}
where $B_{R^{\mu\nu}_{W}}$ is a prefactor and the ellipsis indicate terms that are subleading in the limit $T \to \infty$.  For the $D_{s}$ meson $M^{\rm PS}_{\bar{q}q} = M_{\eta_{ss'}} \sim 690\,{\rm MeV}$ (see footnote 4 for the definition of $\eta_{ss'}$) so that the threshold value of $x_{\gamma}= x_{\gamma}^{\mathrm{th}}$ at which the error on $R^{\mu\nu}_{W}(t,\bs{k},0)$ starts to grow asymptotically is given by $x_{\gamma}^{\mathrm{th}} =  2\frac{M_{\eta_{ss'}}}{M_{D_{s}}} \sim 0.7$, in agreement with our numerical results.  Notice that in Eq.\,(\ref{eq:scaling_error}) $\sigma_{{R}_{W}^{\mu\nu}}(t,\bs{k},0)$ is finite only because of the finite temporal extent $T$ of the lattice, and the signal-to-noise (S/N) problem is thus amplified on large lattices. In addition, the S/N problem becomes much more severe for heavy-light mesons with a $u$ or $d$ valence quark,
such as $P=D$ or $B$ mesons, where $x_{\gamma}^{\mathrm{th}}$ is proportional to the ratio between the pion mass and $M_{P}$. Large errors are therefore to be expected even for rather small values of $x_{\gamma}$. A way forward to mitigate this problem is briefly discussed at the end of Appendix\,\ref{app:A}. We note that the approach that we propose in Appendix.\ref{app:A} to tame the S/N problem in Eq.\,(\ref{eq:scaling_error}) has been discussed in great detail in Ref.\,\cite{Giusti:2023pot} where it was called the $3d$ method. In Ref.\,\cite{Giusti:2023pot} the authors provide a detailed comparison, on a single coarse ensemble with $a\simeq 0.11\,{\rm fm}$, $T/a = 64$, and $M_{\pi} \simeq 340~{\rm MeV}$, of the unwanted exponential contamination appearing in the $3d$ method and in the approach we use in the present work, based on the study of the three-point correlation function in Eq.\,(\ref{eq:Cmunudef}) (in Ref.\,\cite{Giusti:2023pot} this approach goes by the name of the $4d$ method). On the single ensemble analyzed, the authors of Ref.~\cite{Giusti:2023pot} find that the $3d$ method gives a better control over the unwanted exponentials. Here, we argue that the $3d$ method can also be helpful to tame the exponential $S/N$ problem for $E_{\gamma} > M_{\bar{q}q}^{\rm PS}$.   \\

\begin{figure}
    \centering
    \includegraphics[scale=0.31]{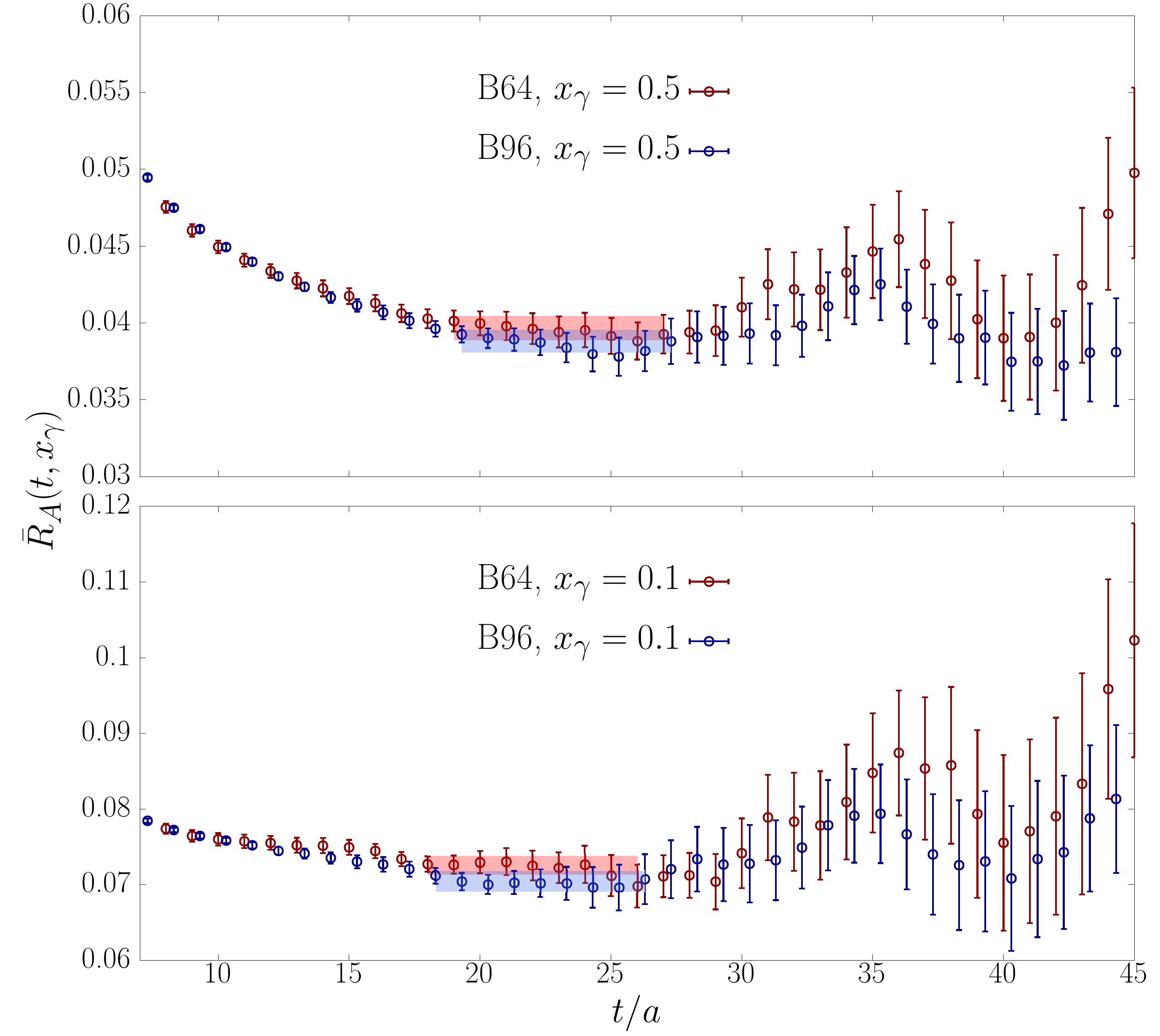}
    \includegraphics[scale=0.31]{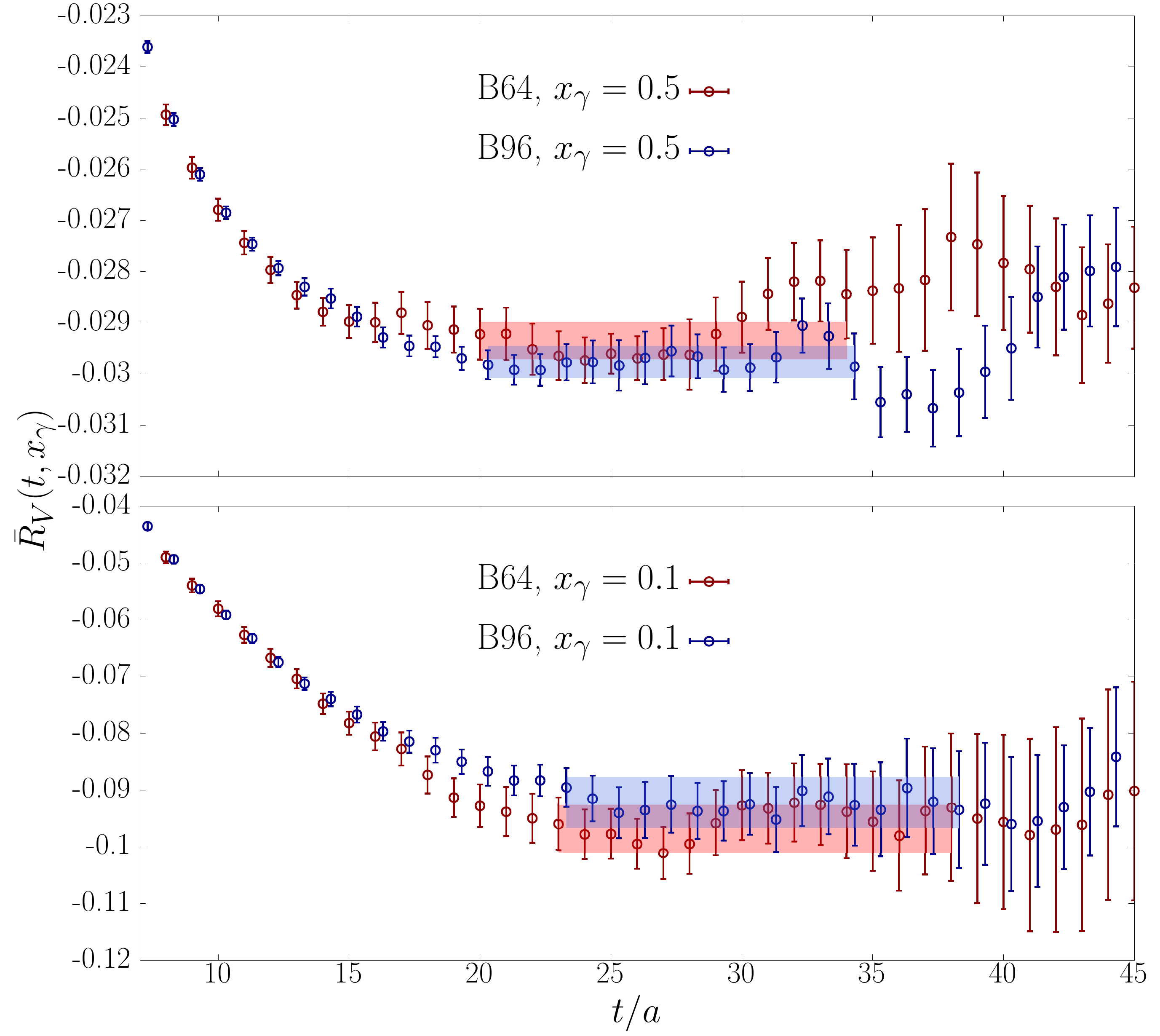}
    \caption{Comparison between the estimators $\bar{R}_{A}(t,x_{\gamma})$ (left panels) and $\bar{R}_{V}(t,x_{\gamma})$ (right panels) as determined on the B64 (red) and B96 (blue) ensembles, for two selected values of the dimensionless variable $x_{\gamma}=0.1$ and $ 0.5$. In each of the four figures the colored bands correspond to the result of a constant fit in the given time interval. The B96 data have been slightly shifted in time for visualization purposes. }
    \label{fig:comp_volumes}
\end{figure}

The ensembles of Table\,\ref{tab:simudetails} all correspond to lattices with a spatial  extent in the range $L \simeq 4.4$\,-\,$5.4\,{\rm fm}$. While these volumes are expected to be large enough for the finite-size effects (FSEs) on $F_{V}$ and $F_{A}$ to be small, in order to  estimate the residual FSEs, we have also used a fifth ensemble, the cB211.072.96 ensemble (B96 for short), with the same parameters as the B64 ensemble, except that $L$ is a factor $3/2$ larger. We have measured both $F_{V}$ and $F_{A}$ on the B96 ensemble up to $x_{\gamma}=0.7$ using $100$ gauge configurations\,\footnote{Beyond $x_{\gamma}= x_{\gamma}^{\rm{th}} = 0.7$ the statistical errors on the B96 ensemble are too big for the results to be useful.  Indeed, since on the B96 ensemble $T/a=192$, the exponential increase of the error with the photon energy
$E_{\gamma}$ described by Eq.\,(\ref{eq:scaling_error}), is much faster than the one present
on the B64 ensemble.}. As a conservative estimate of the FSEs, we associate to the values of $F_{W}(x_{\gamma})$ determined on each of the ensembles of Table\,\ref{tab:simudetails} an additional systematic uncertainty $\sigma^{\rm FSE}_{W}(x_{\gamma})$ given by\,\footnote{ For $x_{\gamma} \geq 0.8$ we associate the same relative systematic uncertainty as determined for $x_{\gamma}=0.7$.}
\begin{align}
\frac{\sigma_{W}^{\rm FSE}(x_{\gamma})}{F_{W}(x_{\gamma})} = \bigg|\frac{\Delta F^{L}_{W}(x_{\gamma})}{F_{W}(x_{\gamma}, \rm{B64})}\bigg|\erf{\left(\frac{\Delta F^{L}_{W}(x_{\gamma})}{\sqrt{2}\sigma^{\rm comb}_{W}(x_{\gamma})}\right)}~,  
\end{align}
where the subscript $W$ represents $V$ or $A$ and
\begin{align}
\Delta F^{L}_{W}(x_{\gamma}) \equiv \bigg| F_{W}(x_{\gamma}, {\rm{B96}}) - F_{W}(x_{\gamma}, {\rm{B64}})\bigg|, \qquad \sigma^{\rm comb}_{W}(x_{\gamma}) \equiv \sqrt{ \sigma^{2, \rm stat}_{W}(x_{\gamma}, \rm{B64}) + \sigma^{2, \rm stat}_{W}(x_{\gamma}, \rm{B96})    }~,
\end{align}
which is the relative spread between the results obtained on the B96 and B64 ensembles, weighted by the probability that the spread is not due to a statistical fluctuation.  In 
Fig.\,(\ref{fig:comp_volumes}) we compare the two estimators $\bar{R}_{A}(x_{\gamma})$ and $\bar{R}_{V}(x_{\gamma})$, at two selected kinematic points $x_{\gamma}=0.1$ and $0.5$, determined on the B64 and B96 ensembles. 
Reassuringly, we find that for all values of $x_{\gamma}$ and for both form factors $\sigma_{W}^{\rm FSE}(x_{\gamma})$ is smaller or of a similar size  than the corresponding statistical uncertainty, and the difference between the results on the two volumes are most probably largely due to statistical fluctuations.\\

\begin{figure}
\includegraphics[scale=0.40]{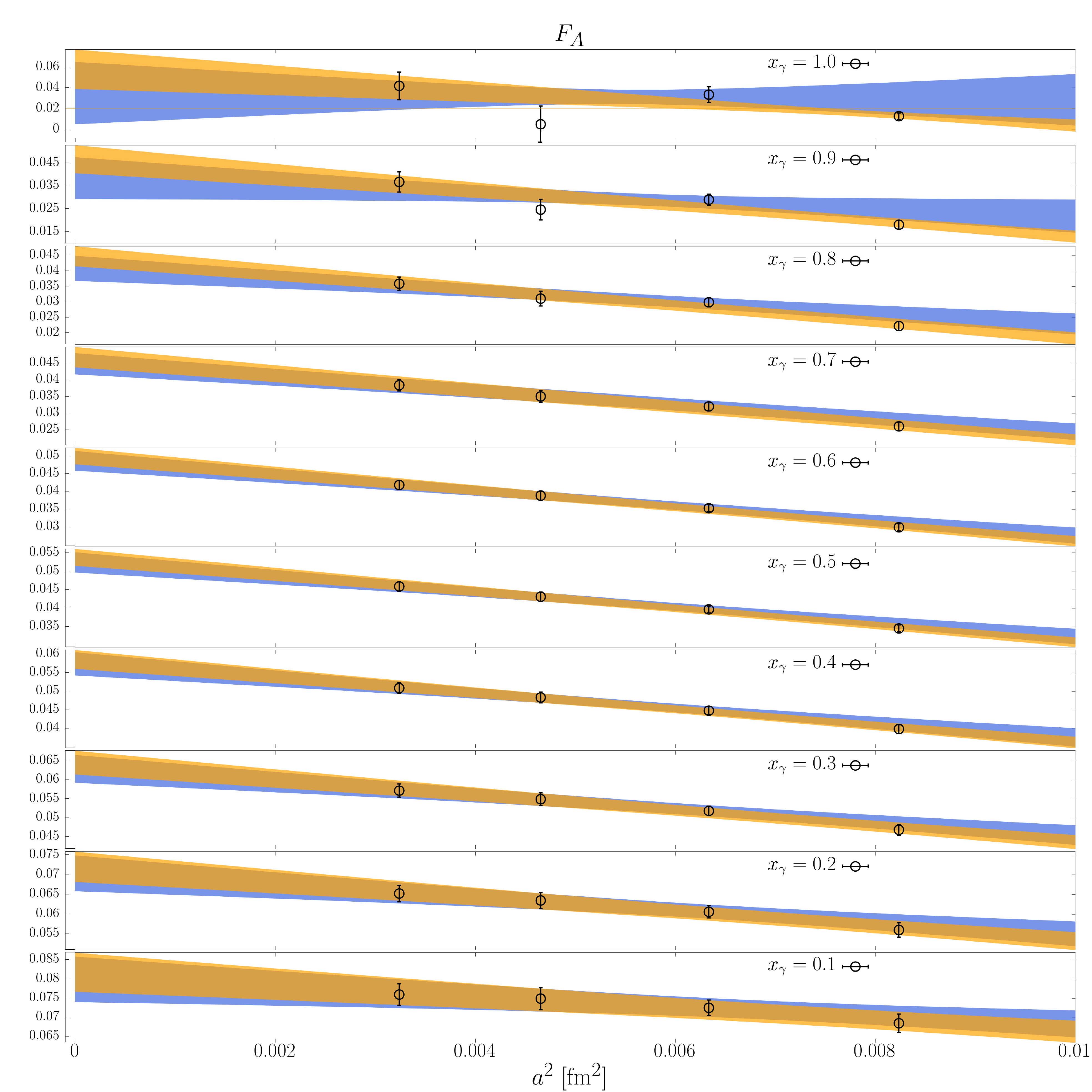}
\caption{The extrapolation of $F_{A}$ to the continuum limit for all ten values of $x_{\gamma}$ considered in this work. The orange and blue bands correspond respectively to the extrapolation curves obtained  including or excluding the data at the coarsest lattice spacing.}
\label{fig:cont_lim_FA}
\end{figure}

\begin{figure}
\includegraphics[scale=0.40]{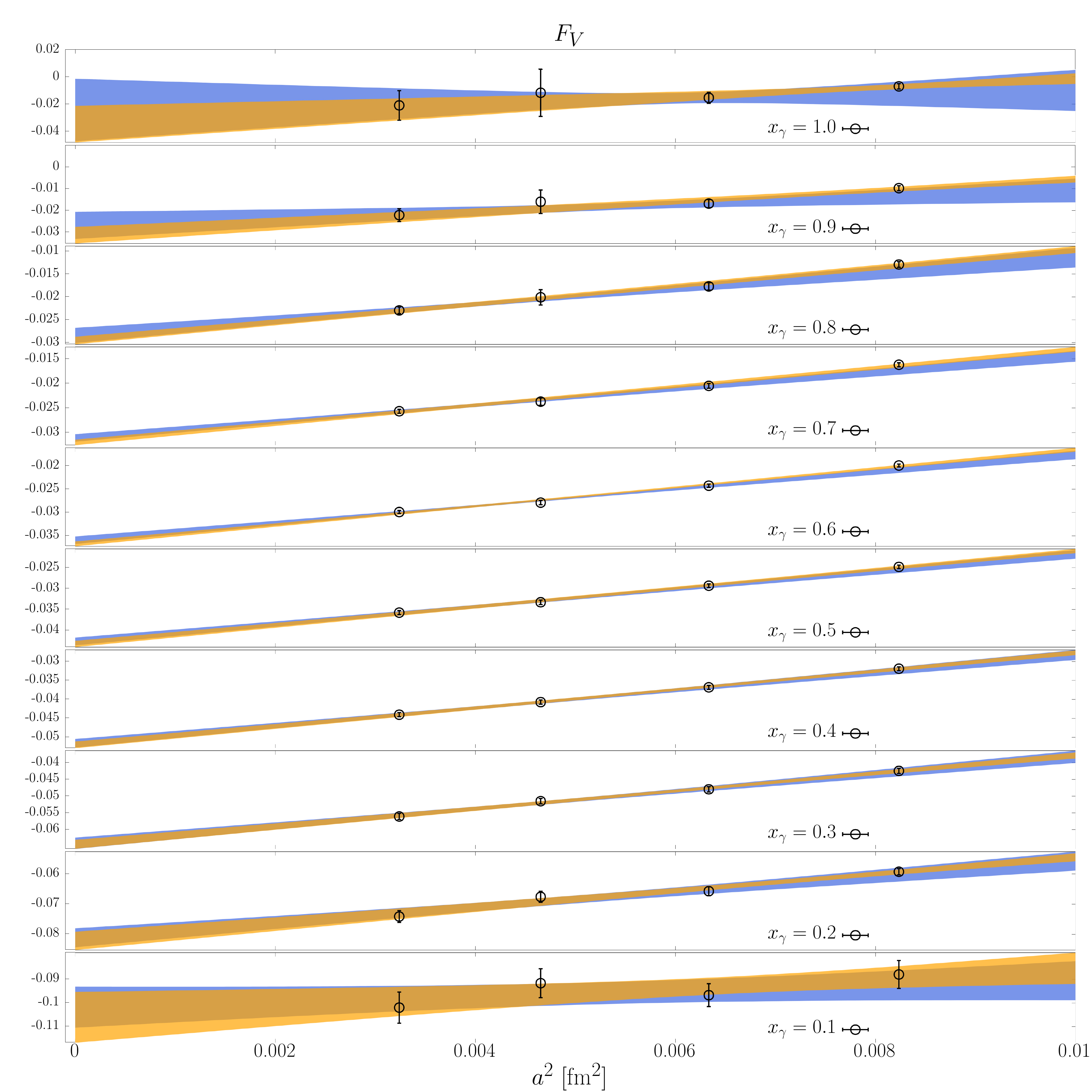}
\caption{The extrapolation of $F_{V}$ to the continuum limit for all ten values of $x_{\gamma}$ considered in this work. The orange and blue bands correspond respectively to the extrapolation curves obtained  including or excluding the data at the coarsest lattice spacing. }
\label{fig:cont_lim_FV}
\end{figure}

Next we consider cut-off effects. For each value of $x_{\gamma}$ the extrapolation to the continuum limit is performed using the following Ansatz
\begin{align}
\label{eq:fit_ansatz}
F_{W}(x_{\gamma}, a) = F_{W}(x_{\gamma})\left( 1 + D_{W}(x_{\gamma})\left(a\Lambda\right)^{2} + D_{2,W}(x_{\gamma})\left(a\Lambda\right)^{4}\right)~,\qquad W= \{ V, A \}~,
\end{align}
with the parameter $\Lambda$ chosen to be $\Lambda = M_{D_{s}} = 1.968\,{\rm GeV}$\,\footnote{With such a choice 
we find that the $D_W(x_\gamma)$ are of $O(1)$ (see Table\,\ref{tab:fit_coefficient} below).}
and $F_{W}(x_{\gamma}),  D_{W}(x_{\gamma})$ and $D_{2,W}(x_{\gamma})$ are dimensionless fit parameters which depend on $x_{\gamma}$ and are different  for the two channels $W = \{ V, A\}$. 
The result of the extrapolation for $F_{A}$ and $F_{V}$, obtained using the 
Ansatz in Eq.\,(\ref{eq:fit_ansatz})
with the fit parameter $D_{2}$ fixed to zero, are shown in Figs.\,\ref{fig:cont_lim_FA} and~\ref{fig:cont_lim_FV}. 
In the two figures, the blue bands correspond to the linear $a^{2}$ extrapolation, performed omitting the measurement at the coarsest value of the lattice spacing. In Table\,\ref{tab:fit_coefficient}, we report the values of the parameters $D_{W}(x_{\gamma})$ obtained from the linear $a^{2}$-fit to the full dataset (orange band in Figs\,\ref{fig:cont_lim_FA} and~\ref{fig:cont_lim_FV}) , along with the corresponding reduced $\chi^{2}$, which is always very good, except for $F_{A}$ at the largest two values of $x_{\gamma}$ ($x_{\gamma}=0.9$ and $1.0$). 
\begin{table}[t]
    \centering
    \resizebox{0.5\textwidth}{!}{
    \begin{tabular}{|c|c|c|c||c|c|c||}
    \hline
    & \multicolumn{3}{c||}{$ F_{A} $} & \multicolumn{3}{c||}{$ F_{V} $}  \\
    \hline
    $x_{\gamma}$ & $D_{A}$ & $\Delta D_{A}$ & $\chi^{2}/$d.o.f. & $D_{V}$ & $\Delta D_{V}$ & $\chi^{2}/$d.o.f. \\
    \hline 
0.1  &  -0.189  &  0.063  &  0.070  &  -0.197  &  0.132  &  0.656 \\ \hline 
0.2  &  -0.262  &  0.057  &  0.105  &  -0.339  &  0.038  &  0.786 \\ \hline 
0.3  &  -0.325  &  0.049  &  0.147  &  -0.414  &  0.024  &  0.396 \\ \hline 
0.4  &  -0.382  &  0.045  &  0.123  &  -0.470  &  0.019  &  0.089 \\ \hline 
0.5  &  -0.430  &  0.040  &  0.179  &  -0.517  &  0.017  &  0.488 \\ \hline 
0.6  &  -0.480  &  0.041  &  0.158  &  -0.551  &  0.016  &  1.597 \\ \hline 
0.7  &  -0.531  &  0.041  &  0.174  &  -0.595  &  0.018  &  0.946 \\ \hline 
0.8  &  -0.594  &  0.055  &  0.750  &  -0.676  &  0.030  &  0.510 \\ \hline 
0.9  &  -0.726  &  0.079  &  1.927  &  -0.824  &  0.063  &  0.756 \\ \hline 
1.0  &  -0.940  &  0.122  &  2.023  &  -0.964  &  0.123  &  0.241 \\ \hline \hline 
    \end{tabular}}
    \caption{\it \small Values of the fit parameters $D_{W}(x_{\gamma})$, their uncertainties $\Delta D_{W}$, and the reduced $\chi^{2}$ obtained in the linear fits to $F_{A}$ and $F_{V}$ for the ten values of $x_{\gamma}$ considered in this work.}
    \label{tab:fit_coefficient} 
\end{table}
For most values of $x_{\gamma}$, the fit parameter $D_{W}$ turns out to be of order $\mathcal{O}(1)$, suggesting the presence of cut-off effects in $F_{V}$ and $F_{A}$ that are of order $\mathcal{O}(a^{2}M_{D_{s}}^{2})$.

We find that including the $D_{2,W}(x_\gamma)(a\Lambda)^4$ terms leads to overfitting without substantially improving the quality of the fit. 
The continuum values obtained using the Ansatz of Eq.\,(\ref{eq:fit_ansatz}) are always consistent within errors with those obtained from linear fits (i.e. with $D_{2,W}(x_\gamma)$ set to 0) shown in Figs.\,\ref{fig:cont_lim_FA} and~\ref{fig:cont_lim_FV}, but have substantially larger statistical uncertainties. Moreover, the coefficient $D_{2,W}(x_\gamma)$ turns out to be always consistent with zero within $1$\,-\,$1.5$ standard deviations, a clear signal of overfitting. Given these observations we have decided to estimate the systematic uncertainty due to the continuum extrapolation, using the two linear extrapolations shown in each of Figs.\,\ref{fig:cont_lim_FA} and~\ref{fig:cont_lim_FV}. Let $f_A$ and $f_B$ represent generically the continuum values of $F_A(x_\gamma)$ or $F_V(x_\gamma)$ at some value of $x_\gamma$ obtained respectively from the linear fit by including or omitting the result at the coarsest lattice spacing.
 We determine the final central value $\bar{f}$ through a weighted average of the form
\begin{align}
\bar{f} = w_{A}~f_{A}~+~w_{B}~f_{B},\qquad w_{A}+w_{B}=1~.
\end{align}
Our estimate of the systematic error, which is added in quadrature to the combined statistical and finite-volume uncertainty, is then obtained using
\begin{align}
\sigma_{\mathrm{syst}}^{2} =  \sum_{i=A,B} w_{i}~( f_{i} - \bar{f})^{2}~.      
\end{align}
The weights $w_{i}$, with  $i=\{A,B\}$, are chosen according to the Akaike Information Criterion\,\cite{Akaike} (AIC) proposed in Ref.\,\cite{Neil:2022joj}, namely
\begin{align}
 \label{eq:AIC}
    w_{i} \propto e^{- \left(\chi_{i}^2 + 2 N^{(i)}_{\rm{pars}} - 2 N^{(i)}_{\rm{data}}\right) / 2} ~ , ~
\end{align}
where $\chi_{i}^{2}$ is the total $\chi^{2}$ obtained in the $i$-th fit, and $N_{\rm{pars}}^{(i)}$ and $N_{\rm{meas}}^{(i)}$ are the corresponding number of fit parameters and measurements\footnote{We have checked that the use of uniform weights, $w_{A}=w_{B}= 1/2$, leads to very similar results.}.\\

\begin{figure}
\includegraphics[scale=0.45]{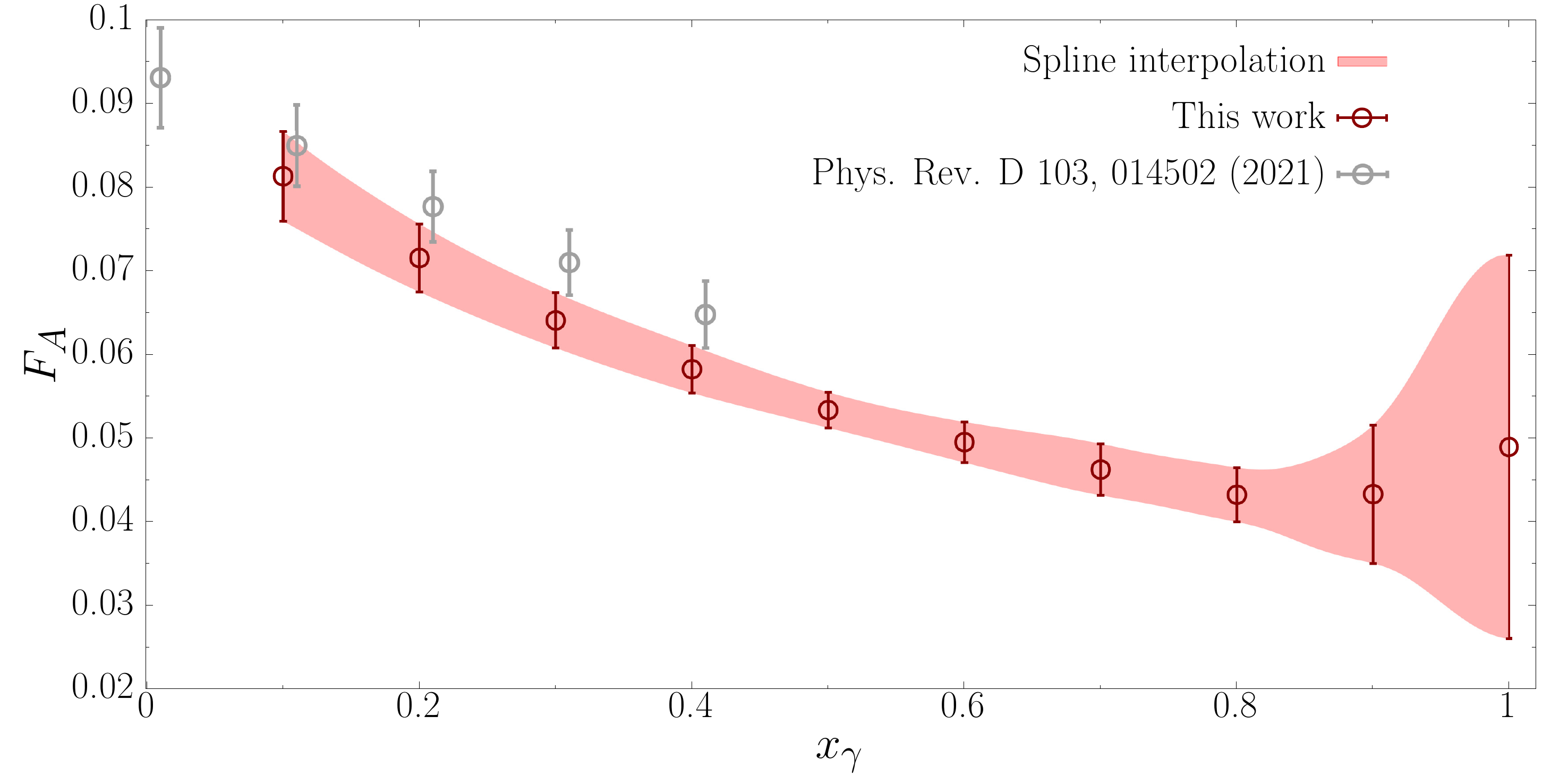}\\
\includegraphics[scale=0.45]{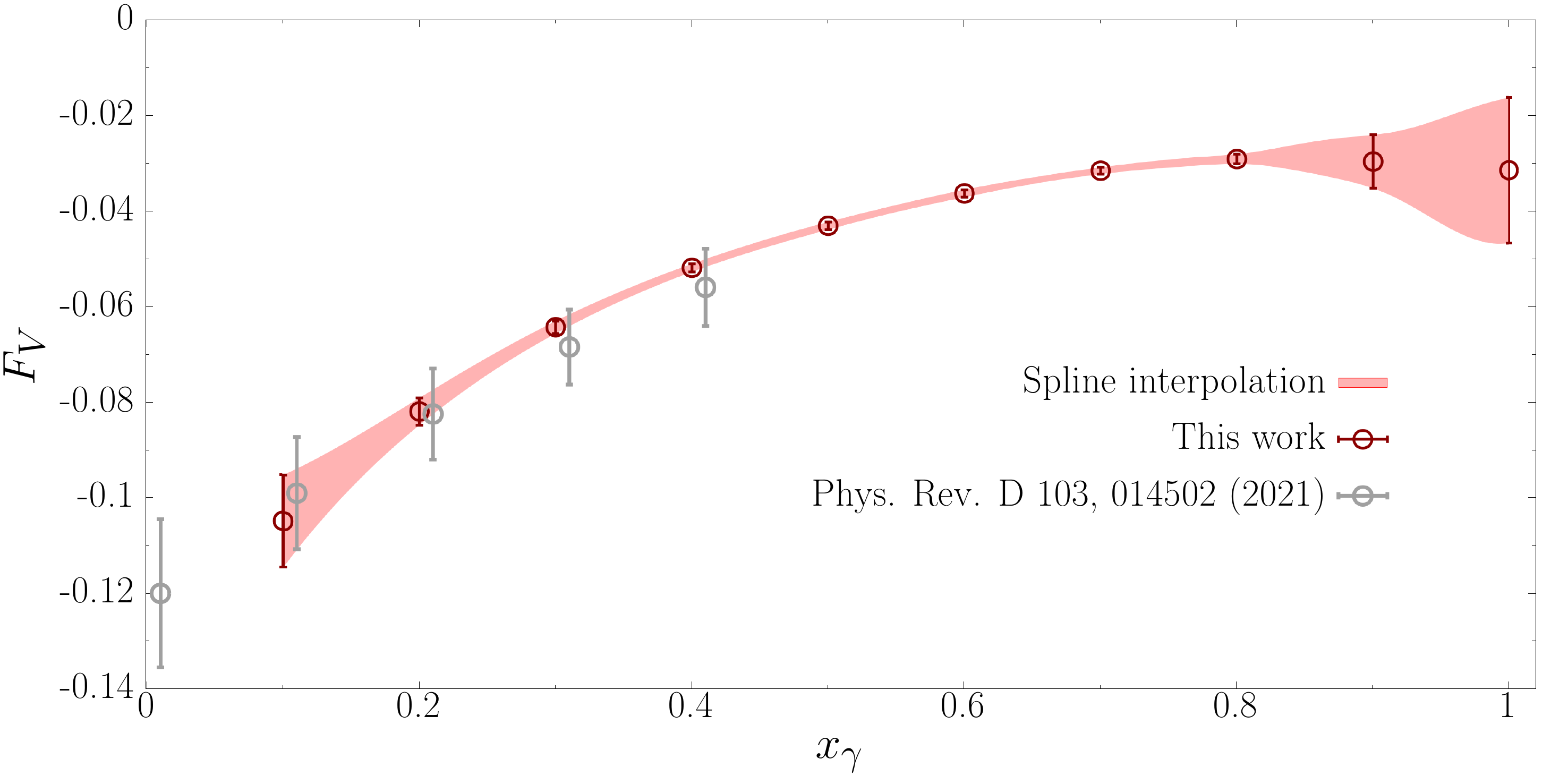}
\caption{The form factors $F_{A}$ (top figure) and $F_{V}$ (bottom figure), obtained after the extrapolation to the continuum limit, shown as a function of the dimensionless variable $x_{\gamma}$. In each of the two figures, the red band is the result of a smooth cubic spline interpolation to our data. The gray data points correspond to the results from Ref.\,\cite{Desiderio:2020oej} which have been slightly shifted horizontally to facilitate comparison.}
\label{fig:cont_spline}
\end{figure}

In Fig.\,\ref{fig:cont_spline} we show our final determination of the axial and vector form factors as a function of $x_{\gamma}$.
The error bars include all the systematic uncertainties discussed above. The results for $F_{A}$ and $F_{V}$ are compared with those of Ref.\,\cite{Desiderio:2020oej}, in which only the phase space region up to $x_{\gamma} \simeq 0.4$ had been explored. As the figures show,  our results are in good agreement with those of Ref.\,\cite{Desiderio:2020oej} for both $F_{A}$ and $F_{V}$, while the statistical uncertainty of the results is significantly improved, particularly for $F_{V}$. In Table\,\ref{tab:result_FA_FV} we collect our final results for the continuum values of $F_{A}$ and $F_{V}$, while in Appendix\,\ref{app:B} we present the full correlation matrix between the form factors evaluated at different values of $x_{\gamma}$, which may be useful for phenomenological analyses.    \\
\begin{table}[t]
    \centering
    \begin{adjustbox}{width=0.4\textwidth}
    \begin{tabular}{|c|c|c|| c|c||}
    \hline
    $x_{\gamma}$ & $ F_{A}  $ & $\Delta F_{A}$ & $ F_{V} $ & $\Delta F_{V
}$  \\
    \hline
   0.1 & 0.0813 & 0.0054 & -0.1048 & 0.0097  \\ \hline 
0.2 & 0.0715 & 0.0041 & -0.0819 & 0.0028  \\ \hline 
0.3 & 0.0641 & 0.0033 & -0.0643 & 0.0013  \\ \hline 
0.4 & 0.0582 & 0.0028 & -0.0519 & 0.0008  \\ \hline 
0.5 & 0.0534 & 0.0021 & -0.0431 & 0.0008  \\ \hline 
0.6 & 0.0495 & 0.0024 & -0.0363 & 0.0008  \\ \hline 
0.7 & 0.0463 & 0.0031 & -0.0316 & 0.0007  \\ \hline 
0.8 & 0.0432 & 0.0032 & -0.0291 & 0.0010  \\ \hline 
0.9 & 0.0433 & 0.0083 & -0.0297 & 0.0056  \\ \hline 
1.0 & 0.0489 & 0.0229 & -0.0315 & 0.0152  \\ \hline 
    \hline
    \end{tabular}
    \end{adjustbox}
    \caption{\it \small Continuum results for $F_{A}$ and $F_{V}$ for the ten values of $x_{\gamma}$ considered in this work.  $\Delta F_{A}$ and $\Delta F_{V}$ are the final errors, which include all systematic uncertainties. The correlations between the form factors at different values of $x_{\gamma}$ are given in Appendix\,\ref{app:B}.}
    \label{tab:result_FA_FV} 
\end{table}

\begin{figure}
\includegraphics[scale=0.55]{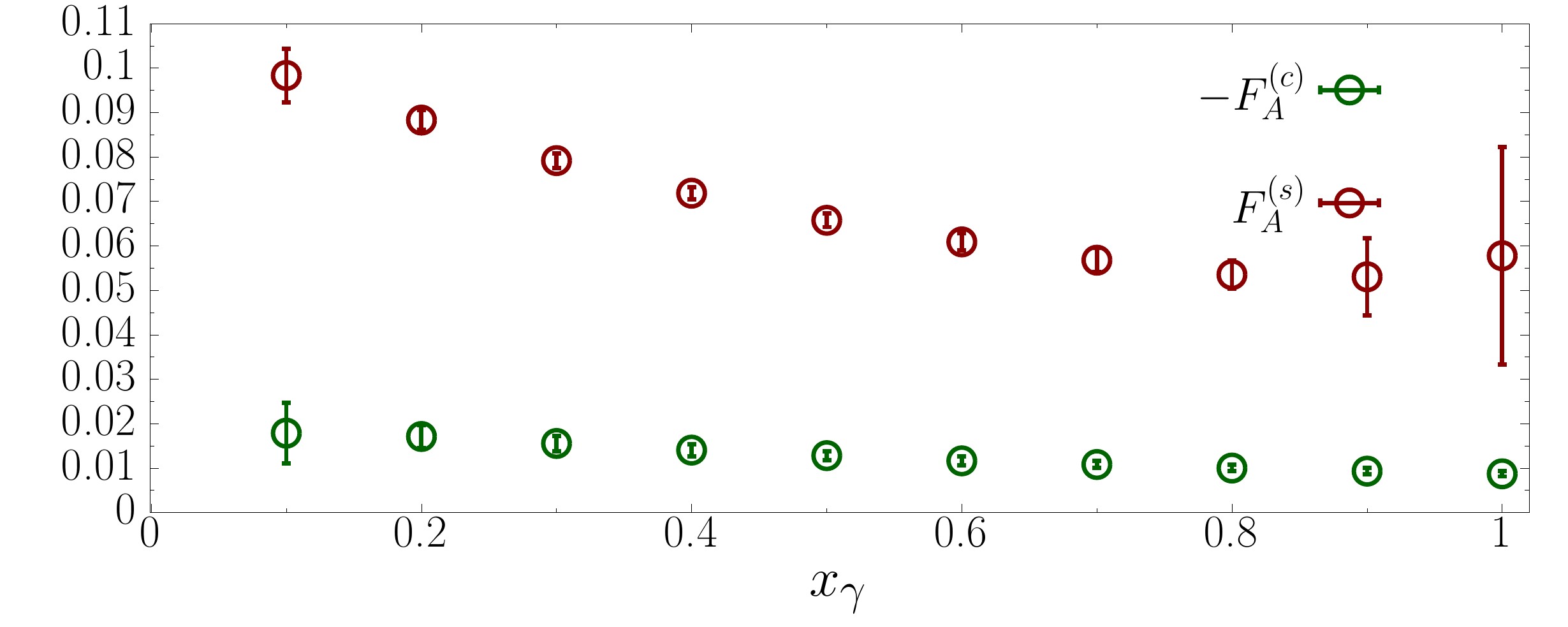}\\
\includegraphics[scale=0.55]{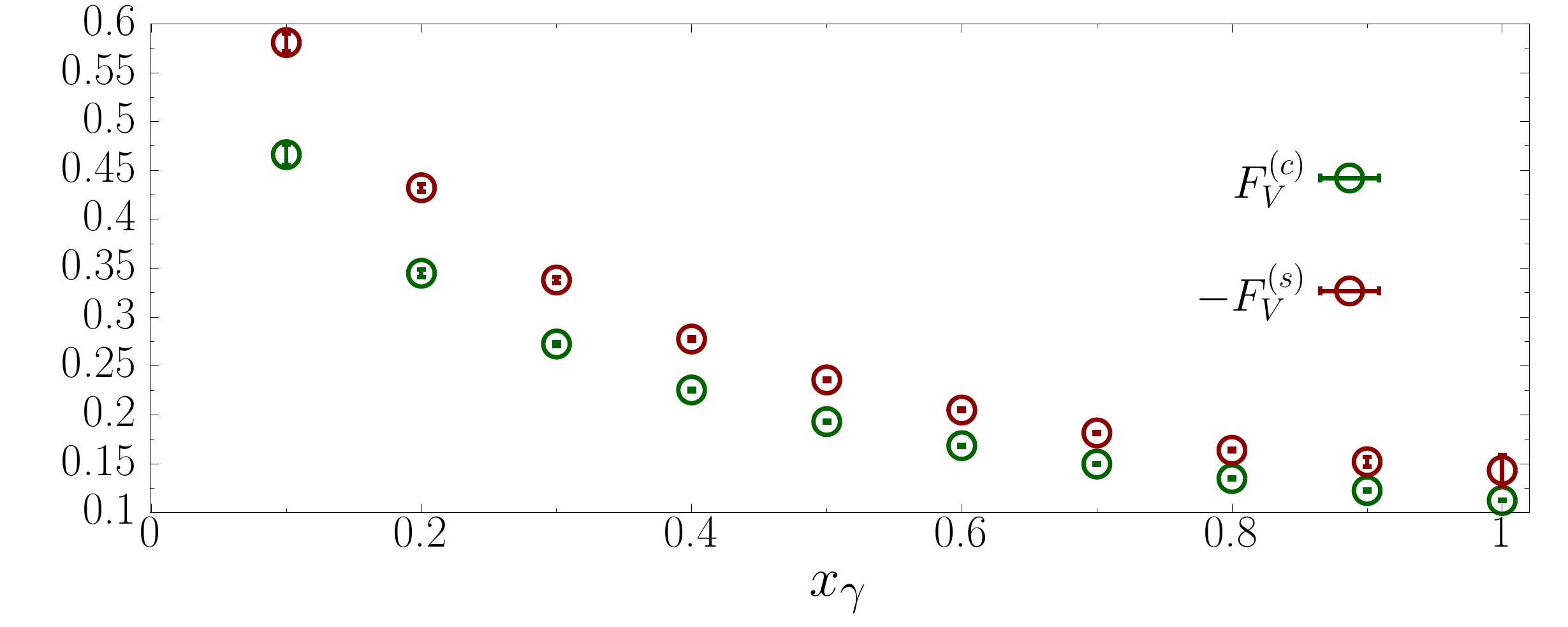}
\caption{The strange- and charm-quark contributions to $F_{A}$ (top figure) and $F_{V}$ (bottom figure) as a function of the dimensionless variable $x_{\gamma}$.}
\label{fig:single_quark_contr}
\end{figure}

We have also determined separately the contributions to the form factors $F_{A,V}$ from the emission of the photon from the strange quark or the charm quark. In practice, the strange-quark (charm-quark) contribution to $R_{A,V}(t,\bs{k})$, indicated in the following by $R_{A,V}^{(s)}(t,\bs{k})$  
($R_{A,V}^{(c)}(t,\bs{k})$), is obtained by setting the electric charge $q_c=0$ ($q_s=0$)
in the evaluation of $C^{\mu\nu}(t,E_{\gamma}, \bs{k}, \bs{p})$ in Eq.\,(\ref{eq:Cmunudef}). The estimators $\bar{R}_{V,A}^{(s)}(t,x_{\gamma})$ and $\bar{R}_{V,A}^{(c)}(t,x_{\gamma})$ are obtained using the usual ratio of RCs $Z_{A}/Z_{V}$ of Table\,\ref{tab:ratios_RCs} and the factor $\bar{Z}_{V}(t)$ of Eq.\,(\ref{eq:est_ZV}) so that 
\begin{align}
R_{A}(t,\bs{k}) = R_{A}^{(s)}(t,\bs{k})+ R_{A}^{(c)}(t,\bs{k})~,\qquad R_{V}(t,\bs{k}) = R_{V}^{(s)}(t,\bs{k})+ R_{V}^{(c)}(t,\bs{k})\,.
\end{align}
In Fig.\,\ref{fig:single_quark_contr} we show the strange- and charm-quark contributions to $F_{A}$ and $F_{V}$, obtained after extrapolating the results to the continuum limit following the same procedure as for the sum.
It is interesting to note that while $F_{A}$ it is completely dominated by the strange-quark contribution, there is a very significant cancellation between the strange and charm contributions to the  vector form factor $F_{V}$. The cancellation between the two contributions becomes more pronounced at large values of $x_{\gamma}$. The implications of this cancellation in the vector form factor will be discussed in the next section. Finally, as is clear from Fig.\,\ref{fig:single_quark_contr},  at large values of $x_{\gamma}$ only the errors on $\bar{R}_{V,A}^{(s)}(t, x_{\gamma})$ increase exponentially, while no degradation of the signal is present for $\bar{R}_{V/A}^{(c)}(t, x_{\gamma})$, as expected from the analysis presented in Appendix\,\ref{app:A}. \\

\subsection{Differential Decay Rate and Branching Fraction}\label{subsec:widths}
From the knowledge of the SD form factors $F_{V}$ and $F_{A}$, the differential decay rate $d\Gamma(D_{s}\to \ell\nu\gamma)/dx_{\gamma}$ can readily be evaluated. The relevant formulae have been derived e.g. in Eqs.\,(1)\,-\,(31) of Ref.\,\cite{Frezzotti:2020bfa}, to which we refer the reader for a more detailed discussion. However, for the sake of completeness we briefly summarize them here. The differential decay rate is expressed as a sum of three different contributions:
\begin{align}
\frac{d\Gamma(D_{s}\to \ell\nu\gamma )}{d x_{\gamma}} = \frac{\alpha_{\rm em}}{4\pi}\Gamma^{(0)}\left\{  \frac{ d R^{\rm pt}}{d x_{\gamma}} +  \frac{ d R^{\rm int}}{d x_{\gamma}} + \frac{ d R^{\rm SD}}{d x_{\gamma}}  \right\} ~,   \end{align}
where $\Gamma^{(0)}$ is the leptonic decay rate in the absence of electromagnetism and is given explicitly by
\begin{align}
\Gamma^{(0)} = \frac{ G_{F}^{2}|V_{cs}|^{2} f_{D_{s}}^{2}}{8\pi}M_{D_{s}}^{3}r_{\ell}^{2}(1-r_{\ell}^{2})^{2}~,  
\end{align}
and the three quantities $dR^{\rm pt}/dx_{\gamma}$, $dR^{\rm int}/dx_{\gamma}$, and $dR^{\rm SD}/dx_{\gamma}$ correspond respectively to the point-like, interference, and SD contribution. The point-like contribution does not depend on the SD form factors, while the interference and the SD contribution depend on $F_{V}$ and $F_{A}$ linearly and quadratically, respectively.  The explicit expression of the three terms is the following ($r_{\ell}\equiv m_{\ell}/M_{D_{s}}$):
\begin{eqnarray}
\label{eq:Rpt}
\frac{ d R^{\rm pt}}{d x_{\gamma}} &=& -\frac{2}{(1 - r_\ell^2)^2}\frac{1}{x_\gamma} \bigg\{ \left[ \frac{(2 - x_\gamma)^2}{1 - x_\gamma} - 4 r_\ell^2 \right] 
                                      (1 - x_\gamma - r_\ell^2) ~ \nonumber \\[2mm]
                                 & - & \left[ 2 (1 - r_\ell^2) (1 + r_\ell^2 - x_\gamma) + x_\gamma^2 \right] 
                                          \log\left( \frac{1 - x_\gamma}{r_\ell^2} \right) \bigg\}~,\\[3mm]
                                          \label{eq:int}
\frac{ d R^{\rm int}}{d x_{\gamma}} &=& - \frac{2 M_{D_{s}}}{f_{D_{s}}(1 - r_\ell^2)^2}\bigg\{  F_{A}\,
x_{\gamma} \left[ \frac{r_\ell^4}{1 - x_\gamma} -1 + x_\gamma + 2 r_\ell^2 \log\left( \frac{1 - x_\gamma}{r_\ell^2} \right) \right]  \nonumber \\[2mm]
&+& (F_{V}-F_{A})\, x_\gamma^2 \left[ \frac{r_\ell^2}{1 - x_\gamma} -1 + \log\left( \frac{1 - x_\gamma}{r_\ell^2} \right) \right]   \bigg\} ~, \\[3mm]
\label{eq:RSD}
\frac{ d R^{\rm SD}}{d x_{\gamma}} &=& \frac{M_{D_{s}}^2}{f_{D_{s}}^{2}}\left( F_{V}^{2}+ F_{A}^{2} \right)\frac{x_{\gamma}^{3}}{ r_\ell^2(1 - r_\ell^2)^2} \frac{(2 + r_\ell^2 - 2 x_\gamma)(1 - x_\gamma - r_\ell^2)^2}{6(1 - x_\gamma)^2}~.
\end{eqnarray}
The total decay rate
\begin{align}
\Gamma_{\ell}(\Delta E_{\gamma}) \equiv \int_{\frac{2\Delta E_{\gamma}}{M_{D_{s}}}}^{1-r_{\ell}^{2}}\,dx_{\gamma}\, \frac{d\Gamma(D_{s}\to \ell\nu\gamma )}{d x_{\gamma}}
\end{align}
can be then evaluated for any desired photon energy cut $\Delta E_{\gamma}$ using the previous formulae and our determination of the form factors $F_{V}$ and $F_{A}$. As Eqs.\,(\ref{eq:Rpt})\,-\,(\ref{eq:RSD}) indicate, the point-like contribution gives rise, in the soft photon limit $\Delta E_{\gamma} \to 0$, to a logarithmically divergent contribution proportional to $\log{\left(\Delta E_{\gamma}\right)}$ and is therefore the dominant contribution in $\Gamma_{\ell}(\Delta E_{\gamma})$ for sufficiently small values of $\Delta E_\gamma$\!
\footnote{ The infrared divergence in the leptonic decay with a real photon in the final state is cancelled by the $\mathcal{O}(\alpha_{\mathrm{em}})$ virtual photon contribution to the purely leptonic decay amplitude, through the Bloch-Nordsieck
mechanism\,\cite{PhysRev.52.54}. The inclusive leptonic decay rate $P\to \ell \nu (\gamma)$ is infrared finite.}.
However, the pointlike contribution is also chirally suppressed with respect to the SD contribution by the factor $r_{\ell}^{2} = (m_{\ell}/M_{D_{s}})^{2}$. Unlike the point-like contribution, the SD contribution to $d\Gamma(D_{s}\to \ell\nu\gamma)/dx_{\gamma}$ is small at small photon energies, then grows reaching a maximum at some value of the photon energy which depends on the specific channel considered, and then decreases to zero at the edge of phase space, i.e. for $x_{\gamma} = 1-r_{\ell}^{2}$. Therefore, for a sufficiently large photon energy cut-off $\Delta E_{\gamma}$ and a small value of $r_{\ell}$, the SD contribution to $\Gamma_{\ell}(\Delta E_{\gamma})$ is the dominant one.

\begin{figure}
    \centering
    \includegraphics[scale=0.5]{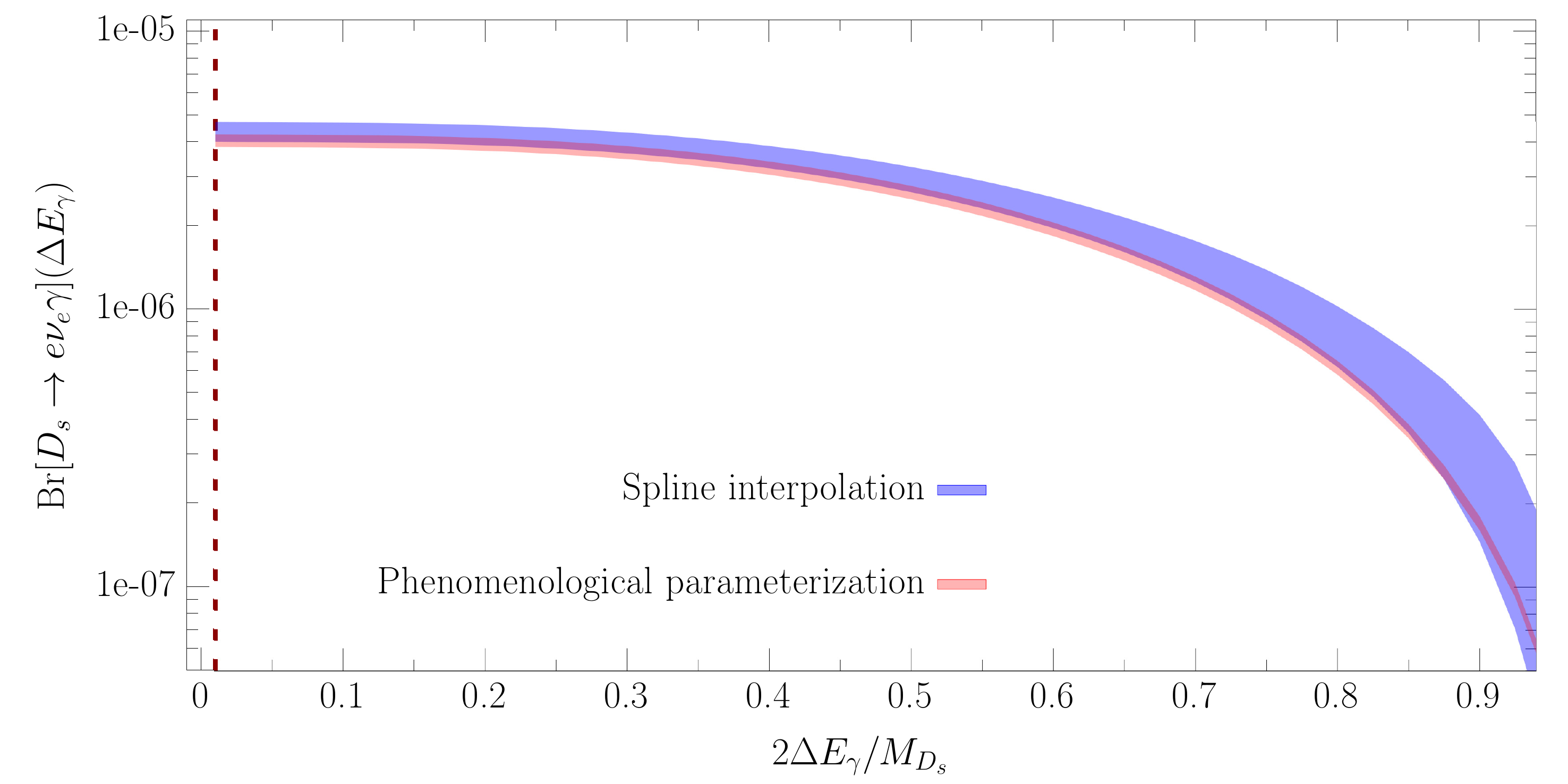}
    \caption{The branching fraction ${\rm Br}[D_{s}\to e \nu_{e}\gamma](\Delta E_{\gamma})$ for different values of the lower cut-off $\Delta E_{\gamma}$ on the photon energy. The red vertical dashed line represents the experimental cut-off $\Delta E_{\gamma} = 10\,{\rm MeV}$ imposed in the BESIII experiment. The blue and red bands correspond respectively to the branching fraction as obtained by employing the spline interpolation of the lattice results for the form factors or their phenomenological parametrization obtained by fitting the Ansatz of Eq.\,(\ref{eq:ansatz}) (see also Sec.\,\ref{sec:phenomenology} and Tab.\,\ref{tab:phen_fitparameters} for more details).}
    \label{fig:decay_rate}
\end{figure}

For the radiative leptonic decays of the $D_{s}$ meson, the only experimental measurement that is currently available is the branching fraction for $D_{s}\to e \nu_{e} \gamma$, for which the BESIII collaboration has given the upper bound at 90\% confidence level\,\cite{BESIII:2019pjk}
\begin{align}
{\rm Br}[D_{s}\to e \nu_{e}\gamma](\Delta E_{\gamma}) \equiv \frac{\Gamma_{e}(\Delta E_{\gamma})}{\Gamma_{\rm tot}}  < 1.3\times 10^{-4}, \qquad  \Gamma_{\rm tot}^{-1} = (5.04 \pm 0.04) \times 10^{-13}~{\rm s}~\text{ \cite{ParticleDataGroup:2020ssz}}~,
\end{align}
including photons with energies $E_{\gamma} > \Delta E_{\gamma} = 10\,{\rm MeV}$. Because of the small mass of the electron, $r_{e} \simeq 2.6\times 10^{-4}$ compared to $r_{\mu} \simeq 5.4\times 10^{-2}$ and $r_{\tau} \simeq 0.9$, the electron channel is the most sensitive to the vector and axial form factors $F_{V}$ and $F_{A}$ and is therefore the most interesting one phenomenologically.
In Fig.\,\ref{fig:decay_rate} we show our determination of the branching fraction as a function of the cut-off on the photon energy, starting from the cut $\Delta E_{\gamma} = 10\,{\rm MeV}$ employed by the BESIII collaboration in Ref.\,\cite{BESIII:2019pjk}, which is indicated in the figure by the dashed red line. 
For this calculation, we used the following values of the CKM matrix element $V_{cs}$ and of the $D_{s}$ decay constant, which we have taken from the 2021 FLAG review~\cite{FlavourLatticeAveragingGroupFLAG:2021npn}
\begin{align}
|V_{cs}| =  0.9741(65)~,\qquad f_{D_{s}} = 249.9(0.5)~{\rm MeV} \qquad  [N_{f}=2+1+1 ~{\rm{averages}}]~.  
\end{align}
We have computed the branching fraction by using the form factors $F_{A,V}$ determined either from the spline interpolation of our numerical lattice results or by fitting to the phenomenological parametrization of the form factors given in Eq.\,(\ref{eq:ansatz}) below and discussed in the next section.
The two different determinations of the branching fraction are represented in Fig.\,\ref{fig:decay_rate} by the blue and red bands respectively. 
The phenomenological parametrization of the form factor leads to a much more precise determination of the form factors in the kinematical region of high values of $x_\gamma$ than the spline interpolation of the lattice data. This is due to the fact that the fit parameters are mainly determined from the most statistically accurate data points, which are the ones at low and intermediate values of $x_\gamma$, while the spline interpolation is designed to always reproduce the actual data points with their corresponding error range. As a consequence, the branching fraction obtained through the phenomenological parametrization of the form factors is more precise than the one obtained by using the spline interpolation, and the difference in the precision increases as the lower cut-off on the photon energy $\Delta E_\gamma$ increases. On the other hand, results obtained by using the spline interpolation of the lattice data are less affected by potential systematic effects due to model dependence and for this reason we conservatively consider these as our final results. Note however, that the two determinations of the branching fraction are always compatible within errors.\\

\begin{figure}
    \centering
    \includegraphics[scale=0.5]{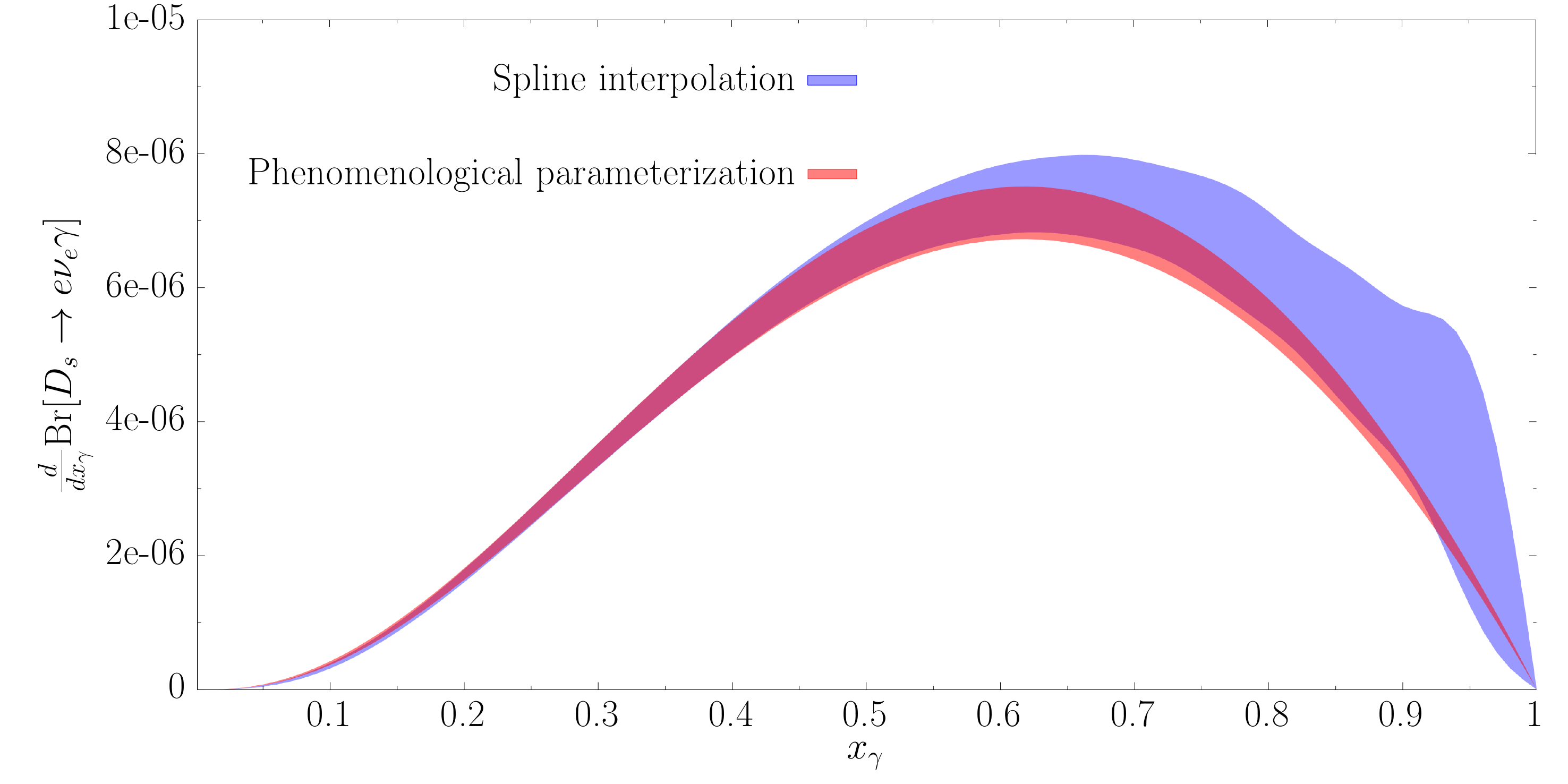}
    \caption{The SD contribution to the differential branching fraction for the decay $D_s\to e\nu_e\gamma$ as a function of $x_{\gamma}$. The blue and red bands correspond respectively to the decay rate obtained by employing the spline interpolation of the lattice results for the form factors or their phenomenological parametrization obtained by fitting the Ansatz of Eq.\,(\ref{eq:ansatz}) (see also Sec.\,\ref{sec:phenomenology} and Tab.\,\ref{tab:phen_fitparameters} for more details).}
    \label{fig:diff_decay_rate}
\end{figure}

From Fig.\,\ref{fig:decay_rate} we see that the results obtained for ${\rm Br}[D_{s}\to e\nu_{e}\gamma](10\,{\rm MeV})$ using either the spline interpolation (${\rm Br}[D_{s}\to e\nu_{e}\gamma](10\,{\rm MeV})=4.4(3) \times 10^{-6}$) or the phenomenological parametrization of the form factors (${\rm Br}[D_{s}\to e\nu_{e}\gamma](10\,{\rm MeV})=4.1(2) \times 10^{-6}$)
are well within the upper bound at 90\% confidence level set by the BESIII collaboration (${\rm Br}[D_{s}\to e\nu_{e}\gamma](10\,{\rm MeV})< 1.3\times 10^{-4}$). Moreover, they are also much smaller than the quark-model predictions of Refs.~\cite{Geng:2000if,Lu:2002mn} which estimate a branching fraction of order $10^{-5}-10^{-4}$, and of the determination of Ref.~\cite{Korchemsky:1999qb} where a branching fraction of order $10^{-3}$ is obtained combining perturbative QCD with the heavy-quark effective theory. We find that, already for $\Delta E_{\gamma} = 10\,{\rm MeV}$ and even more so for higher-energy cuts, the decay rate is completely dominated by the SD term. The point-like contribution is always below one percent and the interference contribution is even smaller. For comparison, adopting the same cut on the photon energy ($\Delta E_{\gamma} = 10~{\rm MeV}$), the corresponding branchings into muon or $\tau$ are respectively $1.86~(3)\times 10^{-4}$ and $ 1.20~(2)\times 10^{-6}$, and in both cases are dominated by the point-like contribution. Finally, in Fig.\,\ref{fig:diff_decay_rate} we provide our determination of the SD contribution to the differential branching fraction in the electron channel, which has a maximum for a value of $x_{\gamma}$ of about $0.6-0.7$. The blue and red bands in Fig.\,\ref{fig:diff_decay_rate} represent respectively the results obtained by using the spline interpolation of the lattice results for the form factors or the phenomenological parametrization based on their fit to the Ansatz of Eq.\,(\ref{eq:ansatz}) below. As before, in the region of high values of $x_\gamma$ the results based on the phenomenological parametrization of the form factors become much more precise than the ones based on their spline interpolation. However, the two determinations are always compatible within errors.
Additional branching fractions, with specific cuts on the photon and/or lepton energies, are available on request from the authors.

\section{Phenomenological parameterization of the form factors} \label{sec:phenomenology}
In this section we present a parameterization of the form factors  inspired by single-pole dominance. In this approximation the form factor is described in terms of the propagation of the nearest resonance. We find that a good description of the momentum dependence of our lattice data is obtained by employing the following Ansatz:
\bea
F_{W}(x_\gamma)=\frac{C_{W}}{\sqrt{R_{W}^2+\dfrac{x_\gamma^2}{4}}\left(\sqrt{R_{W}^2+\dfrac{x_\gamma^2}{4}}+\dfrac{x_\gamma}{2}-1\right)}+B_{W}\,,\quad W=\{A,V\}\label{eq:ansatz},
\eea
in which the single-pole dominance approximation corresponds to fixing $R_W=M^\prime/M_{D_s}$ where $M^\prime$ is the mass of the nearest resonance, and setting $B_W=0$ . Here instead $C_{W}$, $R_{W}$ and $B_{W}$ are three free parameters to be determined from the fit. The constant term $B_{W}$ represents the leading, non-singular correction in the Laurent expansion of a function around a pole. We refer the reader to Appendix\,\ref{app:C} for a more detailed analysis of the different Ans\"atze which we have examined. 

The Ansatz of Eq.\,(\ref{eq:ansatz}) has also been used to fit 
separately the contributions from the emission of the photon from the charm and strange valence quarks. These are simply obtained by setting the charge of the other quark to zero.
We label these separate contributions by $F_{A,V}^{(c)}$ and 
$F_{A,V}^{(s)}$, where the superscript indicates the quark from which the photon has been emitted. 
Interestingly, we find that including the parameter $B_W$ is only required to obtain an acceptable fit for the separate contributions $F_V^{(c)}$ and $F_V^{(s)}$. For all the other cases ($F_A^{(c)}$, $F_A^{(s)}$, $F_A$ and $F_V$), $B_W$ can be set to 0 without increasing the $\chi^2$/d.o.f., resulting in good parametrizations of the form factors. 
\begin{table}[t]
    \centering
    \begin{tabular}{ c| c| c| c|c|c|c|c}

     \rule[-2mm]{0mm}{20pt} & $C_W$ & $R_W$ & $B_W$& cor($C_W,R_W$)& cor($C_W,B_W$)& cor($R_W,B_W$)& $\chi^2/$d.o.f.\\
     \hline
      \rule[-2mm]{0mm}{20pt} $\ F_A\ $ & $\ 0.0518(30)\ $ &  $\ 1.413(30)\ $ & $0$ (fixed)& $\ 0.766\ $& $-$& $-$& $0.41$\\
      \hline
    \rule[-2mm]{0mm}{20pt} $\ F^{(c)}_A\ $ & $\ -0.0135(10)\ $ &  $\ 1.453(59)\ $ &  $0$ (fixed)& $-0.732$& $-$& $-$& $0.23$\\
      \hline
    \rule[-2mm]{0mm}{20pt} $\ F^{(s)}_A\ $ & $\ 0.0662(56)\ $ &  $\ 1.423(36)\ $ &  $0$ (fixed)&$\ 0.975\ $& $-$& $-$& $0.24$\\
      \hline 
      \rule[-2mm]{0mm}{20pt} $\ F_V\ $ & $\ -0.01792(76)\ $ &  $\ 1.091(11)\ $ & $0$ (fixed) & $\ -0.936\ $& $-$& $-$& $0.45$\\
    \hline 
    \rule[-2mm]{0mm}{20pt} $\ F^{(c)}_V\ $ & $\ 0.0624(15)\ $ &  $\ 1.0809(43)\ $ & $\ 0.0369(14)\ $ & $\ 0.922\ $& $\ -0.864\ $& $\ -0.819\ $& $0.31$\\
    \hline 
    \rule[-2mm]{0mm}{20pt} $\ F^{(s)}_V\ $ & $\ -0.0792(24)\ $ &  $\ 1.0794(37)\ $ & $\ -0.0367(31)\ $ & $\ -0.911\ $& $\ -0.936\ $& $\ 0.831\ $& $1.8$\\
    \hline 
    \end{tabular}
    \caption{Values of the parameters $C_W$, $R_W$ and $B_W$ for the two form factors $F_A$ and $F_V$, and for the individual contributions $F_{A,V}^{(c)}$ and $F_{A,V}^{(s)}$, obtained by fitting the Ansatz of Eq.\,(\ref{eq:ansatz}). For the axial data, and for the total vector form factor, the parameter $B_W$ is not included in the fit and has been set to zero, since it is not necessary to describe the data. Correlations among the fitted parameters are also reported in this table.}
    \label{tab:phen_fitparameters}
\end{table}
\begin{figure} 
 \hspace{-0.5cm}
	\subfloat{%
		\includegraphics[scale=0.38]{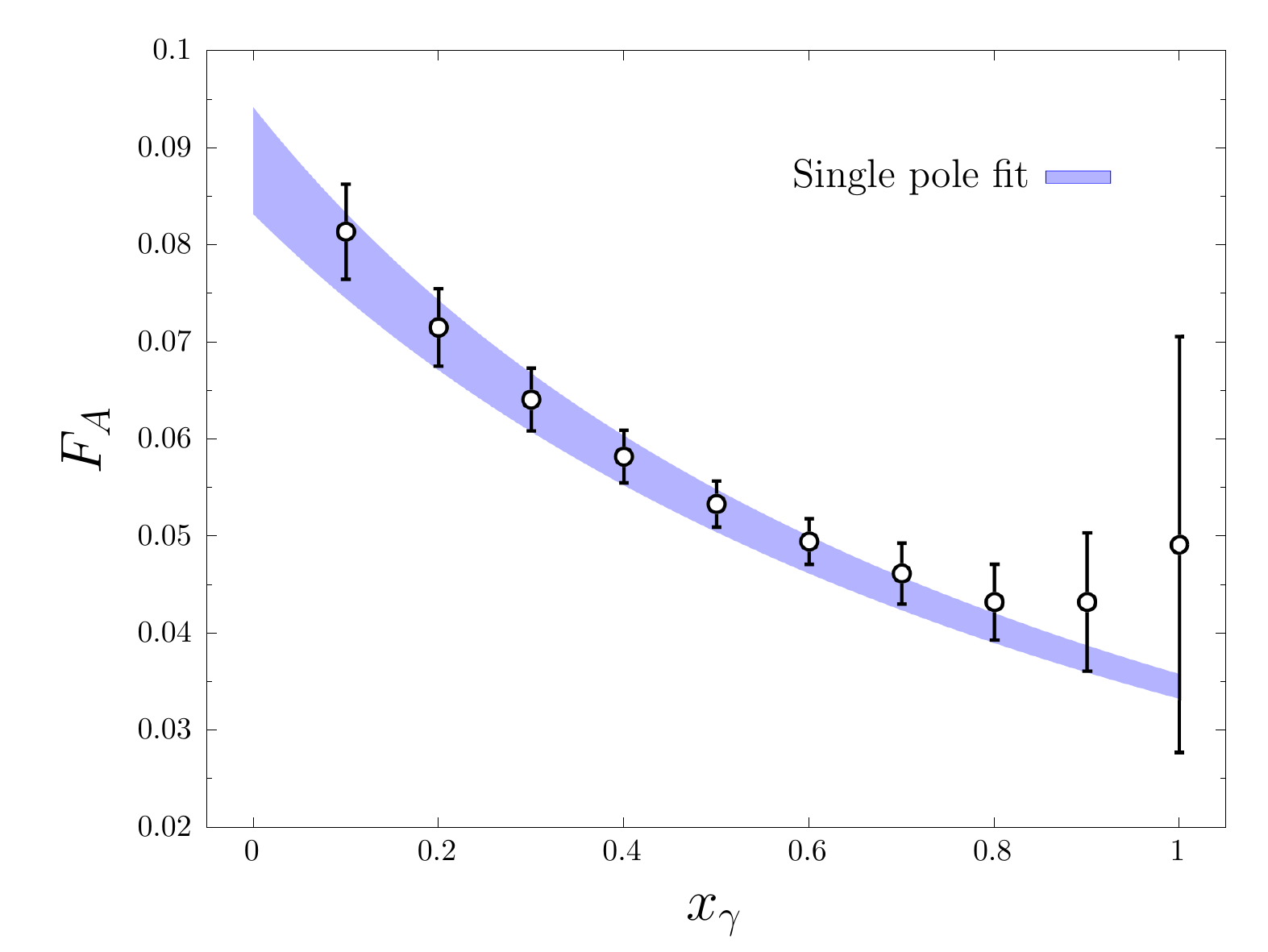}}
	\subfloat{%
		\includegraphics[scale=0.38]{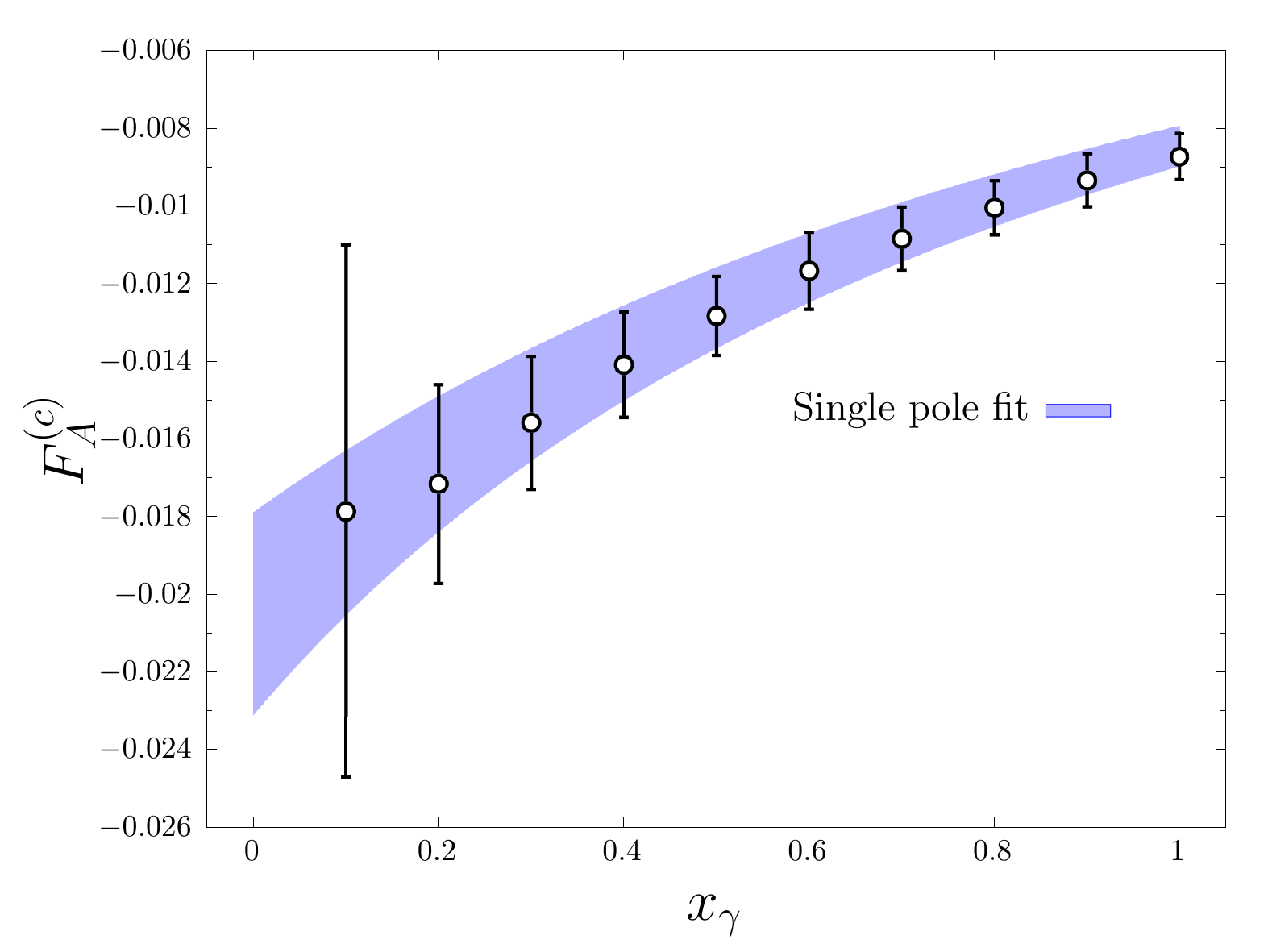}}
	\subfloat{%
		\includegraphics[scale=0.38]{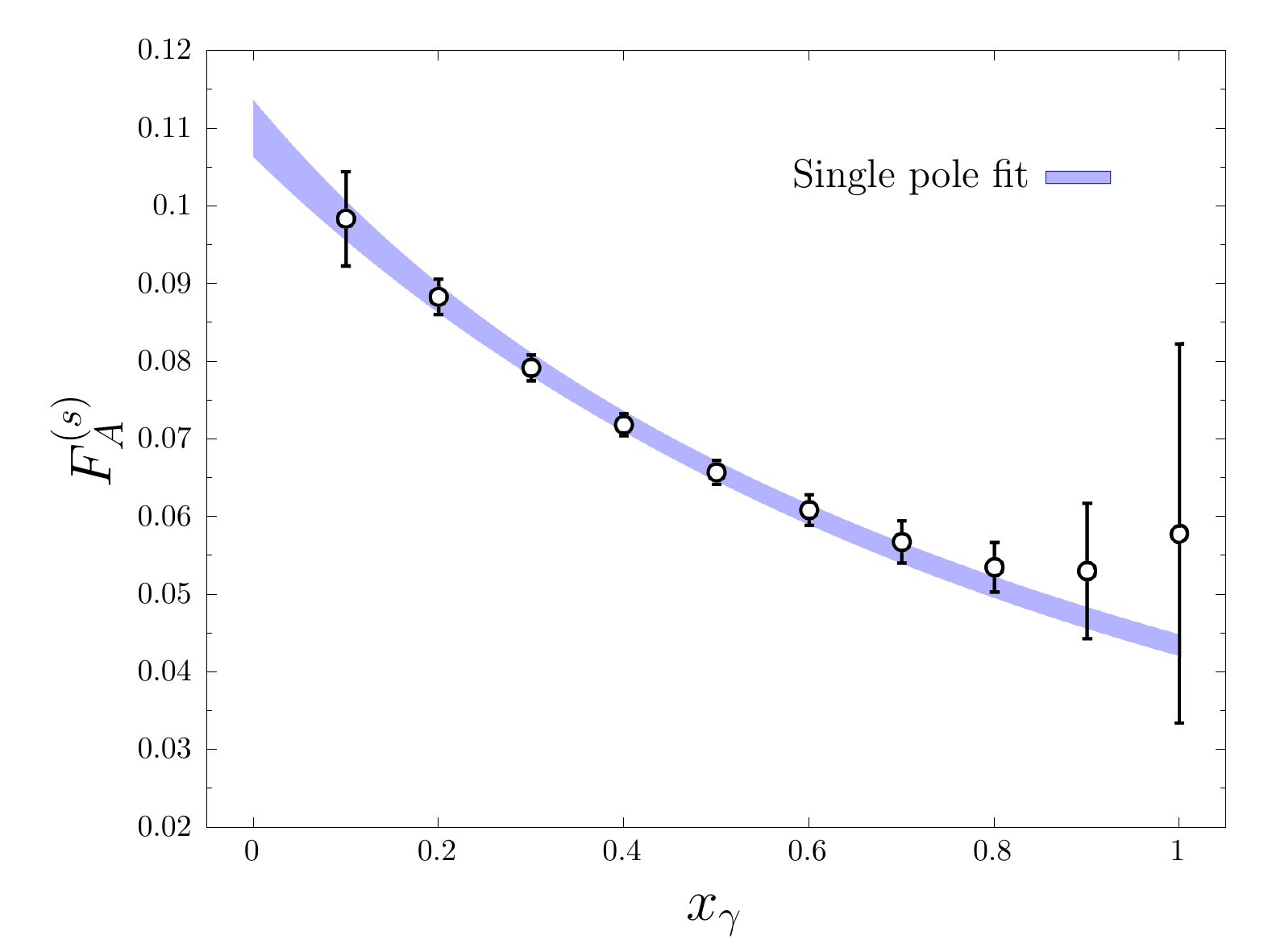}} \\
\hspace{-0.5cm}
	\subfloat{%
		\includegraphics[scale=0.38]{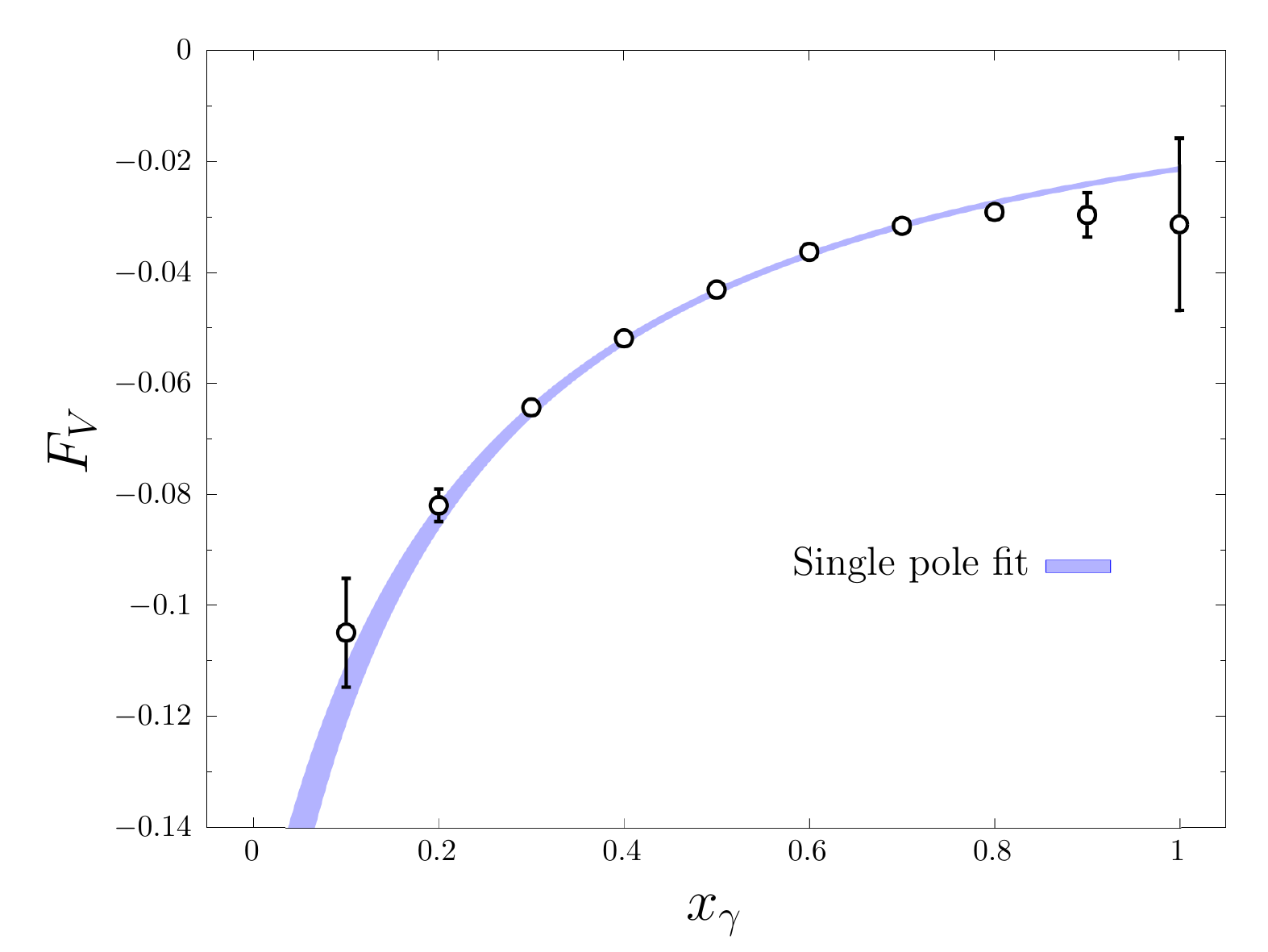}}
	\subfloat{%
		\includegraphics[scale=0.38]{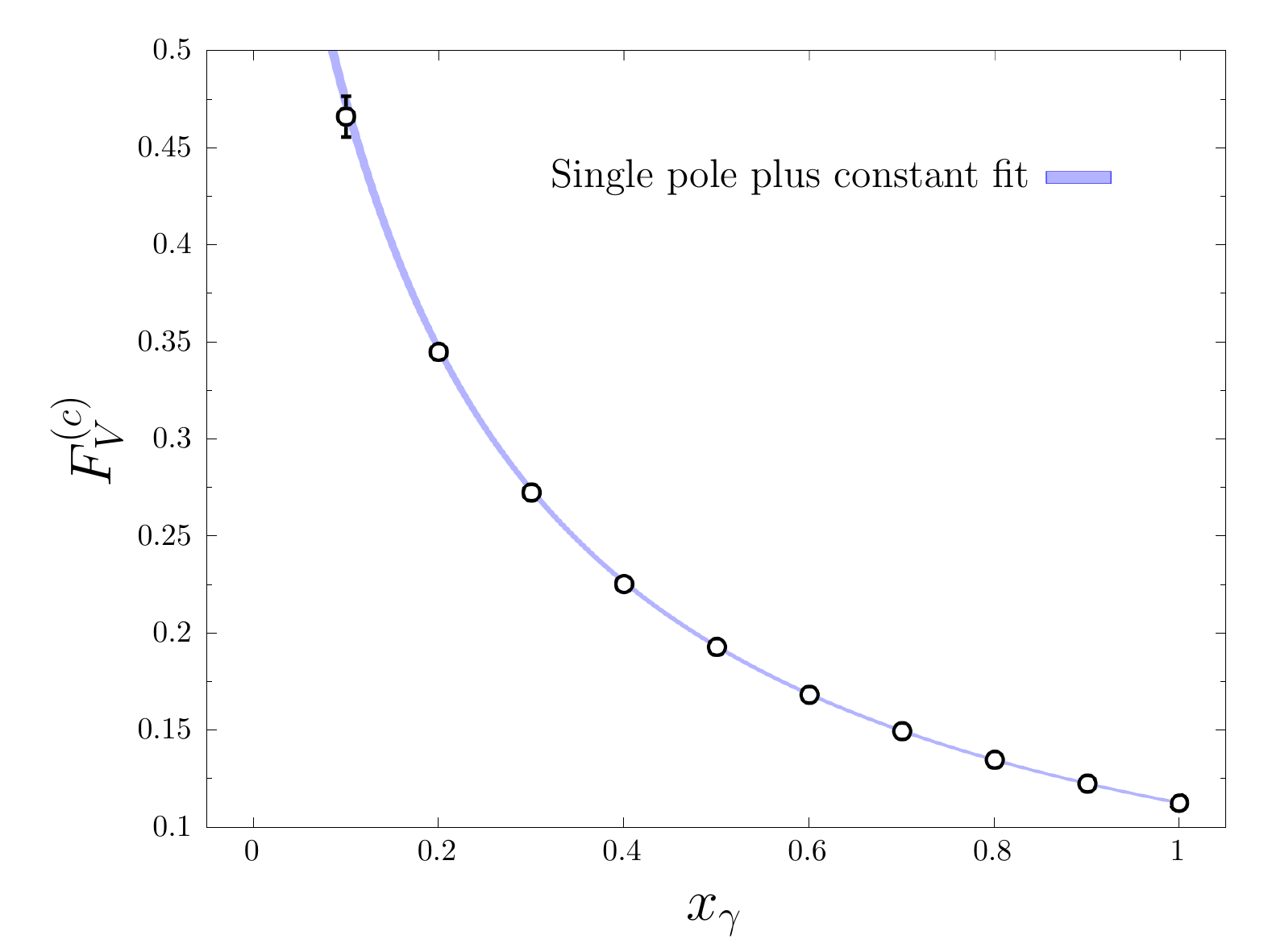}}
	\subfloat{%
		\includegraphics[scale=0.38]{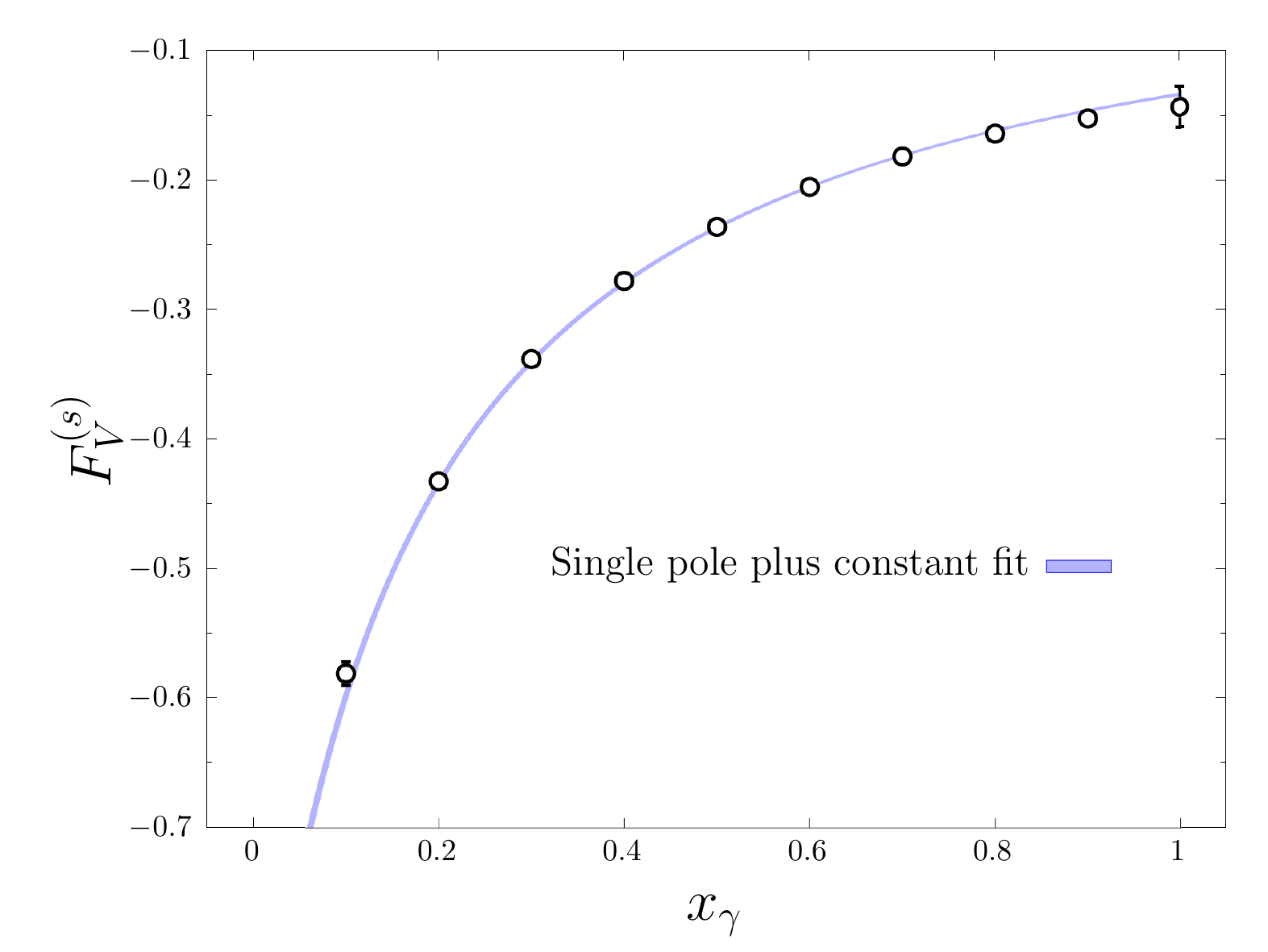}}
    \caption{The fit functions corresponding to the Ansatz of Eq.\,(\ref{eq:ansatz}) are plotted, along with the lattice data, for the axial channel (top panels) and for the vector channel (bottom panels). 
    The "Single pole fit" denotes the fit to the data with $B_W$ fixed to zero, while "Single pole plus constant fit" denotes the fit to the data with all the three parameters $C_W$, $R_W$ and $B_W$ left free. }
    \label{fig:phen_plots}
\end{figure}
The results of the fits are reported in Tab.\,\ref{tab:phen_fitparameters}, while in Fig.\,\ref{fig:phen_plots} we plot the resulting fitting functions together with our lattice data. 
We see from Fig.\,\ref{fig:phen_plots} and Tab.\,\ref{tab:phen_fitparameters} that the fits provide a very good representation of our lattice data and low values of the correlated $\chi^2$, even for the most precisely determined form factors.
The remarkably strong, $\mathcal{O}(90\%)$, cancellation between the obtained values of $B^{(c)}_V$ and $B^{(s)}_V$ in Table\,\ref{tab:phen_fitparameters} explains why $B_V$ can be dropped when fitting the total vector form factor.
The degree of cancellation between $B_V^{(c)}$ and $B_V^{(s)}$, and also between the contributions to the form factor $F_V$ from $F_V^{(c)}$ and $F_V^{(s)}$ in the lower panel of Fig.\,\ref{fig:single_quark_contr}, depends on both the charges and masses of the two valence quarks and should therefore be considered to be accidental.
It will be interesting in the future to observe to what extent these cancellations hold for the decays of the $B$-meson.
Finally, we observe that the parametrization we provide for the form factors is more precise than the lattice data points for kinematics above the threshold value $x_\gamma\simeq 0.8$. This is because the fit parameters are mainly determined from the most statistically accurate data points, which are the ones at low and intermediate values of $x_\gamma$.

Single pole dominance implies that 
the values of $R_{W}$ and $C_{W}$ are related respectively to the masses of the nearest internal resonances contributing to the correlator and to their transition amplitudes to the external states. In the present case, the resonances are the 
$D_s^*$ for the vector channel and the $D_{s1}$ for the axial one.
The values of the amplitudes $C_{W}$ are therefore related to the couplings 
$g_{D_s^*D_s\gamma}$ and $g_{D_{s1}D_s\gamma}$, i.e. to the $D_s^*\to D_s \gamma$ and $D_{s1} \to D_s \gamma$ decay amplitudes respectively. 
However, for such an interpretation of the $C_W$ 
to be considered to be physically meaningful, we first need to check that the fitted values are stable under variations of the fit Ansatz. We perform such an analysis in detail in Appendix\,\ref{app:C}. The outcome is that the data in the axial channel, i.e. for $F_A$, $F_A^{(c)}$ and $F_A^{(s)}$, are well described by any of the different Ans\"atze we employed, but the resulting values for the amplitude $C_A$ are very different depending on the Ansatz which was used. We conclude that the fitted value of  $C_A$ is not stable and is not reliable as an estimate of the coupling $g_{D_{s1}D_s\gamma}$. Note that in the axial channel, there is a second resonance, the $D_{s1}(2536)$, with a mass which is only 
76\,MeV above the nearest resonance, the $D_{s1}(2460)$. Since the difference in the masses 
is so small, the fitted residue $C_A$ may encode contributions from both of these internal states, resulting in an unreliable determination of the coupling $g_{D_{s1} D_s \gamma}$.
For the vector channel on the other hand, we found that the value of $C_V$ is very stable under variations of the fit Ansatz, provided that the corresponding fits result in a low value of $\chi^2/$d.o.f.\,.

Having established in Appendix\,\ref{app:C} that the value of $C_V$ is stable, we obtain the corresponding value of the coupling $g^{\mbox{}}_{D_s^*D_s\gamma}$, using the relation:
\bea
C_V=-\frac{M_{D_s^*}f_{D^*_s}g^{\mbox{}}_{D_s^*D_s\gamma}}{2M_{Ds}}\,,
\eea
where $f_{D^*_s}$ is the decay constant of the $D_s^*$ meson, for which we take the value $f_{D^*_s}=268.8(6.6)$\,MeV obtained from the lattice computation of Ref.\,\cite{Lubicz:2017}.
In Tab.\,\ref{tab:g}, we report our estimate for the coupling $g^{\mbox{}}_{D_s^*D_s\gamma}$, and for the individual contributions from the radiation from the strange and charm quarks. In the quoted uncertainties, we include an estimate of the systematic error due to the 
use of single-pole dominance as a model parameterization of the form factors. This is obtained in Appendix\,\ref{app:C} from the variation of the results in Tab.\,\ref{tab:g2} determined using different Ans\"atze. 
In Table\,\ref{tab:g} we also provide a comparison of our results with the values of the $g^{\mbox{}}_{D_s^*D_s\gamma}$ couplings obtained from a direct lattice computation of the $D^*_s\to D_s\gamma$ decay amplitude\,\cite{Donald:2014}, and from the calculation based on LCSR at next-to-leading order\,\cite{Pullin:2021ebn}.
Our results are in excellent agreement with those of Ref.\,\cite{Donald:2014} and with the value of $g^{(c)}_{D_s^*D_s\gamma}$ obtained in the LCSR calculation\,\cite{Pullin:2021ebn}. However, we find a discrepancy of a factor of about 2 with the value of $g^{(s)}_{D_s^*D_s\gamma}$ obtained in Ref.\,\cite{Pullin:2021ebn} which,
given the strong cancellation between the strange and charm-quark contributions, is
amplified in the total coupling $g^{\mbox{}}_{D_s^*D_s\gamma}$ to a factor of about 5; specifically the value from the LCSR calculation is about five times larger than the ones obtained from the lattice computations\,\footnote{According to Ref.\,\cite{rzwicky} the uncertainty in the $g_{D_s^\ast D_s\gamma}$ coupling given in Ref.\,\cite{Pullin:2021ebn} may be an underestimate because of the significant cancellation between the charm and the strange quark contributions.}

The authors of Ref.\,\cite{Lyon:2012fk} have also provided the values of the radiative form factors $F_A$ and $F_V$ for the $D_s$ meson at a single kinematic point $x_\gamma=0.846$;
$F_A\left(x_\gamma=0.846\right)=-0.44$ and $F_V\left(x_\gamma=0.846\right)=-0.11$. They add that they refrain from estimating the uncertainties on these values as they "only aim to provide rough estimates in order to motivate experimental searches"\,\cite{Lyon:2012fk}. 
Nevertheless, these estimates are in strong disagreement with the values collected in Table\,\ref{tab:result_FA_FV} from our direct lattice computation. In particular, we notice that, around this specific kinematic point, the magnitudes of $F_V$ differ by approximately a factor $4$, while for $F_A$ the results differ by an order of magnitude and have the opposite sign.
This raises some questions about the precision of the approach of Refs.\cite{Pullin:2021ebn} and \cite{Lyon:2012fk}, based on LCSR, for  describing heavy-meson radiative form factors.   

\begin{table}[]
    \centering
    \begin{tabular}{c| c| c| c}
     \hline 
     \rule[-2mm]{0mm}{20pt} & LCSR \cite{Pullin:2021ebn} 
 &HPQCD \cite{Donald:2014}& This paper \\
     \hline
      \rule[-3mm]{0mm}{20pt} $g^{\mbox{}}_{D_s^*D_s\gamma}\ [\textrm{GeV}^{-1}]$ & $\ 0.60(19) \ $ & $\ 0.10(2)\ $ &  $\ 0.118(13)\ $ \\
           \hline
      \rule[-3mm]{0mm}{20pt} $g^{(s)}_{D_s^*D_s\gamma}\ [\textrm{GeV}^{-1}]$ & $\ 1.0\ $ & $\ 0.50(3)\ $ &  $\ 0.532(15)\ $  \\
           \hline
      \rule[-3mm]{0mm}{20pt} $g^{(c)}_{D_s^*D_s\gamma}\ [\textrm{GeV}^{-1}]$ & $\ -0.4\ $ & $\ -0.40(2)\ $ &  $\ -0.415(16)\ $ \\
           \hline
      \rule[-4mm]{0mm}{30pt} $\dfrac{g^{(s)}}{g^{(c)}}$ &  $\ -2.5\ $ &  $\ -1.25(10)\ $ & $\ -1.282(61)\ $ \\
    \hline 
    \end{tabular}
    \caption{Our results for the $g_{D_s^*D_s\gamma}$, $g^{(s)}_{D_s^*D_s\gamma}$ and $g^{(c)}_{D_s^*D_s\gamma}$ couplings are presented and compared to the previous computations of Refs.\,\cite{Donald:2014} and \cite{Pullin:2021ebn}, based on lattice simulations and LCSR at NLO, respectively. Note the indirect nature of our estimate of the couplings, obtained by using an effective single-pole parameterization of the radiative form factors, as compared to the direct lattice computation of the $D^\ast_s\to D_s\gamma$ decay amplitude\,\cite{Donald:2014}.}
    \label{tab:g}
\end{table}

\section{Conclusions} \label{sec:conclusions}
In this paper we have presented a lattice calculation, in the electro-quenched approximation, of the structure-dependent axial and vector form factors, $F_A$ and $F_V$, which contribute to the amplitudes for the radiative leptonic decays $D_s\to\,\ell\,\nu_\ell\,\gamma$. Our results extend and improve the analysis presented in Ref.\,\cite{Desiderio:2020oej}, and are the first lattice predictions for these form factors over the whole physical phase space in the continuum limit. \\

We have also presented the individual contributions to the form factors from the emission of the photon from the charm and strange valence quarks, $F_{V,A}^{(c)}$ and $F_{V,A}^{(s)}$ respectively, with $F_{V,A}=F_{V,A}^{(c)}+F_{V,A}^{(s)}$. 
A remarkable feature is that $F_V^{(c)}(x_\gamma)\approx-F_V^{(s)}(x_\gamma)$, see the lower panel of Fig.\,\ref{fig:single_quark_contr}, so that there is a very significant cancellation in the determination of $F_V(x_\gamma)$. The axial form factor $F_A$ is dominated by $F_A^{(s)}$ and
there is no such cancellation, see the upper panel of Fig.\,\ref{fig:single_quark_contr}.\\

We use our results for the form factors to compute the differential decay rate for the process $D_s\to\,e\,\nu_e\,\gamma$ as a function of the photon energy in the meson rest frame, separating the SD contribution from the point-like one.
By integrating the differential decay rate, we obtain the branching ratio for the $D_s\to\,e\,\nu_e\,\gamma$ decay as a function of the lower cut-off, $\Delta E_{\gamma}$, on the energy of the photon in the rest frame of the decaying meson. Our result for the branching ratio for $\Delta E_{\gamma}=10$\,MeV is $4.4(3)\times 10^{-6}$, well
below the upper bound of $1.3\times 10^{-4}$ set by the BESIII collaboration \cite{BESIII:2019pjk}. Even for as low a value of $\Delta E_\gamma$ as 10\,MeV, we find that the 
the SD contribution dominates the branching ratio due to the strong helicity suppression, by a factor $r_e^2=(m_e/M_{D_s})^2$, of the point-like term.\\

Having determined the form factors, we use the results to investigate the validity and applicability of model-dependent calculations, such as ones based on single-pole dominance or light-cone sum rules.
Such model estimates are commonly used in the analysis of radiative processes involving heavy mesons for which lattice calculations are often not available. We showed that the LCSR computations at next-to-leading order of Refs.\,\cite{Pullin:2021ebn,Lyon:2012fk} fail to reproduce our results for the form factors of the $D_s$ meson, and that a pure VMD parametrization does not always reproduce their momentum behavior. In Eq.\,(\ref{eq:ansatz}) and Tab.\,\ref{tab:phen_fitparameters} we propose a simple parametrization of the form factors, based on an
extension of the single-pole dominance Ansatz,  which reproduces our lattice results very well and may therefore be useful for future phenomenological analyses.\\ 

For $F_V(x_\gamma)$ we find that results for the residue of the pole are very stable, allowing us to interpret the result in terms of the $g_{D_sD_s^\ast\gamma}$ coupling. The result is presented in Tab.\,\ref{tab:g}, where it is also compared to the results from a direct lattice computation of the rate for the decay $D^\ast\to D\gamma$\,\cite{Donald:2014} (we find good agreement) and to the LCSR calculation of Ref.\cite{Pullin:2021ebn} (we disagree significantly).\\

A non-perturbative, model-independent theoretical prediction for the
amplitudes of
real photon emission in leptonic decays is important for testing the Standard Model and for searches for new physics. 
Indeed, such results are required in order to include 
$O(\alpha_{\textrm{em}})$ corrections in the determination of fundamental SM parameters such as the CKM matrix elements. 
In addition, the SD contribution to $P\to\ell\nu_\ell\gamma$ decays probes the internal structure of the decaying meson and by comparing SM results for the form factors to experimental measurements one can test for hypothetical new physics effects. 
This is especially true for the decays of heavy mesons into an electron and its neutrino, where the SD contribution dominates the rate already at low photon energies such as 10\,MeV, which are included in some current experimental studies. 
First-principles lattice computations are particularly important for heavy mesons since chiral perturbation theory does not apply in that case.
For this reason, in the future we plan to compute the radiative SD form factors also for the $D$ and $B$ mesons. 
When applying the strategy that we have presented in this work to these mesons however, the presence of a light valence quark will significantly lower the threshold value of the photon energy above which statistical fluctuation start to grow exponentially. 
We have identified the origin of this issue in Appendix\,\ref{app:A}, where we also briefly discuss a possible way to mitigate this problem
based on the different lattice approach proposed in Ref.\,\cite{Giusti:2023pot}, where it is called the $3d$ method.

\section{Acknowledgements}
We thank all members of the ETMC for the most enjoyable collaboration. We thank Roman Zwicky for interesting correspondence and discussions and for suggesting that we compute the contributions to the form factors from the emission of the photon from the strange and charm quarks separately.
We acknowledge CINECA for the provision of CPU time under the specific
initiative INFN-LQCD123 and IscrB\_S-EPIC. F.S. G.G and S.S. are supported by the Italian Ministry
of University and Research (MIUR) under grant PRIN20172LNEEZ. F.S. and G.G are supported by
INFN under GRANT73/CALAT. C.T.S. was partially supported by an Emeritus Fellowship from the
Leverhulme Trust and by STFC (UK) grant ST/T000775/1.
F.S. is supported by ICSC – Centro Nazionale di Ricerca in High Performance
Computing, Big Data and Quantum Computing, funded by European Union –
NextGenerationEU.

\bibliography{biblio}

\providecommand{\href}[2]{#2}\begingroup\raggedright\begin{thebibliography}{10}

\bibitem{Bychkov:2008ws}
M.~Bychkov et~al., \emph{{New Precise Measurement of the Pion Weak Form Factors
  in $pi^+ \to e^+ \nu \gamma$ Decay}},
  \href{https://doi.org/10.1103/PhysRevLett.103.051802}{\emph{Phys. Rev. Lett.}
  {\bfseries 103} (2009) 051802}
  [\href{https://arxiv.org/abs/0804.1815}{{\ttfamily 0804.1815}}].

\bibitem{E787:2000ehx}
{\scshape E787} collaboration, \emph{{Measurement of structure dependent K+
  ---\ensuremath{>} muon+ neutrino(muon) gamma decay}},
  \href{https://doi.org/10.1103/PhysRevLett.85.2256}{\emph{Phys. Rev. Lett.}
  {\bfseries 85} (2000) 2256}
  [\href{https://arxiv.org/abs/hep-ex/0003019}{{\ttfamily hep-ex/0003019}}].

\bibitem{KLOE:2009urs}
{\scshape KLOE} collaboration, \emph{{Precise measurement of $\Gamma(K \to e
  \nu(\gamma)) / \Gamma(K \to \mu \nu(\gamma))$ and study of $K \to e \nu
  \gamma$}}, \href{https://doi.org/10.1140/epjc/s10052-009-1217-6}{\emph{Eur.
  Phys. J. C} {\bfseries 64} (2009) 627}
  [\href{https://arxiv.org/abs/0907.3594}{{\ttfamily 0907.3594}}].

\bibitem{OKA:2019gav}
{\scshape OKA} collaboration, \emph{{Measurement of the
  $K^+\rightarrow{\mu^+}{\nu_{\mu}}{\gamma}$ decay form factors in the OKA
  experiment}},
  \href{https://doi.org/10.1140/epjc/s10052-019-7145-1}{\emph{Eur. Phys. J. C}
  {\bfseries 79} (2019) 635}
  [\href{https://arxiv.org/abs/1904.10078}{{\ttfamily 1904.10078}}].

\bibitem{ISTRA:2010smy}
{\scshape ISTRA+} collaboration, \emph{{Extraction of Kaon Formfactors from
  $K^- \to \mu^- \nu_\mu \gamma$ Decay at ISTRA+ Setup}},
  \href{https://doi.org/10.1016/j.physletb.2010.10.043}{\emph{Phys. Lett. B}
  {\bfseries 695} (2011) 59} [\href{https://arxiv.org/abs/1005.3517}{{\ttfamily
  1005.3517}}].

\bibitem{JPARCE36:2021yvz}
{\scshape J-PARC E36} collaboration, \emph{{Measurement of structure dependent
  radiative $K^+ \to e^+ \nu \gamma$ decays using stopped positive kaons}},
  \href{https://doi.org/10.1016/j.physletb.2022.136913}{\emph{Phys. Lett. B}
  {\bfseries 826} (2022) 136913}
  [\href{https://arxiv.org/abs/2107.03583}{{\ttfamily 2107.03583}}].

\bibitem{BESIII:2017whk}
{\scshape BESIII} collaboration, \emph{{Search for the radiative leptonic decay
  $D^{+}\to \gamma e^{+} {\nu}_{e}$}},
  \href{https://doi.org/10.1103/PhysRevD.95.071102}{\emph{Phys. Rev. D}
  {\bfseries 95} (2017) 071102}
  [\href{https://arxiv.org/abs/1702.05837}{{\ttfamily 1702.05837}}].

\bibitem{BESIII:2019pjk}
{\scshape BESIII} collaboration, \emph{{Search for the decay $D_s^+\rightarrow
  \gamma e^+\nu_e$}},
  \href{https://doi.org/10.1103/PhysRevD.99.072002}{\emph{Phys. Rev. D}
  {\bfseries 99} (2019) 072002}
  [\href{https://arxiv.org/abs/1902.03351}{{\ttfamily 1902.03351}}].

\bibitem{Belle:2015mpp}
{\scshape Belle} collaboration, \emph{{Search for $B^+ \to \ell^+ \nu_\ell
  \gamma$ decays with hadronic tagging using the full Belle data sample}},
  \href{https://doi.org/10.1103/PhysRevD.91.112009}{\emph{Phys. Rev. D}
  {\bfseries 91} (2015) 112009}
  [\href{https://arxiv.org/abs/1504.05831}{{\ttfamily 1504.05831}}].

\bibitem{Belle:2018jqd}
{\scshape Belle} collaboration, \emph{{Search for the rare decay of $B^+ \to
  \ell^{\,+} \nu_{\ell} \gamma$ with improved hadronic tagging}},
  \href{https://doi.org/10.1103/PhysRevD.98.112016}{\emph{Phys. Rev. D}
  {\bfseries 98} (2018) 112016}
  [\href{https://arxiv.org/abs/1810.12976}{{\ttfamily 1810.12976}}].

\bibitem{Desiderio:2020oej}
A.~Desiderio et~al., \emph{{First lattice calculation of radiative leptonic
  decay rates of pseudoscalar mesons}},
  \href{https://doi.org/10.1103/PhysRevD.103.014502}{\emph{Phys. Rev. D}
  {\bfseries 103} (2021) 014502}
  [\href{https://arxiv.org/abs/2006.05358}{{\ttfamily 2006.05358}}].

\bibitem{Frezzotti:2020bfa}
R.~Frezzotti, M.~Garofalo, V.~Lubicz, G.~Martinelli, C.T.~Sachrajda,
  F.~Sanfilippo et~al., \emph{{Comparison of lattice QCD+QED predictions for
  radiative leptonic decays of light mesons with experimental data}},
  \href{https://doi.org/10.1103/PhysRevD.103.053005}{\emph{Phys. Rev. D}
  {\bfseries 103} (2021) 053005}
  [\href{https://arxiv.org/abs/2012.02120}{{\ttfamily 2012.02120}}].

\bibitem{Alexandrou:2018egz}
C.~Alexandrou et~al., \emph{{Simulating twisted mass fermions at physical
  light, strange and charm quark masses}},
  \href{https://doi.org/10.1103/PhysRevD.98.054518}{\emph{Phys. Rev. D}
  {\bfseries 98} (2018) 054518}
  [\href{https://arxiv.org/abs/1807.00495}{{\ttfamily 1807.00495}}].

\bibitem{ExtendedTwistedMass:2021qui}
{\scshape Extended Twisted Mass} collaboration, \emph{{Ratio of kaon and pion
  leptonic decay constants with Nf=2+1+1 Wilson-clover twisted-mass fermions}},
  \href{https://doi.org/10.1103/PhysRevD.104.074520}{\emph{Phys. Rev. D}
  {\bfseries 104} (2021) 074520}
  [\href{https://arxiv.org/abs/2104.06747}{{\ttfamily 2104.06747}}].

\bibitem{ExtendedTwistedMass:2021gbo}
{\scshape Extended Twisted Mass} collaboration, \emph{{Quark masses using
  twisted-mass fermion gauge ensembles}},
  \href{https://doi.org/10.1103/PhysRevD.104.074515}{\emph{Phys. Rev. D}
  {\bfseries 104} (2021) 074515}
  [\href{https://arxiv.org/abs/2104.13408}{{\ttfamily 2104.13408}}].

\bibitem{Alexandrou:2022amy}
C.~Alexandrou et~al., \emph{{Lattice calculation of the short and intermediate
  time-distance hadronic vacuum polarization contributions to the muon magnetic
  moment using twisted-mass fermions}},
  \href{https://arxiv.org/abs/2206.15084}{{\ttfamily 2206.15084}}.

\bibitem{Giusti:2023pot}
D.~Giusti, C.F.~Kane, C.~Lehner, S.~Meinel and A.~Soni, \emph{{High-precision
  determination of radiative-leptonic-decay form factors using lattice QCD: a
  study of methods}},  \href{https://arxiv.org/abs/2302.01298}{{\ttfamily
  2302.01298}}.

\bibitem{Korchemsky:1999qb}
G.P.~Korchemsky, D.~Pirjol and T.-M.~Yan, \emph{{Radiative leptonic decays of B
  mesons in QCD}},
  \href{https://doi.org/10.1103/PhysRevD.61.114510}{\emph{Phys. Rev. D}
  {\bfseries 61} (2000) 114510}
  [\href{https://arxiv.org/abs/hep-ph/9911427}{{\ttfamily hep-ph/9911427}}].

\bibitem{Atwood:1994za}
D.~Atwood, G.~Eilam and A.~Soni, \emph{{Pure leptonic radiative decays B+-,
  D(s) $\to$ lepton neutrino gamma and the annihilation graph}},
  \href{https://doi.org/10.1142/S0217732396001090}{\emph{Mod. Phys. Lett. A}
  {\bfseries 11} (1996) 1061}
  [\href{https://arxiv.org/abs/hep-ph/9411367}{{\ttfamily hep-ph/9411367}}].

\bibitem{Yang:2012jp}
J.-C.~Yang and M.-Z.~Yang, \emph{{Radiative Leptonic Decays of the charged $B$
  and $D$ Mesons Including Long-Distance Contribution}},
  \href{https://doi.org/10.1142/S0217732312501209}{\emph{Mod. Phys. Lett. A}
  {\bfseries 27} (2012) 1250120}
  [\href{https://arxiv.org/abs/1204.2383}{{\ttfamily 1204.2383}}].

\bibitem{Donald:2014}
{\scshape HPQCD Collaboration} collaboration, \emph{Prediction of the
  ${D}_{s}^{*}$ width from a calculation of its radiative decay in full lattice
  qcd}, \href{https://doi.org/10.1103/PhysRevLett.112.212002}{\emph{Phys. Rev.
  Lett.} {\bfseries 112} (2014) 212002}.

\bibitem{Pullin:2021ebn}
B.~Pullin and R.~Zwicky, \emph{{Radiative decays of heavy-light mesons and the
  $ {f}_{H,{H}^{\ast },{H}_1}^{(T)} $ decay constants}},
  \href{https://doi.org/10.1007/JHEP09(2021)023}{\emph{JHEP} {\bfseries 09}
  (2021) 023} [\href{https://arxiv.org/abs/2106.13617}{{\ttfamily
  2106.13617}}].

\bibitem{Lyon:2012fk}
J.~Lyon and R.~Zwicky, \emph{{$A_{\textrm{CP}}[D_{(s)}^{0,+} \to V \gamma]$
  from Large ${\cal O}_8$}},
  \href{https://doi.org/10.1103/PhysRevD.106.053001}{\emph{Phys. Rev. D}
  {\bfseries 106} (2022) 053001}
  [\href{https://arxiv.org/abs/1210.6546}{{\ttfamily 1210.6546}}].

\bibitem{Guadagnoli:2023zym}
D.~Guadagnoli, C.~Normand, S.~Simula and L.~Vittorio, \emph{{From $D_s \to
  \gamma$ in lattice QCD to $B_s \to \mu \mu \gamma$ at high $q^2$}},
  \href{https://arxiv.org/abs/2303.02174}{{\ttfamily 2303.02174}}.

\bibitem{Carrasco:2015xwa}
N.~Carrasco, V.~Lubicz, G.~Martinelli, C.T.~Sachrajda, N.~Tantalo, C.~Tarantino
  et~al., \emph{{QED Corrections to Hadronic Processes in Lattice QCD}},
  \href{https://doi.org/10.1103/PhysRevD.91.074506}{\emph{Phys. Rev. D}
  {\bfseries 91} (2015) 074506}
  [\href{https://arxiv.org/abs/1502.00257}{{\ttfamily 1502.00257}}].

\bibitem{Bijnens:1992en}
J.~Bijnens, G.~Ecker and J.~Gasser, \emph{{Radiative semileptonic kaon
  decays}}, \href{https://doi.org/10.1016/0550-3213(93)90259-R}{\emph{Nucl.
  Phys. B} {\bfseries 396} (1993) 81}
  [\href{https://arxiv.org/abs/hep-ph/9209261}{{\ttfamily hep-ph/9209261}}].

\bibitem{Gagliardi:2022szw}
G.~Gagliardi, F.~Sanfilippo, S.~Simula, V.~Lubicz, F.~Mazzetti, G.~Martinelli
  et~al., \emph{{Virtual photon emission in leptonic decays of charged
  pseudoscalar mesons}},
  \href{https://doi.org/10.1103/PhysRevD.105.114507}{\emph{Phys. Rev. D}
  {\bfseries 105} (2022) 114507}
  [\href{https://arxiv.org/abs/2202.03833}{{\ttfamily 2202.03833}}].

\bibitem{Sachrajda:2004mi}
C.T.~Sachrajda and G.~Villadoro, \emph{{Twisted boundary conditions in lattice
  simulations}},
  \href{https://doi.org/10.1016/j.physletb.2005.01.033}{\emph{Phys. Lett. B}
  {\bfseries 609} (2005) 73}
  [\href{https://arxiv.org/abs/hep-lat/0411033}{{\ttfamily hep-lat/0411033}}].

\bibitem{Flynn:2007ess}
J.M.~Flynn, A.~Juttner, C.T.~Sachrajda, P.A.~Boyle and J.M.~Zanotti,
  \emph{{Hadronic form factors in Lattice QCD at small and vanishing momentum
  transfer}}, \href{https://doi.org/10.1088/1126-6708/2007/05/016}{\emph{JHEP}
  {\bfseries 05} (2007) 016}
  [\href{https://arxiv.org/abs/hep-lat/0703005}{{\ttfamily hep-lat/0703005}}].

\bibitem{Iwasaki:1985we}
Y.~Iwasaki, \emph{{Renormalization Group Analysis of Lattice Theories and
  Improved Lattice Action: Two-Dimensional Nonlinear O(N) Sigma Model}},
  \href{https://doi.org/10.1016/0550-3213(85)90606-6}{\emph{Nucl. Phys. B}
  {\bfseries 258} (1985) 141}.

\bibitem{Frezzotti:2000nk}
{\scshape Alpha} collaboration, \emph{{Lattice QCD with a chirally twisted mass
  term}}, \href{https://doi.org/10.1088/1126-6708/2001/08/058}{\emph{JHEP}
  {\bfseries 08} (2001) 058}
  [\href{https://arxiv.org/abs/hep-lat/0101001}{{\ttfamily hep-lat/0101001}}].

\bibitem{Frezzotti:2003ni}
R.~Frezzotti and G.C.~Rossi, \emph{{Chirally improving Wilson fermions. 1. O(a)
  improvement}},
  \href{https://doi.org/10.1088/1126-6708/2004/08/007}{\emph{JHEP} {\bfseries
  08} (2004) 007} [\href{https://arxiv.org/abs/hep-lat/0306014}{{\ttfamily
  hep-lat/0306014}}].

\bibitem{Frezzotti:2004wz}
R.~Frezzotti and G.C.~Rossi, \emph{{Chirally improving Wilson fermions. II.
  Four-quark operators}},
  \href{https://doi.org/10.1088/1126-6708/2004/10/070}{\emph{JHEP} {\bfseries
  10} (2004) 070} [\href{https://arxiv.org/abs/hep-lat/0407002}{{\ttfamily
  hep-lat/0407002}}].

\bibitem{ParticleDataGroup:2016lqr}
{\scshape Particle Data Group} collaboration, \emph{{Review of Particle
  Physics}}, \href{https://doi.org/10.1088/1674-1137/40/10/100001}{\emph{Chin.
  Phys. C} {\bfseries 40} (2016) 100001}.

\bibitem{Osterwalder:1977pc}
K.~Osterwalder and E.~Seiler, \emph{{Gauge Field Theories on the Lattice}},
  \href{https://doi.org/10.1016/0003-4916(78)90039-8}{\emph{Annals Phys.}
  {\bfseries 110} (1978) 440}.

\bibitem{Borsanyi:2020mff}
S.~Borsanyi et~al., \emph{{Leading hadronic contribution to the muon magnetic
  moment from lattice QCD}},
  \href{https://doi.org/10.1038/s41586-021-03418-1}{\emph{Nature} {\bfseries
  593} (2021) 51} [\href{https://arxiv.org/abs/2002.12347}{{\ttfamily
  2002.12347}}].

\bibitem{ParticleDataGroup:2020ssz}
{\scshape Particle Data Group} collaboration, \emph{{Review of Particle
  Physics}}, \href{https://doi.org/10.1093/ptep/ptaa104}{\emph{PTEP} {\bfseries
  2020} (2020) 083C01}.

\bibitem{Hatton:2020qhk}
{\scshape HPQCD} collaboration, \emph{{Charmonium properties from lattice
  $QCD$+QED : Hyperfine splitting, $J/\psi$ leptonic width, charm quark mass,
  and $a^c_\mu$}},
  \href{https://doi.org/10.1103/PhysRevD.102.054511}{\emph{Phys. Rev. D}
  {\bfseries 102} (2020) 054511}
  [\href{https://arxiv.org/abs/2005.01845}{{\ttfamily 2005.01845}}].

\bibitem{Zhang:2021xrs}
R.~Zhang, W.~Sun, F.~Chen, Y.~Chen, M.~Gong, X.~Jiang et~al.,
  \emph{{Annihilation diagram contribution to charmonium masses}},
  \href{https://doi.org/10.1088/1674-1137/ac3d8c}{\emph{Chin. Phys. C}
  {\bfseries 46} (2022) 043102}
  [\href{https://arxiv.org/abs/2110.01755}{{\ttfamily 2110.01755}}].

\bibitem{Carrasco:2014uya}
N.~Carrasco et~al., \emph{{$D^0-\bar{D}^0$ mixing in the standard model and
  beyond from $N_f$ =2 twisted mass QCD}},
  \href{https://doi.org/10.1103/PhysRevD.90.014502}{\emph{Phys. Rev. D}
  {\bfseries 90} (2014) 014502}
  [\href{https://arxiv.org/abs/1403.7302}{{\ttfamily 1403.7302}}].

\bibitem{Akaike}
H.~Akaike, \emph{{A new look at the statistical model identification}},
  {\emph{IEEE Transactions on Automatic Control} {\bfseries 19} (1974) 716}.

\bibitem{Neil:2022joj}
E.T.~Neil and J.W.~Sitison, \emph{{Improved information criteria for Bayesian
  model averaging in lattice field theory}},
  \href{https://arxiv.org/abs/2208.14983}{{\ttfamily 2208.14983}}.

\bibitem{PhysRev.52.54}
F.~Bloch and A.~Nordsieck, \emph{Note on the radiation field of the electron},
  \href{https://doi.org/10.1103/PhysRev.52.54}{\emph{Phys. Rev.} {\bfseries 52}
  (1937) 54}.

\bibitem{FlavourLatticeAveragingGroupFLAG:2021npn}
{\scshape Flavour Lattice Averaging Group (FLAG)} collaboration, \emph{{FLAG
  Review 2021}},
  \href{https://doi.org/10.1140/epjc/s10052-022-10536-1}{\emph{Eur. Phys. J. C}
  {\bfseries 82} (2022) 869}
  [\href{https://arxiv.org/abs/2111.09849}{{\ttfamily 2111.09849}}].

\bibitem{Geng:2000if}
C.Q.~Geng, C.C.~Lih and W.-M.~Zhang, \emph{{Study of radiative leptonic D meson
  decays}}, \href{https://doi.org/10.1142/S021773230000267X}{\emph{Mod. Phys.
  Lett. A} {\bfseries 15} (2000) 2087}
  [\href{https://arxiv.org/abs/hep-ph/0012066}{{\ttfamily hep-ph/0012066}}].

\bibitem{Lu:2002mn}
C.-D.~Lu and G.-L.~Song, \emph{{Radiative leptonic decays of D+-(s) and D+-
  mesons}}, \href{https://doi.org/10.1016/S0370-2693(03)00549-5}{\emph{Phys.
  Lett. B} {\bfseries 562} (2003) 75}
  [\href{https://arxiv.org/abs/hep-ph/0212363}{{\ttfamily hep-ph/0212363}}].

\bibitem{Lubicz:2017}
{\scshape ETM Collaboration} collaboration, \emph{Masses and decay constants of
  ${D}_{(s)}^{*}$ and ${B}_{(s)}^{*}$ mesons with ${N}_{f}=2+1+1$ twisted mass
  fermions}, \href{https://doi.org/10.1103/PhysRevD.96.034524}{\emph{Phys. Rev.
  D} {\bfseries 96} (2017) 034524}.

\bibitem{rzwicky}
R.~Zwicky. private communication.

\bibitem{Parisi:1983ae}
G.~Parisi, \emph{{The Strategy for Computing the Hadronic Mass Spectrum}},
  \href{https://doi.org/10.1016/0370-1573(84)90081-4}{\emph{Phys. Rept.}
  {\bfseries 103} (1984) 203}.

\bibitem{Lepage:1989hd}
G.P.~Lepage, \emph{{The Analysis of Algorithms for Lattice Field Theory}},  in
  \emph{{Theoretical Advanced Study Institute in Elementary Particle Physics}},
  6, 1989.

\bibitem{Workman:2022ynf}
{\scshape Particle Data Group} collaboration, \emph{{Review of Particle
  Physics}}, \href{https://doi.org/10.1093/ptep/ptac097}{\emph{PTEP} {\bfseries
  2022} (2022) 083C01}.

\end{thebibliography}\endgroup
\bibliographystyle{JHEP}


\appendix

\section{Behaviour of the signal-to-noise ratio for $\mathbf{F_{V}}$ and $\mathbf{F_{A}}$ at large $\mathbf{x_{\gamma}}$}
\label{app:A}
In this appendix, we show why the intrinsic statistical fluctuations of $C^{\mu\nu}_{W}(t,E_{\gamma},\bs{k}, \bs{p})$ become exponentially large for small values of $t$ and large values of $x_{\gamma}$. For this discussion we make a certain number of simplifications, but these have no impact on the main conclusions. We consider only the case $\bs{p}=\bs{0}$, i.e. we always work in the decaying hadron's reference frame. We shall also use continuum notation throughout this appendix, and replace lattice sums by definite integrals. Also, when considering the Euclidean three-point correlation function in Eq.\,(\ref{eq:Cmunudef}) we only consider the term with $t_y < T/2$,
which is the dominant one in the limit $T\to \infty$. We discuss the case of an arbitrary pseudoscalar meson $P= \bar{U}\gamma^{5}D$ made of an up- and a down-type quark.  \\

We choose the 4 momentum $k=(|\bs{k}|,\bs{k})$ and $p=(m_P,\bs{0})$, therefore we now denote the correlation function $C^{\mu\nu}_W(t;k,p)$ in Eq.\,(\ref{eq:Cmunudef}) simply by $C^{\mu\nu}_{W}(t,\bs{k})$.
For $t ,t_y< T/2$ it is given by
\begin{equation}\label{eq:CmunuApp}
C^{\mu\nu}_W(t;\bs{k})=-i\int_0^{T/2}dt_y\, e^{\hspace{1pt}E_\gamma t_y}~
\bra 0\hat{\mathrm{T}}\,[j^\nu_W(t,\bs{0})\hat{j}_\mathrm{em}^\mu(t_y,\bs{k})\hat{\phi}^\dagger_{P}(0,\bs{0})]\ket 0\,,
\end{equation}
where
\begin{equation}
\hat\phi^\dagger_P(t_x,\bs{p})=\int\dthree x\,\phi^\dagger_P(t_x,\bs{x})\,e^{i\bs{p}\cdot\bs{x}},\qquad
\hat{j}^\mu_\mathrm{em}(t_y,\bs{k})=\int\dthree y\,j^\mu_\mathrm{em}(t_y,\bs{y})\,e^{-i\bs{k}\cdot\bs{y}}
\end{equation}
and $E_\gamma=|\bs{k}|$.
We now separate the contribution from the region with $t_y<t$ from that with $t_y>t$:
\begin{eqnarray}
C^{\mu\nu}_W(t;\bs{k}) \equiv C^{\mu\nu; 1}_{W}(t,\bs{k}) + C^{\mu\nu;2}_{W}(t,\bs{k}) = &-i&\int_0^{t}\hspace{-3pt}dt_y\, e^{\hspace{1pt}E_\gamma t_y}
\bra 0j^\nu_W(t,\bs{0})\hat{j}_\mathrm{em}^\mu(t_y,\bs{k})\hat{\phi}^\dagger_{P}(0,\bs{0})\ket 0 \nonumber\\
&-i&\int_t^{T/2}\hspace{-8pt}dt_y\,e^{\hspace{1pt}E_\gamma t_y}
\bra 0\hat{j}_\mathrm{em}^\mu(t_y,\bs{k})j^\nu_W(t,\bs{0})\hat{\phi}^\dagger_{P}(0,\bs{0})]\ket 0\,.
\end{eqnarray}

The time ordering relevant for our discussion is $t < t_y < T/2$, i.e.\!\! with the weak current acting before the electromagnetic current; this 
corresponds to the second contribution to the correlation function, $C^{\mu\nu;2}_{W}(t,\bs{k})$. We also
distinguish the contributions to $C^{\mu\nu; 2}_{W}(t,\bs{k})$ from the emission of the real photon from the up-type (U) or down-type (D) valence quark, and define
\begin{equation}
C^{\mu\nu;2}_{W}(t, \bs{k})\equiv C^{\mu\nu;2}_{U,W}(t,\bs{k}) + C^{\mu\nu;2}_{D,W}(t,\bs{k})\,,
\end{equation}
where
\begin{equation}
C^{\mu\nu;2}_{f,W}(t,\bs{k}) 
\equiv -i\int_{t}^{T/2}dt_y ~e^{\hspace{1pt}E_\gamma t_y}
\bra 0\hat{j}_{f,\mathrm{em}}^\mu(t_y,\bs{k})j^\nu_W(t,\bs{0})\hat{\phi}^\dagger_{P}(0,\bs{0})]\ket 0\,,\qquad f=(U,D)
\end{equation}
in terms of the single flavour contribution to the current,
\begin{equation}
\hat{j}^{\mu}_{f,{\rm{em}}}(t_y,\boldsymbol{k}) \equiv q_{f}\int\dthree y\,\,e^{-i\bs{k}\cdot\bs{y}}\,
\bar{\psi}_{f}(t_y,\bs{y}) \gamma^{\mu}\psi_{f}(t_y,\boldsymbol{y})\,.
\end{equation}
Since we aim at understanding the behaviour of the error at small $t$, we focus on the case in which the weak current $j^{\nu}_{W}$ and the interpolating operator $\phi$ are separated by a time distance of a few lattice spacings. In this case, we can interpret the quantity
\begin{align}
O^{\nu}_{W}(t) \equiv  j_{W}^{\nu}(t,\bs{0})\, \hat{\phi}^\dagger_P(0,\bs{0})
\end{align}
as a non-local interpolating operator for vector states (in both cases $W=V,A$), and 
\begin{align}
M^{\mu\nu}_{f, W}(t_y-t, t, \bs{k}) \equiv \bra{0} \hat{j}^{\mu}_{f,{\rm{em}}}(t_y,\boldsymbol{k}) 
j^\nu_W(t,\bs{0})\hat{\phi}^\dagger_P(0,\bs{0})\ket{0} 
= \bra{0} \hat{j}^{\mu}_{f,{\rm{em}}}(t_y,\boldsymbol{k})\,O^{\nu}_{W}(t)\ket{0}\;,
\end{align}
as a standard two-point correlation function, where vector states propagate between Euclidean time $t$ and Euclidean time $t_y > t$. Ignoring finite volume interactions, a standard application of the Parisi\,\cite{Parisi:1983ae} and Lepage\,\cite{Lepage:1989hd} argument, shows that at fixed time $t$ and large-time separations $t_y - t$, the variance $\sigma^2_{M^{\mu\nu}_{f,W}}(t_y -t, t, \bs{k})$ of $M^{\mu\nu}_{f, W}(t_y-t, t, \bs{k})$ decreases exponentially as
\begin{align}
\label{eq:varM_scaling}
\sigma^2_{M^{\mu\nu}_{f,W}}(t_y -t, t, \bs{k}) \propto e^{-2M_{\bar{f}f}^{{\rm{PS}}} (t_{y} - t)}~,
\end{align}
with $M_{\bar{f}f}^{{\rm{PS}}}$ the mass of the lightest pseudoscalar $\bar{f}\gamma^{5}f$ state. Instead, the signal $M^{\mu\nu}_{f, W}(t_y-t, t, \bs{k})$ scales asymptotically as
\begin{align}
\label{eq:M_scaling}
M^{\mu\nu}_{f, W}(t_y-t, t, \bs{k}) \propto e^{-E_{V}(\bs{k}) (t_y - t)}~,\qquad  E_{f,V}(\bs{k}) = \sqrt{ M_{f,V}^{2} + |\bs{k}|^{2}}~,
\end{align}
where $M_{f,V}$ is the mass of the lightest vector state interpolated by $\hat{j}^{\mu}_{f,{ \rm{em}}}(t_{y},\bs{k})$. This implies the following asymptotic scaling of the signal-to-noise (S/N) ratio of $M^{\mu\nu}_{f, W}(t_y-t, t, \bs{k})$
\begin{align}
\label{eq:SN_scaling}
\frac{ M_{f,W}^{\mu\nu}(t_y - t,t,\bs{k})}{\sigma_{M_{f,W}^{\mu\nu}}(t_y - t, t, \bs{k})} \sim e^{-(E_{f,V}(
\bs{k}) - M_{\bar{f}f}^{{\rm{PS}}}) (t_{y} - t)}\,.
\end{align}
Eqs.\,(\ref{eq:varM_scaling})-(\ref{eq:SN_scaling}) enable us to understand the scaling of the error as we discuss in the following subsection. 
When $P=D_{s}$, we have $M_{s,V}= M_{\phi} \simeq 1\,{\rm GeV}$, $M_{c,V} = M_{J/\Psi} \simeq 3.1\,{\rm GeV}$, $M_{\bar{s}s}^{{\rm{PS}}} = M_{\eta_{ss'}} \simeq 0.69\,{\rm GeV}$ and $M^{{\rm{PS}}}_{\bar{c}c} = M_{\eta_{c}}\simeq 2.98\,{\rm GeV}$. We underline that Eqs.\,(\ref{eq:varM_scaling})-(\ref{eq:SN_scaling}) only hold for large time separations $t_{y} - t$. 

\subsection{ Analysis of the scaling of the signal-to-noise ratio of $\mathbf{C_{f,W}^{\mu\nu ;2}}$  }
We now have all the necessary ingredients to understand the scaling of the S/N ratio of $C_{f,W}^{\mu\nu;2}(t)$. To this end, we define a time, $t_{\mathrm{cut}}$, such that for $t_{y} - t > t_\mathrm{cut}$ the asymptotic formulae in Eqs.\,(\ref{eq:varM_scaling})-(\ref{eq:SN_scaling}) hold, i.e. both $M_{f,W}^{\mu\nu}(t_y-t, t,\bs{k})$ and $\sigma^{2}_{M_{f,W}^{\mu\nu}}(t_y-t, t, \bs{k})$ are dominated by the contributions from the lowest-energy vector and pseudoscalar intermediate states respectively:
\begin{align}
\label{repr_C_b}
M_{f,W}^{\mu\nu}(t_y-t, t,\bs{k})
&\equiv \sum_{n=1} A_{f, n} ~ e^{- E_{f,n}(\bs{k}) (t_y - t)} + A_{f,0} ~ e^{- E_{f,V}(\bs{k})( t_y -t)} 
\underset{t_y -t > t_\mathrm{cut}}{\simeq} A_{f,0} ~ e^{- E_{V}(\bs{k})( t_y -t)} \\[10pt]
\sigma^{2}_{M_{f,W}^{\mu\nu}}(t_y-t, t, \bs{k}) &\equiv \sum_{n=1} B_{f,n}^{2}~e^{-M_{f,n}(t_y - t)} + B_{f,0}^{2}~e^{- 2M_{\bar{f}f}^{{\rm{PS}}}( t_y -t)}
\underset{t_y -t > t_\mathrm{cut}}{\simeq} B_{f,0}^{2} ~e^{- 2M_{\bar{f}f}^{{\rm{PS}}}( t_y -t)}\,,
\end{align}
where $E_{f,n}(\bs{k}) > E_{f,V}(\bs{k})$ and $M_{f,n} > 2M_{\bar{f}f}^{{\rm{PS}}}$ for $n\ge 1$. We thus have\,\footnote{In writing Eq.\,(\ref{eq:scaling_varC}) we are assuming that the values of $M_{f,W}^{\mu\nu}(t_y-t, t,\bs{k})$ at different times $t_{y}$ are fully correlated, which is a fairly good assumption given that the different times are typically evaluated using the same set of gauge configurations. However, the main result obtained in this appendix, namely the exponential growth of the error in Eq.~(\ref{eq:asymptotic_scaling_var}),  does not depend upon this assumption.}
\begin{align}
\label{eq:scaling_C}
C_{f, W}^{\mu\nu; 2}(t, \bs{k}) &\simeq -i \left[ \sum_{n=1} A_{f,n} ~ e^{E_{f,n}(\bs{k})t}\int_{t}^{t_\mathrm{cut}}dt_y~ e^{- (E_{f,n}(\bs{k}) - E_\gamma) t_y} + A_{f,0} ~ e^{E_{f,V}(\bs{k})t}\int_{t}^{T/2}dt_y ~ e^{-(E_{f,V}(\bs{k}) - E_\gamma)t_y}\right] \\[12pt]
\label{eq:scaling_varC}
\sigma_{C_{f,W}^{\mu\nu ;2}}(t,\bs{k}) &= \int_{t}^{T/2 } dt_y ~ e^{E_\gamma t_{y}} ~ \sigma_{M_{f,W}^{\mu\nu}}(t_y-t,t , \bs{k})~  \nonumber \\
&\simeq  \int_{t}^{t_\mathrm{cut}}dt_y ~ \sqrt{\sum_{n=0} \left[ B_{f,n}^{2} ~ e^{M_{f,n}t} ~ e^{-( M_{f,n} - 2E_\gamma)t_{y}}\right]} + B_{f,0} ~ e^{M_{\bar{f}f }^{{\rm{PS}}} t} \int_{t_{\mathrm{cut}}}^{T/2} dt_y ~ e^{-( M_{\bar{f}f }^{{\rm{PS}}} - E_\gamma)t_{y}}\,.
\end{align}
Since for each value of the photon's energy, $E_\gamma=|\bs{k}|$, one has $E_{f,n}(\bs{k}) > E_{f,V}(\bs{k}) > E_\gamma$, the integral over $t_y$ in  Eq.\,(\ref{eq:scaling_C}) is always convergent and dominated by the time region where $t_y$ is close to $t$. Indeed, there are no intermediate states lighter than the energy of the external states. However, this is not always the case for the standard deviation $\sigma_{C_{f,W}^{\mu\nu ;2}}(t,\bs{k})$. When passing the threshold value $E_\gamma = M_{\bar{f}f}^{ {\rm{PS}}}$ the leading exponential contribution in Eq.\,(\ref{eq:scaling_varC}) (the term proportional to $B_{f,0}$) grows asymptotically with $t_y$ and is only regularized by the finite time extent $T$ of the lattice. In this case, from the leading exponential term in Eq.~(\ref{eq:scaling_varC}) one has that the divergent part of the error for $E_{\gamma} > M_{\bar{f}f}^{ {\rm{PS}}}$ is given by
\begin{align}
\label{eq:asymptotic_scaling_var}
\sigma_{C_{f,W}^{\mu\nu;2}}(t, \bs{k}) \simeq B_{f,0} ~ e^{M_{\bar{f}f }^{{\rm{PS}}} t} \int_{t}^{T/2} dt_y ~ e^{-( M_{\bar{f}f }^{{\rm{PS}}} - E_\gamma)t_{y}} 
 = B_{f,0}~\frac{e^{E_\gamma t}}{(E_\gamma-M_{\bar{f}f}^{{\rm{PS}}})}\left[ e^{(E_\gamma - M_{\bar{f}f}^{{\rm{PS}}})(T/2 -t )} - 1     \right] .
\end{align}
The prefactor $e^{E_{\gamma}t}$ in Eq.\,(\ref{eq:asymptotic_scaling_var}) is irrelevant since it does not contribute to $R^{\mu\nu}_{W}$ (see Eq.~(\ref{eq:Rinf})) and thus to the hadronic tensor $H^{\mu\nu}_{W}$.  \\

The reason behind the behaviour described by Eqs.~(\ref{eq:scaling_C})-(\ref{eq:asymptotic_scaling_var}) is that the kernel function $e^{E_{\gamma}t_y}$, accounting for the propagation of the photon, weights the different regions in $t_y$ in different ways, giving an exponential enhancement at large times $t_y$, which are therefore noisier. For real photon emission, the kernel $e^{E_{\gamma}t_y}$   never gives rise to a divergent integral in Eq.\,(\ref{eq:scaling_C}), since the propagating vector states have non-zero three-momentum $\bs{k}$ so that $E_{f,V}(\bs{k}) > E_{\gamma}$ (see Eq.\,(\ref{eq:M_scaling})). However, the states propagating in 
$\sigma^{2}_{M_{f,W}^{\mu\nu}}(t_y-t,t, \bs{k} )$ are at rest, and when $E_{\gamma} > M_{\bar{f}f}^{{\rm{PS}}}$, the leading exponential contribution proportional to $\int dt_{y} ~ e^{(E_{\gamma} - M_{\bar{f}f}^{{\rm{PS}}})t_{y}}$ in Eq.\,(\ref{eq:asymptotic_scaling_var}) becomes divergent in the limit $T \to\infty$.

\subsection{Numerical checks}
For the $D_{s}$ meson studied in this paper, the threshold value of $x_{\gamma} = 2E_{\gamma}/M_{D_{s}}$  above which the error starts to grow asymptotically is, according to Eq.\,(\ref{eq:asymptotic_scaling_var}), given by
\begin{align}
x_{\gamma}^{th} = 2\frac{M_{\eta_{ss'}}}{M_{D_{s}}} \simeq 0.7~. 
\end{align} 
For $x_{\gamma} > x_{\gamma}^{th}$ the error will  increase only in the contribution to $C^{\mu\nu}_{W}(t,\bs{k})$  where the photon is emitted from the strange quark, because for the emission from the charm quark, one has $M_{\bar{c}c}^{{\rm{PS}}} = M_{\eta_{cc'}} \simeq 3\,{\rm GeV}$, and the corresponding threshold value of $x_{\gamma}$ is well beyond the physical region explored $x_{\gamma} \leq 1$.

\begin{figure}
    \centering
    \includegraphics[scale=0.58]{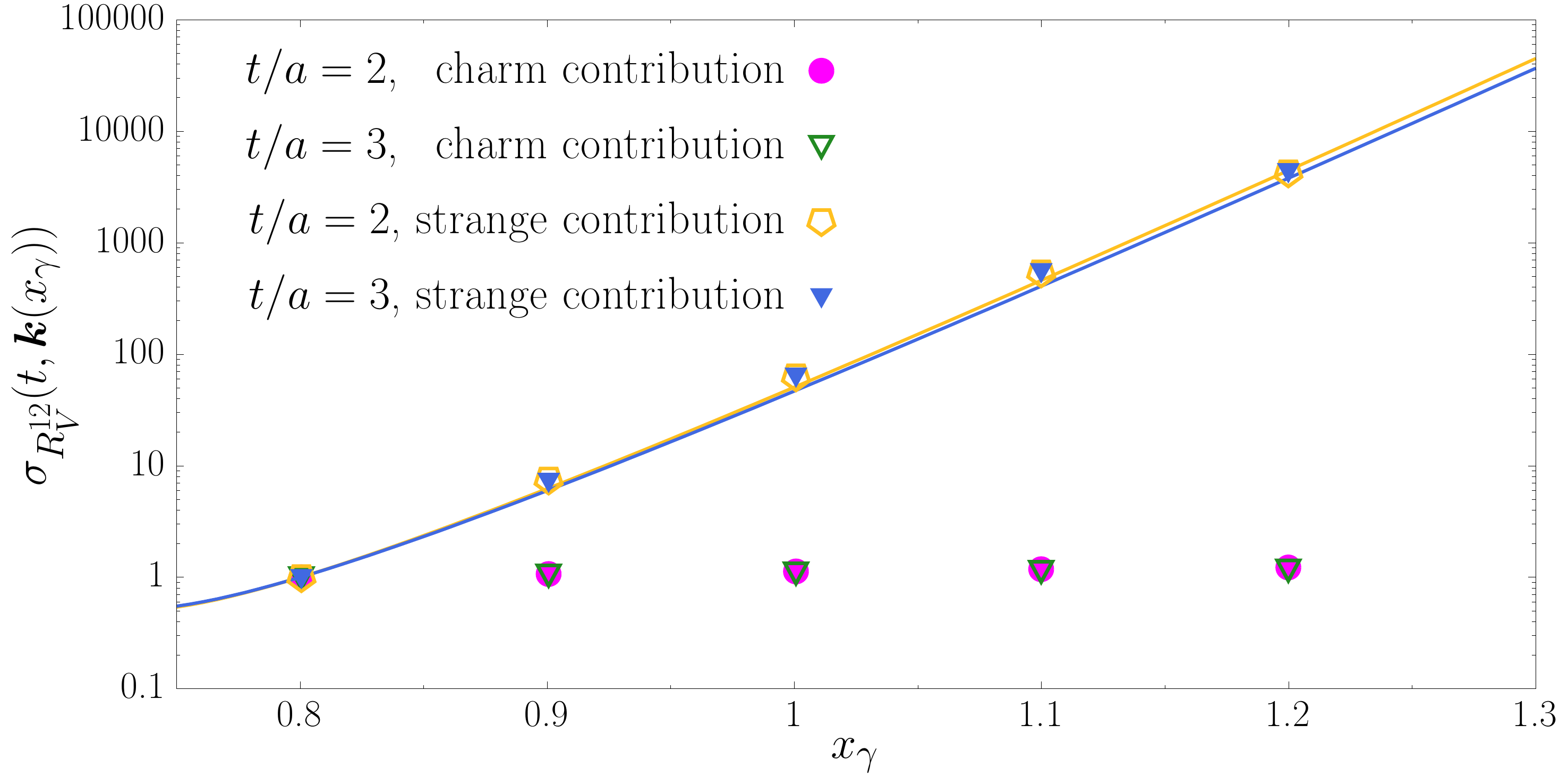}
    \caption{The statistical error on the strange- and charm-quark contribution to $R_{W}^{\mu\nu}(t,\bs{k})$, with $\mu=1, \nu=2$ and $W=V$, is plotted as a function of $x_{\gamma}$ for two different times $t/a = 2, 3$. The orange and blue curves correspond to the theoretical prediction of Eq.~(\ref{scaling_final_error}) with $A_{R_{W}^{\mu\nu}}=0$. Both the theoretical predictions and the numerical data have been rescaled in such a way that at $x_{\gamma} = 0.8$ the error is exactly one. }
    \label{fig1}
\end{figure}

The total error on the strange-quark contribution to $R^{\mu\nu}_{W}(t,\bs{k})\equiv R^{\mu\nu}_{W}(t,\bs{k},\bs{0})$ (see Eq.~(\ref{eq:redef_R})) for small times $t$ can be modelled as:
\begin{align}
\label{scaling_final_error}
\sigma _{R_{W}^{\mu\nu}}(t, \bs{k}) = A_{R_{W}^{\mu\nu}} + \frac{B_{R_{W}^{\mu\nu}}}{|E_{\gamma} - M_{\eta_{ss'}}|}e^{(E_{\gamma}- M_{\eta_{ss'}})(T/2-t)}~, 
\end{align}
where $A_{R_{W}^{\mu\nu}}$ is a \textit{background noise} term, which we take as being independent from $E_{\gamma}$. The contribution $A_{R_{W}^{\mu\nu}}$ to the noise arises from the non-divergent contributions to the noise in Eq.~(\ref{eq:scaling_C}) as well as those coming from the first time ordering, $t > t_y$, in Eq.\,(\ref{eq:CmunuApp}). Assuming that at $x_{\gamma} = 0.8$ the error is large enough such that the term $A_{R_{W}^{\mu\nu}}$ is negligible compared to the one proportional to $B_{R_{W}^{\mu\nu}}$, we can directly test Eq.~(\ref{scaling_final_error}) against our numerical data. This is shown in Fig.~(\ref{fig1}) where the error on the strange-quark contribution to $R^{12}_{V}(t, x_{\gamma}) \equiv R^{12}_{V}(t, \bs{k}(x_{\gamma}))$ is plotted as a function of $x_{\gamma} \geq 0.8$ for two different times  $t/a = 2,3$. As it is clear from the figure, the data are in remarkably good agreement with the theoretical prediction.\\

We conclude this appendix with a remark concerning the possibility of extending the calculation of $F_{V}$ and $F_{A}$  to the decays of $D_d$ and $B_u$ mesons over the full kinematical range, which is of even greater interest for phenomenology. In light of the above discussion, for these heavy-light mesons the threshold value of the photon energy $E_{\gamma}$ above which the errors will start to exhibit the exponential behaviour shown in Eq.\,(\ref{eq:asymptotic_scaling_var}), is given by the pion mass $M_{\pi}$. This means that intrinsically large statistical fluctuations are to be expected in $C^{\mu\nu}_{W}(t,\bs{k})$ already at very small values of $x_{\gamma}$. In this case, a possible step towards avoiding the S/N problem, consists in evaluating the integral over $t_{y}$ in Eq.\,(\ref{eq:CmunuApp}) on a reduced time interval $t_y \in [0 , t_\mathrm{cut} ]$, and then checking for convergence of the result as a function of $t_\mathrm{cut}$. In this way one can expect to avoid including in the integration large values of $t_{y}$ which do not contribute substantially to the signal (which is dominated by the region of times $t_{y}$ close to $t$) but which are responsible for the exponential increase of the error.

However, such approach requires the computation of the Euclidean three points function  
\begin{equation}
\label{eq:Mmunudef}
M_W^{\mu\nu}(t_y,t;\boldsymbol{k},\boldsymbol{p})=\bra{0} \hat{\mathrm{T}}\,[\hspace{1pt}j^\nu_W(t)\,\hat{j}^\mu_{\mathrm{em}}(t_y,\boldsymbol{k})\,\hat{\phi}^\dagger_{P}(0,\boldsymbol{p})]\ket{0}\,,
\end{equation}
for different values of $t$ and for all values $t_{y}$. In this way it is later possible to perform the integral over $t_{y}$ for each value of $t$. 
Computations at several values of $t$ need to be performed in order to verify that the ground state has been isolated.
This makes the approach more expensive than computing $C^{\mu\nu}_{W}$ in Eq.\,(\ref{eq:Cmunudef}) directly, which can be done for all values of $t$ at the cost of a single \textit{sequential propagator}. Whether such an extra cost for heavy-light mesons is offset by a significant improvement in accuracy remains to be seen. We plan to investigate this in the future.

\section{Results and correlation matrices for $F_{A}(x_{\gamma})$ and $F_{V}(x_{\gamma})$. }
\label{app:B}
\begin{eqnarray} && \qquad \qquad \qquad \qquad   \qquad \qquad \qquad \qquad \qquad   F_A \quad {\rm Correlation} \quad  {\rm  Matrix} \nonumber \\ &&\footnotesize{
\begin{tabular}{|c|c|c|}
\hline
 $x_\gamma$  & $F_A $  & $\Delta_{F_A} $\\ \hline
0.1  &  0.08129  &  0.00538 \\ \hline 
0.2  &  0.07153  &  0.00406 \\ \hline 
0.3  &  0.06408  &  0.00330 \\ \hline 
0.4  &  0.05824  &  0.00282 \\ \hline 
0.5  &  0.05337  &  0.00213 \\ \hline 
0.6  &  0.04953  &  0.00243 \\ \hline 
0.7  &  0.04626  &  0.00309 \\ \hline 
0.8  &  0.04325  &  0.00325 \\ \hline 
0.9  &  0.04332  &  0.00827 \\ \hline 
1.0  &  0.04893  &  0.02291 \\ \hline 
\end{tabular}
\quad \left(\begin{tabular}{c|c|c|c|c|c|c|c|c|c}
1.00000  &  0.93275  &  0.90319  &  0.87687  &  0.83165  &  0.76844  &  0.65813  &  0.39596  &  0.19820  &  0.05885  \\ 
0.93275  &  1.00000  &  0.92220  &  0.91694  &  0.86874  &  0.80104  &  0.65870  &  0.40917  &  0.19818  &  0.06649  \\ 
0.90319  &  0.92220  &  1.00000  &  0.91785  &  0.88262  &  0.82681  &  0.67405  &  0.43075  &  0.20483  &  0.04460  \\ 
0.87687  &  0.91694  &  0.91785  &  1.00000  &  0.89968  &  0.84780  &  0.67133  &  0.42214  &  0.22359  &  0.06804  \\ 
0.83165  &  0.86874  &  0.88262  &  0.89968  &  1.00000  &  0.86177  &  0.70806  &  0.46356  &  0.21540  &  0.04650  \\ 
0.76844  &  0.80104  &  0.82681  &  0.84780  &  0.86177  &  1.00000  &  0.78497  &  0.54501  &  0.27262  &  0.08763  \\ 
0.65813  &  0.65870  &  0.67405  &  0.67133  &  0.70806  &  0.78497  &  1.00000  &  0.65900  &  0.36929  &  0.16734  \\ 
0.39596  &  0.40917  &  0.43075  &  0.42214  &  0.46356  &  0.54501  &  0.65900  &  1.00000  &  0.60339  &  0.35453  \\ 
0.19820  &  0.19818  &  0.20483  &  0.22359  &  0.21540  &  0.27262  &  0.36929  &  0.60339  &  1.00000  &  0.59821  \\ 
0.05885  &  0.06649  &  0.04460  &  0.06804  &  0.04650  &  0.08763  &  0.16734  &  0.35453  &  0.59821  &  1.00000  \\

\end{tabular}\right) \nonumber
  }\end{eqnarray}

\begin{eqnarray} && \qquad \qquad \qquad \qquad   \qquad \qquad \qquad \qquad \qquad   F_V \quad {\rm Correlation} \quad  {\rm  Matrix} \nonumber \\ &&\footnotesize{
\begin{tabular}{|c|c|c|}
\hline
 $x_\gamma$  & $F_V $  & $\Delta_{F_V} $\\ \hline
0.1  &  -0.10483  &  0.00966 \\ \hline 
0.2  &  -0.08188  &  0.00284 \\ \hline 
0.3  &  -0.06428  &  0.00131 \\ \hline 
0.4  &  -0.05187  &  0.00085 \\ \hline 
0.5  &  -0.04307  &  0.00080 \\ \hline 
0.6  &  -0.03632  &  0.00078 \\ \hline 
0.7  &  -0.03157  &  0.00071 \\ \hline 
0.8  &  -0.02913  &  0.00101 \\ \hline 
0.9  &  -0.02966  &  0.00559 \\ \hline 
1.0  &  -0.03147  &  0.01521 \\ \hline 
\end{tabular}
\quad \left(\begin{tabular}{c|c|c|c|c|c|c|c|c|c}
1.00000  &  0.88397  &  0.76281  &  0.62931  &  0.41384  &  0.24756  &  0.20298  &  0.08518  &  -0.01028  &  -0.02871  \\ 
0.88397  &  1.00000  &  0.92314  &  0.76805  &  0.49098  &  0.31933  &  0.19618  &  0.08374  &  0.01045  &  -0.02545  \\ 
0.76281  &  0.92314  &  1.00000  &  0.88045  &  0.59188  &  0.43306  &  0.23283  &  0.09787  &  0.03337  &  -0.01943  \\ 
0.62931  &  0.76805  &  0.88045  &  1.00000  &  0.82870  &  0.61088  &  0.44651  &  0.20618  &  0.08193  &  0.02915  \\ 
0.41384  &  0.49098  &  0.59188  &  0.82870  &  1.00000  &  0.68718  &  0.59022  &  0.29135  &  0.13671  &  0.08414  \\ 
0.24756  &  0.31933  &  0.43306  &  0.61088  &  0.68718  &  1.00000  &  0.61811  &  0.38090  &  0.21946  &  0.14966  \\ 
0.20298  &  0.19618  &  0.23283  &  0.44651  &  0.59022  &  0.61811  &  1.00000  &  0.69088  &  0.45154  &  0.29618  \\ 
0.08518  &  0.08374  &  0.09787  &  0.20618  &  0.29135  &  0.38090  &  0.69088  &  1.00000  &  0.79170  &  0.56277  \\ 
-0.01028  &  0.01045  &  0.03337  &  0.08193  &  0.13671  &  0.21946  &  0.45154  &  0.79170  &  1.00000  &  0.75941  \\ 
-0.02871  &  -0.02545  &  -0.01943  &  0.02915  &  0.08414  &  0.14966  &  0.29618  &  0.56277  &  0.75941  &  1.00000  \\

\end{tabular}\right) \nonumber
  }\end{eqnarray}

\section{Analysis of single-pole parameterizations for $\mathbf{F_{A}(x_{\gamma})}$ and $\mathbf{F_{V}(x_{\gamma})}$}
\label{app:C}
Single-pole dominance, sometimes called Vector Meson Dominance (VMD) in the literature, is a model used to describe the momentum behavior of form factors, as determined by the propagation of the nearest internal resonance contributing to the amplitude.
In this appendix we fit the data from our non-perturbative lattice computation of $F_{A}(x_{\gamma})$ and $F_{V}(x_{\gamma})$, to check the validity of such a parameterization. 
After demonstrating that a pure VMD Ansatz is not consistent with our results for the form factors, we propose a simple extension of the parameterization that can fit our data with good precision over the whole kinematical range. This parameterization provides a simple and practical description of our data which can be used in future phenomenological analyses.
We also check the stability of the fitted values of the residues of the singular pole terms in the parameterization (i.e. the $C_W$ in Eq.(\ref{eq:ansatz})). This is a necessary test in order to assess the validity of relating the fitted $C_W$ to the $D_s^*\to D_s \gamma$ and $D_{s1}\to D_s \gamma$ decay amplitudes for the vector and axial channels respectively, as predicted by single-pole dominance.

VMD predictions are obtained by inserting a sum over intermediate states between the two operators in the correlation function defining $H_W^{\mu\nu}(k, \bs{p})$ in Eq.\,(\ref{eq:H_munu}) and approximating this sum by the contribution from the nearest state. 
This approximation is also appropriately called single-pole dominance. 
For the vector and axial components of the weak current the nearest internal states are the $D_s^*$ and $D_{s1}$ mesons respectively, contributing to the time-ordering in which the electromagnetic current acts on the initial meson $D_s$ at an earlier time than that at which the weak current is inserted, i.e. $t_y<0$ in Eq.\,(\ref{eq:Cmunudef}). 
Thus, by assuming single-pole dominance, we obtain the following parameterization for the momentum behaviour of the form factors\,\footnote{Here and in the following, we employ the reference frame in which the initial $D_s$ meson is at rest.}:
\bea
F_A(\bs{k})&=&\frac{C_A'}{E_{D_{s1}}(\bs{k})\left(E_{D_{s1}}(\bs{k})+E_\gamma-E\right)}=\frac{C_A'}{\sqrt{M_{D_{s1}}^2+|\bs{k}|^2}\left(\sqrt{M_{D_{s1}}^2+|\bs{k}|^2}+|\bs{k}|-M_{D_s}\right)}\label{eq:FA_vmd}\,,\\
F_V(\bs{k})&=&\frac{C_V'}{E_{D_{s}^*}(\bs{k})\left(E_{D_{s}^*}(\bs{k})+E_\gamma-E\right)}=\frac{C_V'}{\sqrt{M_{D_{s}^*}^2+|\bs{k}|^2}\left(\sqrt{M_{D_{s}^*}^2+|\bs{k}|^2}+|\bs{k}|-M_{D_s}\right)}\label{eq:FV_vmd}\,,
\eea
where $M_{D_{s1}}$ and $M_{D^*_{s}}$ are the masses of the $D^{\mbox{}}_{s1}$ and $D^\ast_s$ mesons respectively and
$C_A'$ and $C_V'$ are constant coefficients with the dimension of energy squared.  
By dividing both the numerator and denominator of Eqs.\,(\ref{eq:FA_vmd}) and (\ref{eq:FV_vmd}) by $M^2_{D_s}$ and expressing $|\bs{k}|$ in terms of $x_\gamma$ as $|\bs{k}|=\frac{x_\gamma M_{D_s}}{2}$, we obtain
\bea
F_A(x_\gamma)&=&\frac{C_A}{\sqrt{R_{D_{s1}}^2+\dfrac{x_\gamma^2}{4}}\left(\sqrt{R_{D_{s1}}^2+\dfrac{x_\gamma^2}{4}}+\dfrac{x_\gamma}{2}-1\right)}\label{eq:FA_vmdXg}\,,\\
F_V(x_\gamma)&=&\frac{C_V}{\sqrt{R_{D_{s}^*}^2+\dfrac{x_\gamma^2}{4}}\left(\sqrt{R_{D_{s}^*}^2+\dfrac{x_\gamma^2}{4}}+\dfrac{x_\gamma}{2}-1\right)}\label{eq:FV_vmdXg}\,,
\eea
where $C_{\{A,V\}}=\frac{C_{\{A,V\}}'}{M^2_{D_s}}$ and $R_{\{D_{s1},D_{s}^*\} }=\frac{M_{\{D_{s1},D_{s}^*\} }}{M_{D_s}}$ are now dimensionless parameters.
\begin{figure} 
 \hspace{-0.5cm}
	\subfloat{%
		\includegraphics[scale=0.38]{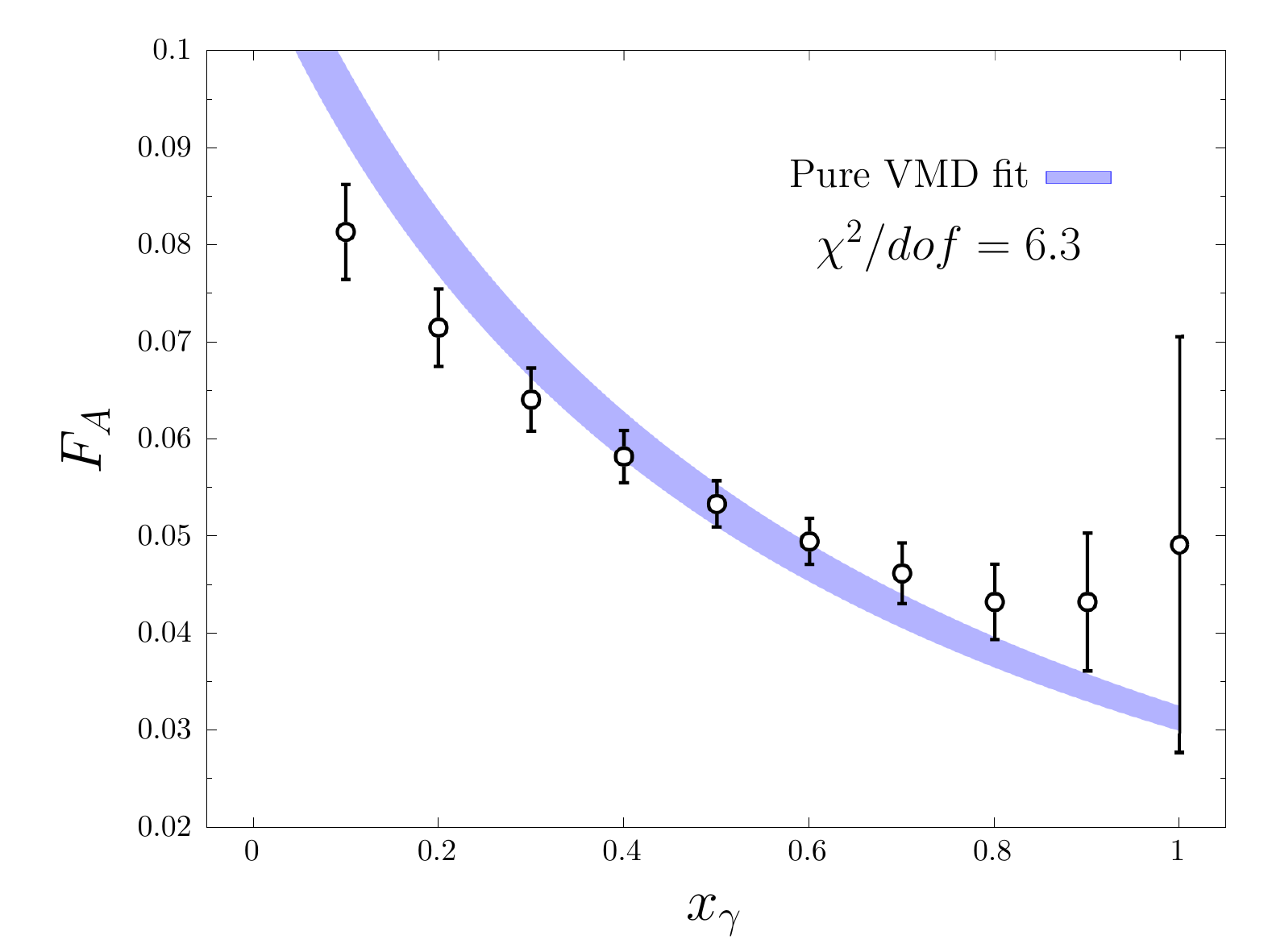}}
	\subfloat{%
		\includegraphics[scale=0.38]{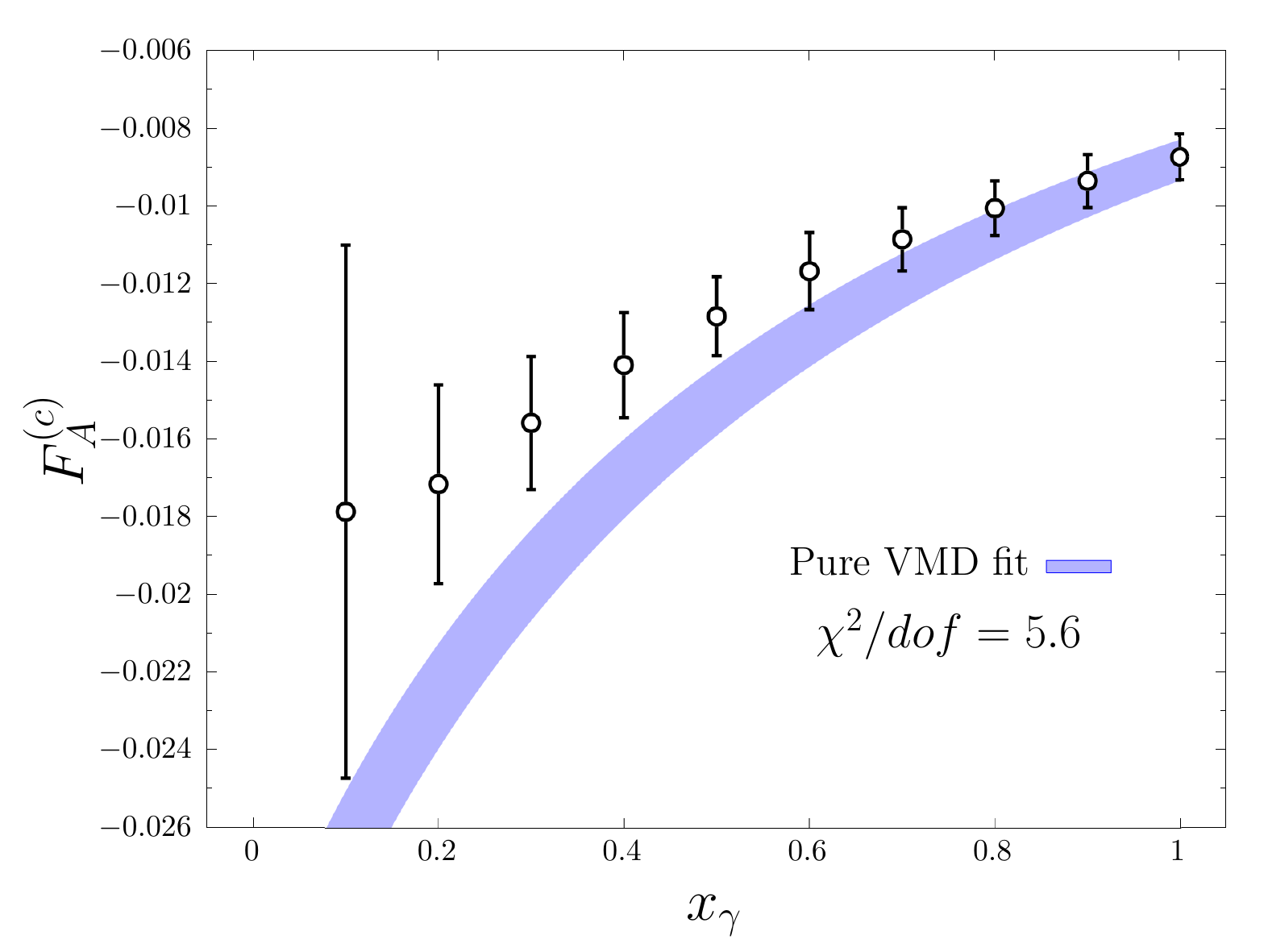}}
	\subfloat{%
		\includegraphics[scale=0.38]{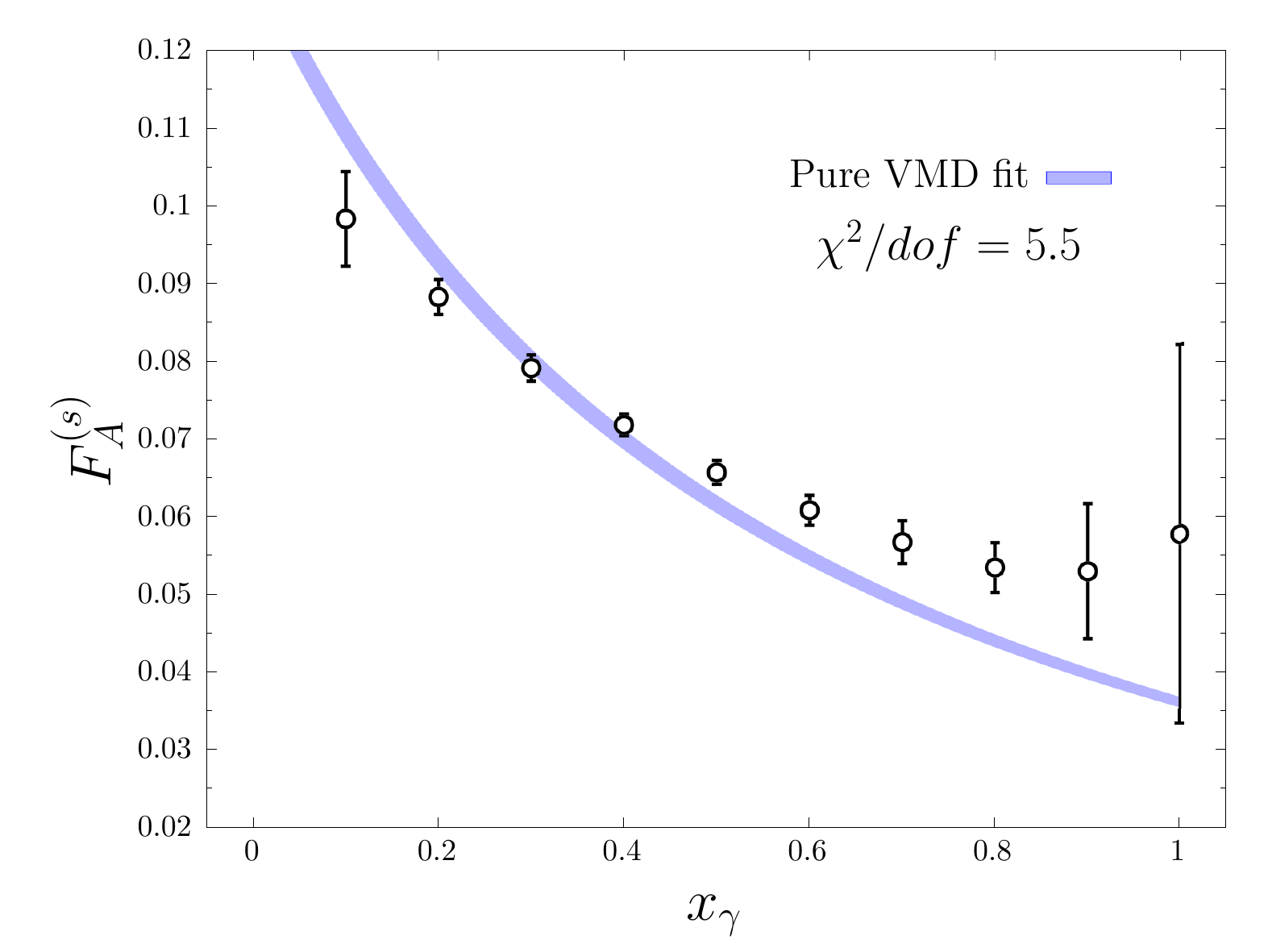}} \\
\hspace{-0.5cm}
	\subfloat{%
		\includegraphics[scale=0.38]{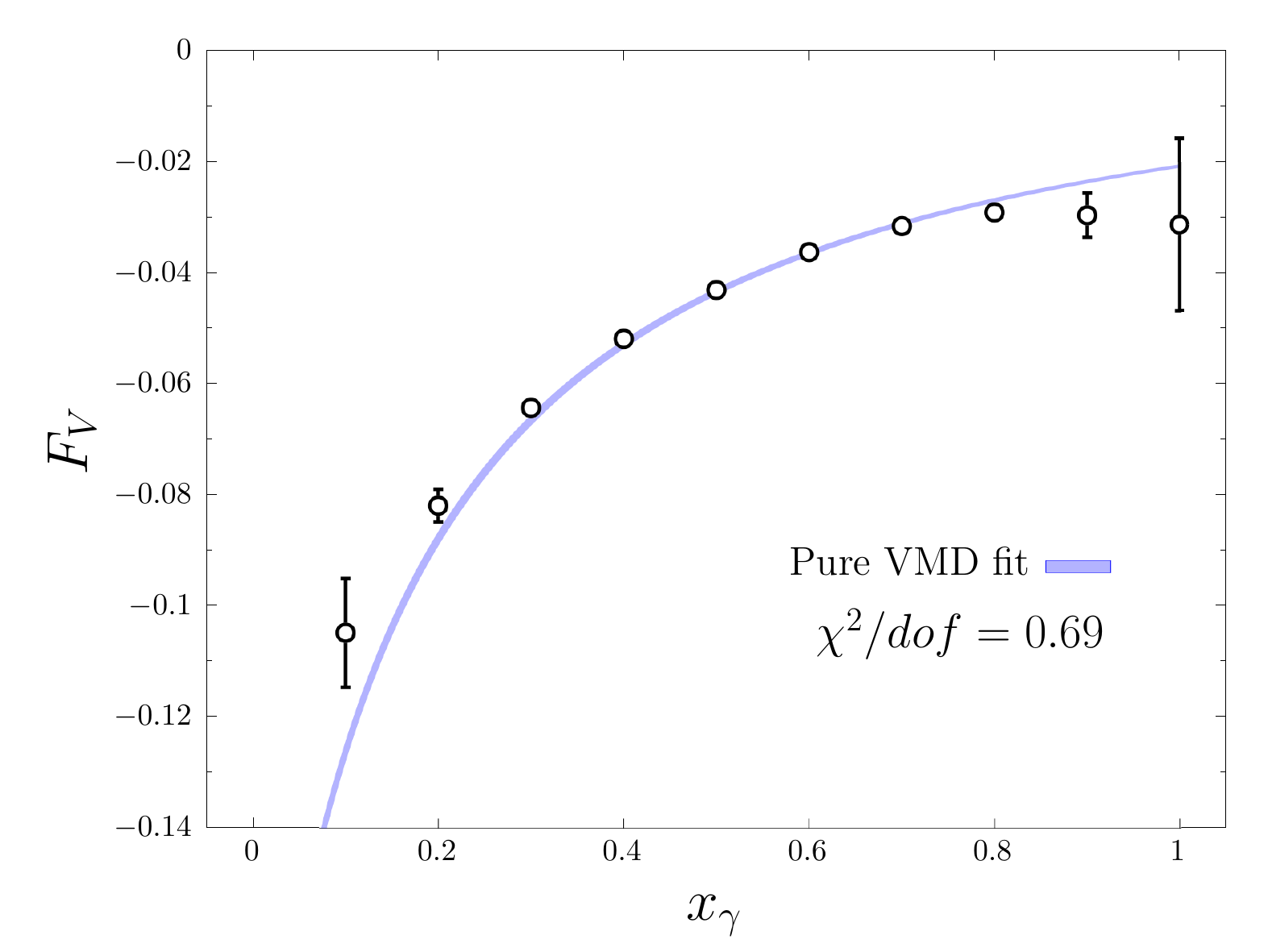}}
	\subfloat{%
		\includegraphics[scale=0.38]{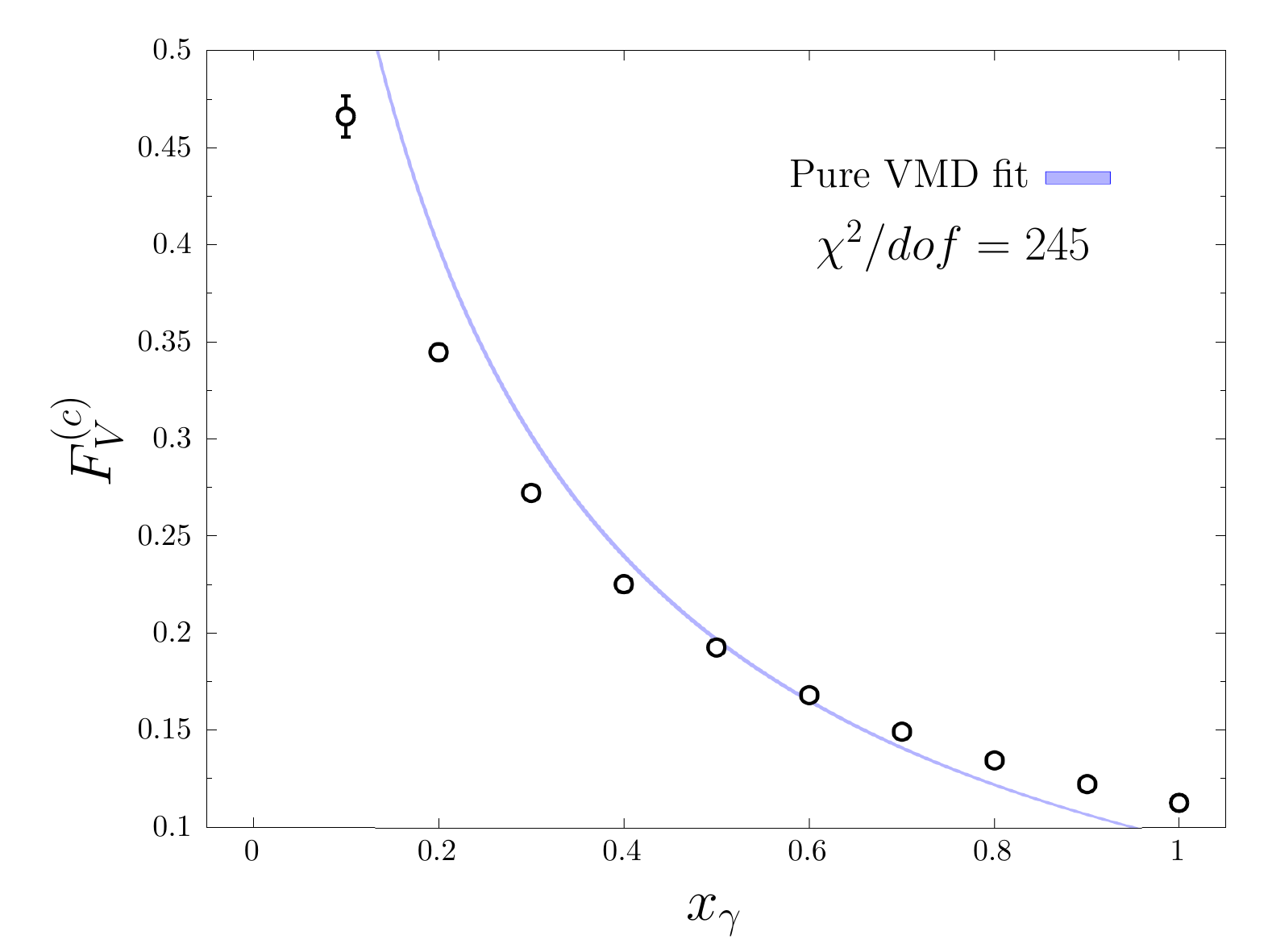}}
	\subfloat{%
		\includegraphics[scale=0.38]{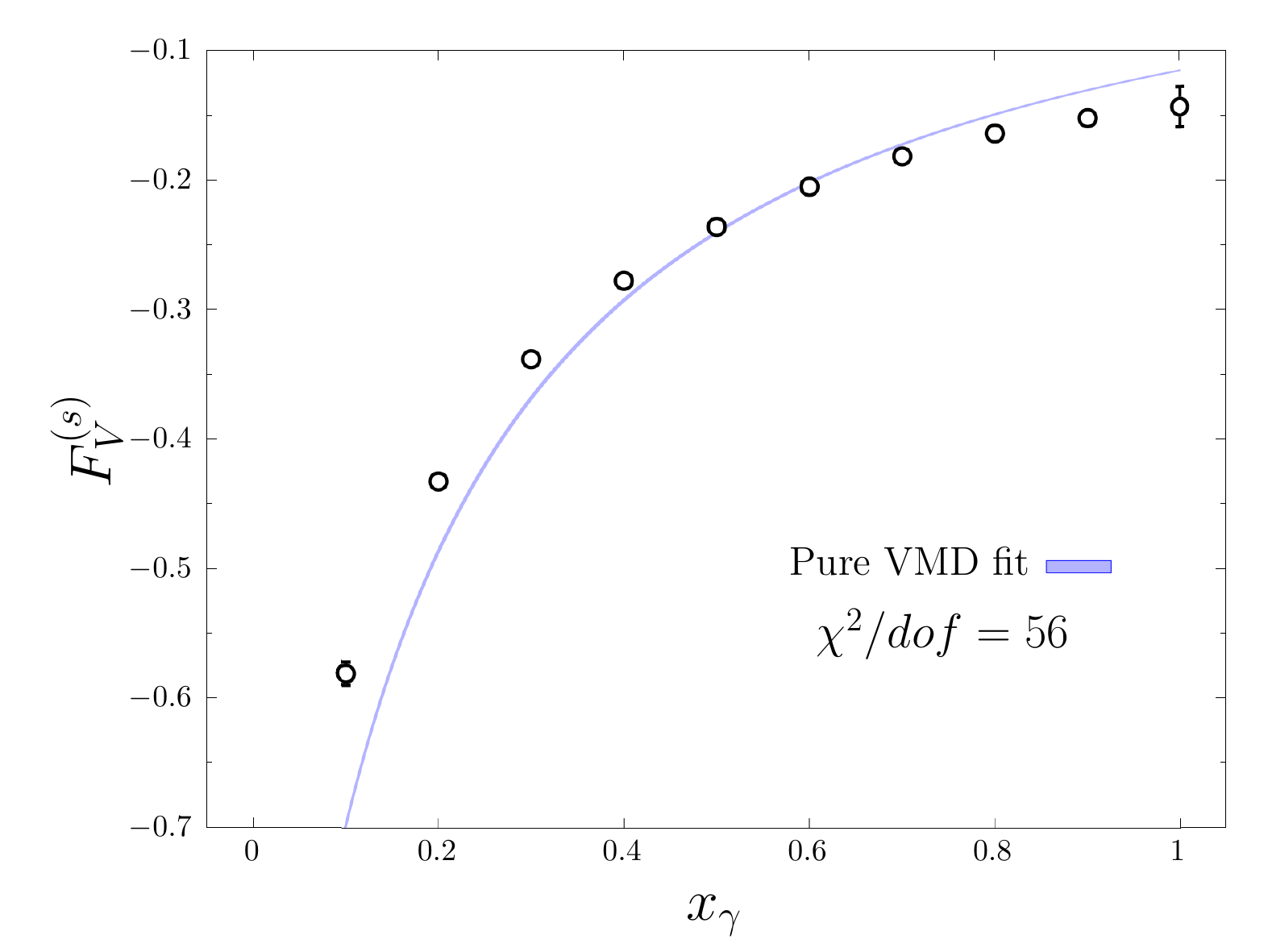}}
    \caption{The fit functions, corresponding to the pure VMD Ansatz of Eqs.\,(\ref{eq:FA_vmdXg})-(\ref{eq:FV_vmdXg}), are plotted, along with the lattice data, for the axial channel (top panels), and for the vector channel (bottom panels).}
    \label{fig:pure_vmd_fit}
\end{figure}
By inserting the values $M_{D_{s1}}=2460$\,MeV, $M_{D_{s}^*}=2112$\,MeV and $M_{D_s}=1968$\,MeV, taken from the PDG\,\cite{Workman:2022ynf}, we obtain the ratios $R_{D_{s1}}=1.25$ and $R_{D_{s}^*}= 1.073$.
Thus, Eqs.\,(\ref{eq:FA_vmdXg}) and (\ref{eq:FV_vmdXg}) describe the momentum dependence, as predicted by VMD, for the axial and vector form factors, and also for the contributions corresponding to the emission of the photon from the charm and strange quarks separately.
In order to check the validity of the VMD prediction, we fit our lattice results to the parameterization of Eqs.\,(\ref{eq:FA_vmdXg}) and (\ref{eq:FV_vmdXg}), with $C_{A}$ and $C_{V}$ taken as free parameters to be determined from the fit\footnote{All the fits of this analysis have been performed by minimizing the \emph{correlated} $\chi^2$.}.
The results of these fits are shown in the plots in Fig.\,\ref{fig:pure_vmd_fit}. 
It is clear from the figure, and confirmed by the corresponding high values of $\chi^2/$d.o.f., that the results of the fit are in general very poor. The only exception is the VMD fit for the total vector form factor $F_V(x_\gamma)$, which proves to be a good fit of our data.

Since the pure VMD fit, with only one free parameter, fails to describe our lattice results for the form factors, we now introduce a more general Ansatz; one that represents the Laurent expansion of a function around a pole:
\bea
 F_{W}(x_\gamma)=\frac{C_{W}}{\sqrt{R_{{W}}^2+\dfrac{x_\gamma^2}{4}}\left(\sqrt{R_{{W}}^2+\dfrac{x_\gamma^2}{4}}+\dfrac{x_\gamma}{2}-1\right)}+B_{W}+D_{W}\,x_\gamma\,,\label{eq:laurent}
\eea
where we have included corrections up to linear terms in $x_\gamma$, with $C_{W}$, $R_{W}$, $B_{W}$ and $D_{W}$ being free parameters to be determined from the fit.
In Eq.\,(\ref{eq:laurent}), the difference of the parameters $R_{W}$ from the VMD values $R_{D_{s1}}=1.25$ and $R_{D_{s}^*}= 1.073$, 
partially accounts for the contributions from heavier internal states. The free parameters $B_{W}$ and $D_{W}$, that describe the first two non-singular terms of the Laurent expansion of a function around a pole, are also expected to encode non-negligible contributions that are not included in the pure VMD description.
\begin{table}[]
    \centering
    \begin{tabular}{ c| c| c| c|c|c}
    \hline
     \multicolumn{4}{c}{\rule[-2mm]{0mm}{15pt} $F_A$ fitted parameters}  \\
     \hline 
     \rule[-2mm]{0mm}{15pt} & $C_A$ & $R_A$ & $B_A$&$D_A$&$\chi^2/$d.o.f.\\
     \hline
      \rule[-2mm]{0mm}{15pt} $\ CR$ fit & $\ 0.0518(30)\ $ &  $\ 1.413(30)\ $ & $\ 0\ $ (fixed) & $\ 0\ $ (fixed)& $\ 0.41\ $ \\
      \hline
    \rule[-2mm]{0mm}{15pt} $\ CRB$ fit & $\ 0.0229(76)\ $ &  $\ 1.242(59)\ $ & $\ 0.0185(67)\ $& $\ 0\ $ (fixed)&$\ 0.04\ $ \\
      \hline
    \rule[-2mm]{0mm}{15pt} $\ CB$ fit & $\ 0.0239(20)\ $ &  $\ 1.25\ $ (fixed) &  $\ 0.0176(18) $&$\ 0\ $ (fixed&$\ 0.03\ $ \\
      \hline
      \rule[-2mm]{0mm}{15pt} $\ CBD$ fit & $\ 0.0246(51)\ $ &  $\ 1.25\ $ (fixed)& $\ 0.016(11)\ $& $\ 0.002(10)\ $ & $\ 0.04\ $ \\
      \hline
   \hline
     \multicolumn{4}{c}{\rule[-2mm]{0mm}{17pt}$F_A^{(c)}$ fitted parameters}  \\
    \hline 
     \rule[-2mm]{0mm}{15pt} & $C_A$ & $R_A$ & $B_A$&$D_A$&$\chi^2/$d.o.f.\\
      \hline
      \rule[-2mm]{0mm}{15pt} $\ CR$ fit & $\ -0.0135(10)\ $ &  $\ 1.453(59)\ $ & $\ 0\ $ (fixed) & $\ 0\ $ (fixed)& $\ 0.23\ $ \\
      \hline
    \rule[-2mm]{0mm}{15pt} $\ CRB$ fit & $\ -0.0075(26)\ $ &  $\ 1.27(11)\ $ & $\ -0.0025(14)\ $& $\ 0\ $ (fixed)&$\ 0.08\ $ \\
      \hline
    \rule[-2mm]{0mm}{15pt} $\ CB$ fit & $\ -0.00696(88)\ $ &   $\ 1.25\ $ (fixed) &  $\ -0.00280(47)\ $& $\ 0\ $ (fixed)&$\ 0.07\ $ \\
    \hline
    \rule[-2mm]{0mm}{15pt} $\ CBD$ fit & $\ -0.0068(23)\ $ &  $\ 1.25\ $ (fixed) &  $\ -0.0030(37)\ $&$\ 0.0002(22)\ $ &$\ 0.08\ $ \\
      \hline
      \hline
     \multicolumn{4}{c}{\rule[-2mm]{0mm}{17pt}$F_A^{(s)}$ fitted parameters}  \\
    \hline 
     \rule[-2mm]{0mm}{15pt} & $C_A$ & $R_A$ & $B_A$&$D_A$&$\chi^2/$d.o.f.\\
          \hline
      \rule[-2mm]{0mm}{15pt} $\ CR$ fit & $\ 0.0662(56)\ $ &  $\ 1.423(36)\ $ & $\ 0\ $ (fixed) & $\ 0\ $ (fixed)& $\ 0.24\ $ \\
      \hline
    \rule[-2mm]{0mm}{15pt} $\ CRB$ fit & $\ 0.031(17)\ $ &  $\ 1.26(10)\ $ & $\ 0.021(13)\ $& $\ 0\ $ (fixed)&$\ 0.09\ $ \\
      \hline
    \rule[-2mm]{0mm}{15pt} $\ CB$ fit & $\ 0.0298(18)\ $ & $\ 1.25\ $ (fixed)  &  $\ 0.0215(31)\ $ &$\ 0\ $ (fixed)&$\ 0.08\ $ \\
      \hline
    \rule[-2mm]{0mm}{15pt} $\ CBD$ fit & $\ 0.0309(87)\ $ &  $\ 1.25\ $ (fixed)  &  $\ 0.018(22)\ $ &$\ 0.003(18)\ $ &$\ 0.08\ $ \\
      \hline
      \hline
     \multicolumn{4}{c}{\rule[-2mm]{0mm}{15pt}$F_V$ fitted parameters}  \\
    \hline 
     \rule[-2mm]{0mm}{15pt} & $C_V$ & $R_V$ & $B_V$&$D_V$&$\chi^2/$d.o.f.\\
          \hline
      \rule[-2mm]{0mm}{15pt} $\ CR$ fit & $\ -0.01792(76)\ $ &  $\ 1.091(11)\ $ & $\ 0\ $ (fixed) & $\ 0\ $ (fixed)& $\ 0.45\ $ \\
      \hline
    \rule[-2mm]{0mm}{15pt} $\ CRB$ fit & $\ -0.0193(23)\ $ &  $\ 1.100(19)\ $ & $\ 0.0018(28)\ $& $\ 0\ $ (fixed)&$\ 0.47\ $ \\
      \hline
    \rule[-2mm]{0mm}{15pt} $\ CB$ fit & $\ -0.01619(58)\ $ &  $\ 1.073\ $ (fixed)&  $\ -0.0017(16)\ $&$\ 0\ $ (fixed) &$\ 0.66\ $ \\
      \hline
      \rule[-2mm]{0mm}{15pt} $\ CBD$ fit & $\ -0.0153(16)\ $ &  $\ 1.073\ $ (fixed) & $\ -0.0064(77)\ $& $\ 0.0045(71)\ $ & $\ 0.75\ $ \\
      \hline
      \hline
     \multicolumn{4}{c}{\rule[-2mm]{0mm}{17pt}$F_V^{(c)}$ fitted parameters}  \\
    \hline 
     \rule[-2mm]{0mm}{15pt} & $C_V$ & $R_V$ & $B_V$&$D_V$&$\chi^2/$d.o.f.\\
          \hline
      \rule[-2mm]{0mm}{15pt} $\ CR$ fit & $\ 0.1144(13)\ $ &  $\ 1.2001(41)\ $ & $\ 0\ $ (fixed) & $\ 0\ $ (fixed)& $\ 53\ $ \\
      \hline
    \rule[-2mm]{0mm}{15pt} $\ CRB$ fit & $\ 0.0624(15)\ $ &  $\ 1.0809(43)\ $ & $\ 0.0369(14)\ $& $\ 0\ $ (fixed)&$\ 0.31\ $ \\
      \hline
    \rule[-2mm]{0mm}{15pt} $\ CB$ fit & $\ 0.05971(55)\ $ &  $\ 1.073\ $ (fixed) &  $\ 0.03886(82)\ $&$\ 0\ $ (fixed)&$\ 0.76\ $ \\
      \hline
      \rule[-2mm]{0mm}{15pt} $\ CBD$ fit & $\ 0.0579(13)\ $ &  $\ 1.073\ $ (fixed) & $\ 0.0466(45)\ $& $\ -0.0055(33)\ $ & $\ 0.43\ $ \\
      \hline
      \hline
     \multicolumn{4}{c}{\rule[-2mm]{0mm}{17pt}$F_V^{(s)}$ fitted parameters}  \\
    \hline 
     \rule[-2mm]{0mm}{15pt} & $C_V$ & $R_V$ & $B_V$&$D_V$&$\chi^2/$d.o.f.\\
          \hline
      \rule[-2mm]{0mm}{15pt} $\ CR$ fit & $\ -0.1099(11)\ $ &  $\ 1.1245(29)\ $ & $\ 0\ $ (fixed) & $\ 0\ $ (fixed)& $\ 11\ $ \\
      \hline
    \rule[-2mm]{0mm}{15pt} $\ CRB$ fit & $\ -0.0792(24)\ $ &  $\ 1.0794(37)\ $ & $\ -0.0367(31)\ $& $\ 0\ $ (fixed)&$\ 1.8\ $ \\
      \hline
    \rule[-2mm]{0mm}{15pt} $\ CB$ fit & $\ -0.07571(96)\ $ &  $\ 1.073\ $ (fixed) &  $\ -0.0410(17)\ $&$\ 0\ $ (fixed)&$\ 1.8\ $ \\
      \hline
      \rule[-2mm]{0mm}{15pt} $\ CBD$ fit & $\ -0.0759(16)\ $ &  $\ 1.073\ $ (fixed) & $\ -0.0399(69)\ $& $\ -0.0011(67)\ $ & $\ 2.1\ $ \\
      \hline
    \end{tabular}
    \caption{Values of the fit parameters for each form factor, and for their individual charm and strange quark contributions, as determined from the different fits based on the Ansatz of Eq.\,(\ref{eq:laurent}).}
    \label{tab:fitparameters}
\end{table}
\begin{table}[]
    \centering
    \begin{tabular}{c| c| c| c| c}
     \hline 
     \rule[-2mm]{0mm}{15pt} & $CR$ fit
 & $CRB$ fit
 &$CB$ fit & $CBD$ fit \\
     \hline
      \rule[-3mm]{0mm}{20pt} $g^{\mbox{}}_{D_s^*D_s\gamma}~[\textrm{GeV}^{-1}]$ & $\ 0.1223(51) \ $ & $\ 0.130(14)\ $ &  $\ 	
0.1123(49)\ $&  $\ 0.106(11)\ $ \\
           \hline
      \rule[-3mm]{0mm}{20pt} $g^{(s)}_{D_s^*D_s\gamma}\ [\textrm{GeV}^{-1}]$ & \line(1,0){10} & $\ 0.546(20)\ $ & $\ 0.525(15)\ $ &  $\ 0.526(17)\ $  \\
           \hline
      \rule[-3mm]{0mm}{20pt} $g^{(c)}_{D_s^*D_s\gamma}\ [\textrm{GeV}^{-1}]$ & \line(1,0){10} & $\ -0.429(14)\ $ &  $\ -0.414(11)\ $&  $\ -0.402(13)\ $ \\
           \hline
    \end{tabular}
    \caption{Predictions for the $g_{D_s^*D_s\gamma}$ coupling, and for its individual charm and strange contributions, as obtained from our various fits, based on the Ansatz of Eq.\,(\ref{eq:laurent}). Only results from fits with a low value of $\chi^2$/d.o.f. have been included.}
    \label{tab:g2}
\end{table}
In order to check for the stability of the residues $C_A$ and $C_V$, and hence their interpretation in terms of the $D_{s1}^{\mbox{}}\to D_s\gamma$ and $D^\ast_{s}\to D_s\gamma$ decay amplitudes, we have performed several fits of our data, based on the Ansatz of Eq.\,(\ref{eq:laurent}), fixing on each occasion some of its parameters to their VMD values\,
\footnote{With the limited number of points for which we have results, it is not possible to perform fits with all four parameters left free to be determined by the fits as this results in overfitting.}. 
Specifically, the $CR$ fit is performed by fixing $B_W=D_W=0$, while the $CRB$ fit is performed by setting only $D_W=0$. 
The $CB$ fit is obtained by fixing $D_W=0$ and $R_W=R_{D_s^*}$ for the vector channel and $R_W=R_{D_{s1}}$ for the axial one.
Finally the $CBD$ fit, obtained by setting $R_W=R_{D_s^*}$ for the vector channel and $R_W=R_{D_{s1}}$ for the axial one, with the remaining 3 parameters determined by the fits. The results of these fits, for each form factor and for their individual charm and strange contributions, are reported in Tab.\,\ref{tab:fitparameters}.

From the table, we see that all the different Ans\"atze provide an adequate fit to our data for the axial channel, but we notice that the fitted values for the residue of the singular term, $C_A$, obtained from the $CR$ fit are very different from those obtained from the fits with $B_A$ as a free parameter. Although all three fits with $B_A$ as a free parameter give consistent results for $C_A$, we avoid relating these results to the $D_{s1}\to D_s \gamma$ decay amplitude for two reasons. Firstly, because the value
of $C_A$ obtained from the $CR$ fit, which is also a good fit to our data, is very different.
Secondly, because of the presence in the axial channel, of another resonance, namely the $D_{s1}(2536)$ meson, with a mass which is only slightly above the nearest resonance, i.e. the $D_{s1}(2460)$ meson. Since the 76\,MeV difference between the masses of the two resonances is so small, the fitted amplitude $C_A$ could encode contributions coming from both of these internal states, resulting in an unreliable determination of the coupling $g_{D_{s1} D_s \gamma}$.

In the vector channel, we note that when fitting our results for $F_V^{(s)}$ and $F_V^{(c)}$ we need to include the presence of the constant terms $B_V^{(s)}$ and $B_V^{(s)}$ to obtain low values of $\chi^2/$d.o.f.\,. 
However, we find that in the sum of the charm and strange-quark contributions to $F_V$, the individual non-singular terms cancel almost exactly, i.e. $B_V^{(c)}\simeq-B_V^{(s)}$. As a result, the simple single-pole Ansatz, with $B_V$ and $D_V$ fixed to zero, already provides a good fit of the data for the total vector form factor $F_V$.
In the fits with a low value of $\chi^2/$d.o.f., i.e. all the fits except the $CR$ fit to $F_V^{(s)}$ and $F_V^{(c)}$, the value of $R_V$ is remarkably close to that from the pure VMD ansatz, i.e. $R_{D^*_s}=1.073$, differing by less than 3\%, and the values of $C_V$ are all very similar. 
The similarities between the values of $R_V$ and $R_{D_s^*}$, and the stability of the values of the $C_V$ parameter allow us to relate the fitted value for the residue of the pole term, $C_V$, to the $D^*_{s}\to D \gamma$ decay amplitude, and to its characteristic coupling $g_{D_s^*D_s\gamma}$, using the pole-dominance relation
\bea
C_V=-\frac{M_{D_s^*}f_{D^*_s}g_{D_s^*D_s\gamma}}{2M_{Ds}}\,.
\eea
In order to determine $g_{D_s^*D_s\gamma}$ and its individual strange and charm-quark contributions, we use the values of $C_V$ from the fits with low values of $\chi^2$/d.o.f. and take the value $f_{D_s^*}=268.8(6.6)$\,MeV from the lattice calculation of Ref.\,\cite{Lubicz:2017}.  The corresponding estimates of the couplings are reported in Table\,\ref{tab:g}.
Averaging the values of the couplings reported in the table, we obtain our final estimates
\bea
g_{D_s^*D_s\gamma}&=&0.1177\pm0.0048_{stat}\pm0.012_{syst}\ \ \textrm{GeV}^{-1}\,,\\
g^{(s)}_{D_s^*D_s\gamma}&=&0.532\pm 0.010_{stat}\pm0.011_{syst}\ \ \textrm{GeV}^{-1}\,,\\
g^{(c)}_{D_s^*D_s\gamma}&=&-0.4150\pm 0.0073_{stat}\pm0.014_{syst}\ \ \textrm{GeV}^{-1}\,,
\eea
where we include half of the maximum dispersion among the values obtained from the different fits as the systematic 
uncertainty. 
In Sec.\,\ref{sec:phenomenology}, we compare these results to two previous estimates of the same quantities, obtained either through a direct lattice computation \cite{Donald:2014} or by using LCSR at next-to-leading order \cite{Pullin:2021ebn}.

\end{document}